\documentclass[prd,aps,a4paper,eqsecnum,nofootinbib,floatfix,twocolumn,superscriptaddress]{revtex4-2}  

\usepackage{graphicx} 
\usepackage[hyperindex,breaklinks]{hyperref}
\usepackage{placeins}
\usepackage{amsmath,amsfonts,amssymb,bm}
\usepackage{tikz}
\usepackage{appendix}
\usepackage{physics}
\usepackage[autostyle]{csquotes}
\usepackage[sort&compress]{natbib}
\usepackage[caption=false,justification=justified]{subfig}
\usepackage[compat=1.1.0]{tikz-feynman}

\newcommand{\beq}{\begin{equation}}
\newcommand{\eeq}{\end{equation}}
\newcommand{\bea}{\begin{eqnarray}}
\newcommand{\eea}{\end{eqnarray}}

\newcommand{\be}{\begin{equation}}      
\newcommand{\ee}{\end{equation}}

\def\nn{\nonumber}

\renewcommand{\quote}[1]{``#1''}

\begin{document}

\title{Hamiltonian Neural Networks approach to fuzzball geodesics}

\author{Andrea Cipriani}
\email{andrea.cipriani@roma2.infn.it}
\affiliation{Dipartimento di Fisica, Universit\`a di Roma Tor Vergata, Via della Ricerca Scientifica 1, 00133 Roma, Italia}
\affiliation{Sezione INFN Roma 2, Via della Ricerca Scientifica 1, 00133, Roma Italia}

\author{Alessandro De Santis}
\email{desantia@uni-mainz.de}
\affiliation{Sezione INFN Roma 2, Via della Ricerca Scientifica 1, 00133, Roma Italia}
\affiliation{Helmholtz-Institut Mainz, Johannes Gutenberg-Universit{\"a}t Mainz, 55099 Mainz, Germany}
\affiliation{GSI Helmholtz Centre for Heavy Ion Research, 64291 Darmstadt, Germany}

\author{Giorgio Di Russo}
\email{gdr794@ucas.ac.cn}
\affiliation{Sezione INFN Roma 2, Via della Ricerca Scientifica 1, 00133, Roma Italia}
\affiliation{School of Fundamental Physics and Mathematical Sciences, Hangzhou Institute for Advanced Study, UCAS, Hangzhou 310024, China}

\author{Alfredo Grillo}
\email{alfredo.grillo89@gmail.com}
\affiliation{Dipartimento di Fisica, Universit\`a di Roma Tor Vergata, Via della Ricerca Scientifica 1, 00133 Roma, Italia}
\affiliation{Sezione INFN Roma 2, Via della Ricerca Scientifica 1, 00133, Roma Italia}

\author{Luca Tabarroni}
\email{luca.tabarroni@roma2.infn.it}
\affiliation{Dipartimento di Fisica, Universit\`a di Roma Tor Vergata, Via della Ricerca Scientifica 1, 00133 Roma, Italia}
\affiliation{Sezione INFN Roma 2, Via della Ricerca Scientifica 1, 00133, Roma Italia}

\date{\today}

\begin{abstract}
The recent increase in computational resources and data availability has led to a significant rise in the use of Machine Learning (ML) techniques for data analysis in physics. However, the application of ML methods to solve differential equations capable of describing even complex physical systems is not yet fully widespread in theoretical high-energy physics. Hamiltonian Neural Networks (HNNs) are tools that minimize a loss function defined to solve Hamilton equations of motion. In this work, we implement several HNNs trained to solve, with high accuracy, the Hamilton equations for a massless probe moving inside a smooth and horizonless geometry known as D1-D5 circular fuzzball. We study both planar (equatorial) and non-planar geodesics in different regimes according to the impact parameter, some of which are unstable. Our findings suggest that HNNs could eventually replace standard numerical integrators, as they are equally accurate but more reliable in critical situations.
\end{abstract}

\maketitle

\section{\label{sec:introduction} Introduction}

Physics-informed neural networks (PINNs) are a widely used tool in today's Machine Learning (ML) landscape. They consist of Neural Networks (NNs) that, during the training phase, learn to solve the differential equations governing the physical laws of a system in a model-independent way. When these differential equations correspond to Hamilton equations of motion, we refer to them as Hamiltonian Neural Networks (HNNs). The HNN paradigm was introduced in \cite{HNN} and in the present work we closely follow the strategy proposed in \cite{Harvard}. The key advantages of HNNs over standard numerical integrators can be summarized as follows:

\begin{itemize}
	\item the predicted solution is analytical in time and not limited to a discrete set of time steps;
	\item conservation laws, symmetries, constraints and prior knowledge of the system can be easily incorporated at the level of the architecture and of the loss function to improve the predictability of the HNN;
	\item the minimization process of the loss function occurs under the constraint that the solution satisfies the system of equations at all times simultaneously and independently, thus avoiding any iterative mechanism.
\end{itemize}

The last point is crucial for systems whose (effective) potential exhibits unstable critical points, for which small changes of the initial conditions lead to completely different behaviors of the solutions, similar to what happens in chaotic dynamical systems. As it is well known, any numerical integrator operating iteratively, building the solution based on the result of the previous time step, accumulates errors and inevitably loses accuracy over long time scales in such situations. HNNs have the potential to overcome this issue.

While HNNs have been applied in various physical contexts\footnote{Due to the rapidly increasing number of ML applications, the literature on this topic is vast, and providing a comprehensive review is beyond the scope of this article. An incomplete selection of applications includes \cite{Crammer:2020,DiHan:2021,Greydanus:2019,deveney:2021,raissi:2017,Wu:2022,Cornell:2022enn,Jain:2022dsh,Luna:2022rql,Luna:2024spo,Geng:2024zzw,Louppe:2017ipp,Nabian:2020,Hashemi:2024azx,Lagaris_1998}.}, their diffusion in high-energy physics remains limited. In this paper, we apply this technique for the first time in the field of String Theory, where ML has only begun to gain attraction in recent years \cite{Cole:2019enn,Cole:2021nnt,MacFadden:2024him,Lanza:2023vee,Lanza:2024mqp,Deen:2020dlf,PhysRevD.111.086007,Ashmore:2019wzb,Jejjala:2020wcc,Larfors:2021pbb,Berglund:2022gvm,Berglund:2023ztk,Anderson:2023viv,Hendi:2024yin,Mirjanic:2024gek,Seong:2023njx,Choi:2023rqg,Seong:2024wkt,He:2024jao}. Specifically, we focus on using HNNs to determine the non-critical and unstable critical geodesics of a massless particle moving in a D1-D5 circular fuzzball geometry by solving the associated equations of motion. The scattering properties, scalar perturbations and tidal deformations of this physical system has already been extensively studied in \cite{Bianchi:2017sds,Bianchi:2022qph,DiRusso:2024hmd}. It serves as a well-suited test case, given that the exact trajectory, to which we will refer to as the \textit{ground truth}, can be explicitly computed. Thus, the results presented in this paper do not add new physical insights into the system under study; rather, the work is methodological in nature and aims to assess the feasibility and performance of HNNs compared to standard numerical methods in a controlled setting. Our investigation, as clearly demonstrated by the numerical results, shows that the HNN performs comparably to numerical integrators for non-critical trajectories and surpasses them for unstable critical trajectories, predicting accurate solutions even over long time scales.

This result suggests that HNNs, and PINNs in general, may offer a robust and reliable solution for numerically determining unstable solutions, not only in String Theory but in all systems where the underlying physics gives rise to complex dynamics. This one, together with other features discussed later on, is one of the reasons that favor these Machine Learning-based algorithms over standard integrators, even when their performances appear comparable. Moreover, the promising outcomes of this paper pave the way for a systematic study of more intricate geometries, where the absence of a ground truth or the lack of physical information necessitates reliable and flexible tools for analysis. HNNs and PINNs are ideal candidates in this regard. As an immediate follow-up, an interesting direction is the study of the so-called D1-D5-P (three-charge) fuzzball, where, due to its intrinsic complexity, an analysis in terms of either geodesic motion or wave propagation has never been completed in full generality. Such a study would also enable a subsequent analysis of the Quasi-Normal Modes (QNMs) spectrum \cite{Bianchi:2023rlt,Bianchi:2023sfs,Cipriani:2024ygw}, which constitutes one of our long-term objectives.

This paper is organized as follows. In Section~\ref{sec:HNN}, we introduce the HNN-based strategy for solving a generic set of Hamiltonian equations. In Section~\ref{sec:fuzzball}, we review the theoretical background of the D1-D5 fuzzball and its geodesics. In Section~\ref{results:planar}, we present numerical results for planar geodesics, while results for the non-planar ones are discussed in Section~\ref{results:non_planar}. Additionally, this work is accompanied by several appendices providing supplemental material to the main text.

\section{\label{sec:HNN}Hamiltonian Neural Networks}

\begin{figure}[t]
    \centering
    \includegraphics[width=\columnwidth]{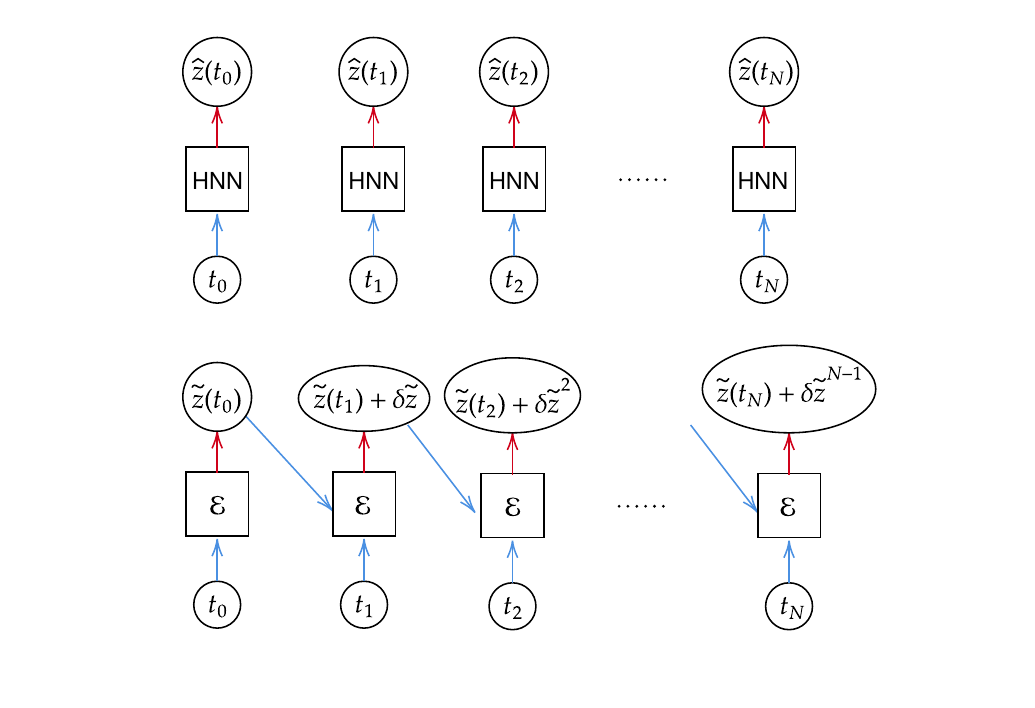}
    \caption{
    \footnotesize{
    Representation of the algorithmic flow for a Hamiltonian Neural Network (HNN, on top) and for the first-order semi-implicit Euler method ($\mathcal{E}$, on bottom).  Blue arrows refer to the inputs at each step, while the red ones to the outputs. As it can be seen, the former produces an output individually for each time without using the information at the previous step, while the latter works iteratively accumulating errors.}}
    \label{fig:Euler_vs_HNN}
\end{figure}

In this Section we provide an introduction to HNNs by summarizing the basic concepts. We refer mostly to \cite{Harvard,HNN}, from which we have taken inspiration for this work.\\

\subsection{Hamiltonian mechanics}\label{sec: Hamiltonian_mechanics}
The phase space in which the  Hamiltonian formalism takes place is defined by the generalized space coordinates $\textbf{q}=(q^1,...,q^n)$ and by the generalized conjugate momenta $\textbf{p}=(p_1,...,p_n)$, where $n \geq 1$ denotes the number of degrees of freedom.  Starting from an initial state at time $t_0$, the evolution of the mechanical system to a final state at $t_1$ is described by

\begin{flalign}\label{eq:time_step_evolution}
    (\textbf{q}_1,\textbf{p}_1)=(\textbf{q}_0,\textbf{p}_0)+\int_{t_0}^{t_1}\textbf{S(q,p)}\, \mathrm{d}t,
\end{flalign}

where

\begin{flalign}\label{time_vector}
\textbf{S(q,p)} = \left(\frac{\mathrm{d}\textbf{q}}
{\mathrm{d}t},\frac{\mathrm{d}\textbf{p}}{\mathrm{d}t}\right)
\end{flalign}

is the time derivatives vector. The core of  Hamiltonian mechanics is the definition of a scalar function $\mathcal{H}(\textbf{q},\textbf{p})$ called Hamiltonian. This function is defined in such a way that Hamilton equations are satisfied \footnote{The notation $\dot{x}$ denotes the time derivative.}

\begin{flalign}\label{eq:generic_hamilton_equations}
\dot{\textbf{q}} = \dfrac{\partial\mathcal{H}}{\partial\textbf{p}}, 
\qquad
\dot{\textbf{p}} = -\dfrac{\partial\mathcal{H}}{\partial\textbf{q}}.
\end{flalign}

This allows to define the \emph{symplectic gradient} from \eqref{time_vector} as
\begin{equation}
    \mathbf{S}_{\mathcal{H}}=\left(\dfrac{\partial\mathcal{H}}{\partial\textbf{p}},-\dfrac{\partial\mathcal{H}}{\partial\textbf{q}}\right)
\end{equation}

and therefore, thanks to (\ref{eq:time_step_evolution}), we can evaluate the evolution of the system. Furthermore, since along the vector $\mathbf{S}_{\mathcal{H}}$ the  Hamiltonian is conserved, we can define the conserved quantity $\mathrm{E}=\mathcal{H}(\textbf{q},\textbf{p})$. In the following we are going to use occasionally the definition of the vector $\textbf{z}\in \mathbb{R}^{2n}$ as
\begin{equation}\label{eq:sympl_coordinates}
    \textbf{z}=(\textbf{q},\textbf{p})^{T}
\end{equation}

\subsection{HNN}\label{sec:HNN_Presentation}
Now we present the strategy based on  Hamiltonian Neural Networks (HNNs) to solve \eqref{eq:generic_hamilton_equations} for a generic system. In this work we will then specialize to the case of the D1-D5 fuzzball geometry whose details are provided in Section~\ref{sec:fuzzball}. In order to start appreciating the differences between machine learning and standard numerical integrators, let us recall the widely used first order semi-implicit Euler method to evaluate a trajectory in the time range $[0,T]$.  This method respects the symplectic structure of the  Hamiltonian system. In particular it automatically conserves a quantity which is the original Hamiltonian shifted by a term proportional to the discretization time step. In fact, the algorithm requires splitting the time interval in $N+1$ points equally spaced with time step $\Delta t$ and such that $T=N\Delta t$.  The solution is then computed  in correspondence of the discrete times $t_j=j\Delta t$, with $j=0,\dots, N$, by updating the generalized coordinates and the corresponding momenta at time step $j+1$ from those at time step $j$ \footnote{The notation $\widetilde{x}$ denotes a quantity obtained with a numerical integrator.},

\begin{flalign}\label{eq:Euler}
    \widetilde{q}_k(t_{j+1})
    &=\widetilde{q}_k(t_j)+\Delta t\dfrac {\partial\mathcal{H}\big(\widetilde{\textbf{q}}(t_j),\widetilde{\textbf{p}}(t_j)\big)}{\partial p_k}, \nonumber \\[4pt]
    \widetilde{p}_k(t_{j+1})&=\widetilde{p}_k(t_j)-\Delta t\dfrac {\partial\mathcal{H}\big(\widetilde{\textbf{q}}(t_{j+1}),\widetilde{\textbf{p}}(t_j)\big)}{\partial q_k}.
\end{flalign}

By employing this iterative method, it is inevitable to accumulate errors through all the input time series, making the prediction of trajectories, especially at large times, usually affected by large discretization errors.  In practice, for particularly complex systems accurate solutions are found only in  the limit of $\Delta t \mapsto 0$ (i.e. $N\mapsto \infty$) where the method is expected to preserve the real Hamiltonian. On the contrary, an HNN provides an analytical solution independently for any input time. This methodological difference is schematically sketched in FIG.~\ref{fig:Euler_vs_HNN}. 
\begin{figure}[t!]
\centering
    \includegraphics[width=\columnwidth]{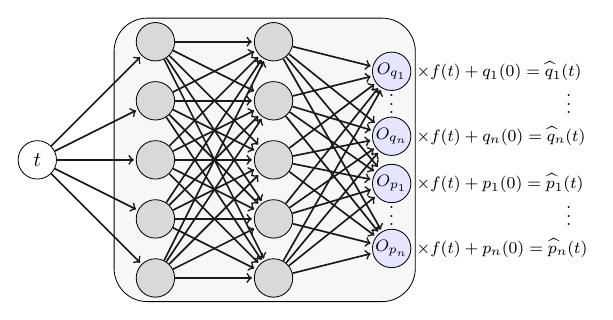}
    \caption{\label{fig:HNN} 
    \footnotesize{The neural network architecture consists in one input node for the time $t$, an arbitrary number of hidden layers and an output layer with a number of nodes (colored light blue) equal to the number of non-trivial Hamilton equations that constitute the system. The network outputs are multiplied by $f(t)=1-e^{-t}$ and added to the initial conditions. Notice that $f(0)=0$ is such that the initial conditions are exaclty satisfied by the predicted solution.}
    } 
\end{figure}

Recalling the definition of Eq.~(\ref{eq:sympl_coordinates}), we can write the HNN solution as \footnote{The notation $\widehat{x}$ denotes a quantity predicted by a neural network.}

\begin{flalign} \label{eq:z_hat}
    \mathbf{\widehat{z}}(t,\textbf{w})=\mathbf{z}_{0}+f(t)\mathbf{O}(t,\textbf{w}),
\end{flalign}

where

\begin{flalign}
\mathbf{O}(t,\textbf{w})=&
\big(O_{q_1}(t,\textbf{w}),\cdots,O_{q_n}(t,\textbf{w}),
\\[4pt] \nonumber
&
O_{p_1}(t,\textbf{w}),\cdots,O_{p_n}(t,\textbf{w}) \big)^T
\end{flalign}

is the output of the NN whose architecture is designed to be a $\mathbb{R}\mapsto \mathbb{R}^{2n}$ map, with the time $t$ being the input, and $\mathbf{z}_0\in \mathbb{R}^{2n}$ is the vector of the initial conditions at $t=0$. \footnote{In case in which some momenta are conserved, due to the fact that the corresponding coordinates are cyclic, then we will not put the nodes associated to them. This is exactly what we mean by non-trivial Hamilton equations in the caption of FIG.\ref{fig:HNN}} The functions $O_{q_k}$ and $O_{p_k}$, with $k=1,\dots,n$, represent single output nodes assuming values in $\mathbb{R}$. The single variables in \eqref{eq:z_hat} are therefore obtained according to

\begin{flalign}\label{eq:qp_hat}
\widehat{q}_k(t,\textbf{w})= q_k(0)+f(t)O_{q_k}(t,\textbf{w})\,,   
\nonumber\\[6pt]
\widehat{p}_k(t,\textbf{w})= p_k(0)+f(t)O_{p_k}(t,\textbf{w})\,,
\end{flalign}

See also FIG.~\ref{fig:HNN} to visualize the network diagram. The symbol $\textbf{w}$ represents the collection of the network weights. We will specify the network architecture in the results' Section, since it is not necessary for the purpose of explaining the method. Let us only clarify that the choice of the network architecture has to be such that the expression $\widehat{\textbf{z}}(t,\textbf{w})$ is analytical and differentiable with respect to the time for any set of weights $\textbf{w}$. We explicitly write in this Section the symbol $\textbf{w}$ for sake of clarity, but we will omit it further on in the text, leaving it understood that whenever a variable is marked by the symbol $\;\widehat{}\;$ it will be dependent on the architecture and weights. The function $f(t)$, which does not contain trainable parameters, is chosen in such a way it satisfies $f(0)=0$ and it is introduced to automatically enforce the initial conditions\footnote{We remark that this might be also achieved by including explicitly the conditions at the level of the loss function. The advantage of the definitions \eqref{eq:qp_hat} is to automatically restrict the space of possible solutions by accelerating and stabilizing the minimization of the loss function.}. In this work we make the same choice as in \cite{Harvard} (where the impact of different choices of $f(t)$ is also illustrated) and define the bounded function 
 
\begin{flalign}\label{eq:ft}
    f(t)=1-e^{-t}.
\end{flalign}

We now explain in detail how the solutions to \eqref{eq:generic_hamilton_equations}, parametrized by \eqref{eq:qp_hat}, can be obtained through the minimization of a  specifically-designed loss function. Let us consider again the time interval $[0,T]$ sampled with $N+1$ points at constant time step $\Delta t$ and indicate $t_j=j\Delta t$ the time after $j$ steps. We define the loss function
\begin{widetext}

\begin{equation} \label{eq:loss_dyn}
    L_\mathrm{dyn}(\textbf{w}) = \sum_{k=1}^{n} \Bigg\{\dfrac{1}{N} \sum_{j=0}^{N} \gamma_{q_k}\Bigg[\left.\Bigg(\dot{\widehat{q}}_k-\widehat{\pdv{\mathcal{H}}{p_{k}}}\Bigg)\right|_{t_j}\Bigg]^2 +\dfrac{1}{N} \sum_{j=0}^{N}\gamma_{p_k}\Bigg[\left.\Bigg(\dot{\widehat{p}}_k+\widehat{\pdv{\mathcal{H}}{q_k}}\Bigg)\right|_{t_j} \Bigg]^2\Bigg\},
\end{equation}

where

\begin{equation}\label{eq:q_dot}
    \left. \dot{\widehat{q}}_k\right|_{t_j} = \left.\dv{f(t)}{t}\right|_{t_j}\cdot O_{q_k}(t_j,\textbf{w})+f(t_j)\cdot \left.\dv{O_{q_k}(t,\textbf{w})}{t}\right|_{t_j}, \qquad\qquad \left. \widehat{\pdv{\mathcal{H}}{p_k}}\right|_{t_j}=\left.\pdv{\mathcal{H}}{p_k}\right|_{\widehat{\textbf{z}}(t_j,\textbf{w})}.
\end{equation}

\end{widetext}
Similarly one can obtain the definitions for $\left.\dot{\widehat{p}}_k\right|_{t_j}$ and $\left.\widehat{\pdv{\mathcal{H}}{q_k}}\right|_{t_j}$ in \eqref{eq:loss_dyn} by replacing $\widehat{q}_k$ with $\widehat{p}_k$ and viceversa. The factors $\gamma_{q_k}$ and $\gamma_{p_k}$ can be chosen to weight a variable more or less relative to the others. Throughout most of the work, we will simply use $\gamma_{q_k}=\gamma_{p_k}=1$.

\eqref{eq:q_dot} makes manifest the dependence of $L_\mathrm{dyn}(\textbf{w})$ on the network weights and on the way $f(t)$ enters the loss function. The subscript \quote{dyn} refers to the fact that,  for a set of weights such that $L_\mathrm{dyn}(\textbf{w})=0$, the network predictions of \eqref{eq:qp_hat} would exactly satisfy the dynamics expressed by the Hamilton equations~(\ref{eq:generic_hamilton_equations}) at the discrete times $t_j$.  The possibility to actually meet the condition  $L_\mathrm{dyn}(\textbf{w})=0$ is supported by the so-called Universal Approximation Theorem \cite{hornik1989multilayer,Goodfellow-et-al-2016}, which states that a neural network large enough is able to approximate any continuous function with arbitrary accuracy. In practice this possibility is spoiled by the fact that a neural network has a finite number of neurons and then, as customary in the Machine Learning field, we can at most search for the set of weights that make the loss function as small as possible, that is we want to solve the equation

\begin{flalign}\label{eq:minimization}
 \pdv{L_\mathrm{dyn}(\textbf{w})}{\textbf{w}}=0   
\end{flalign}

and find an approximation to the solution. We will call \textit{training} the process of minimization of the loss function, but it has to be emphasised that, unlike common regression problem, the training is not data-driven but rather equation-driven. The meaning is that the only information to let the NN fulfill the desired task is codified in the Hamilton equations, which are analytically known after defining the Hamiltonian, and no assumption on the true trajectory is required. From this perspective the HNNs alone offer a model independent solution to the problem and additional informations, e.g. the knowledge of the solution at given times due to experiments or different methods, can be easily incorporated in the training process to improve the accuracy of the predicted solution.

\begin{figure}
\includegraphics[width=\columnwidth]{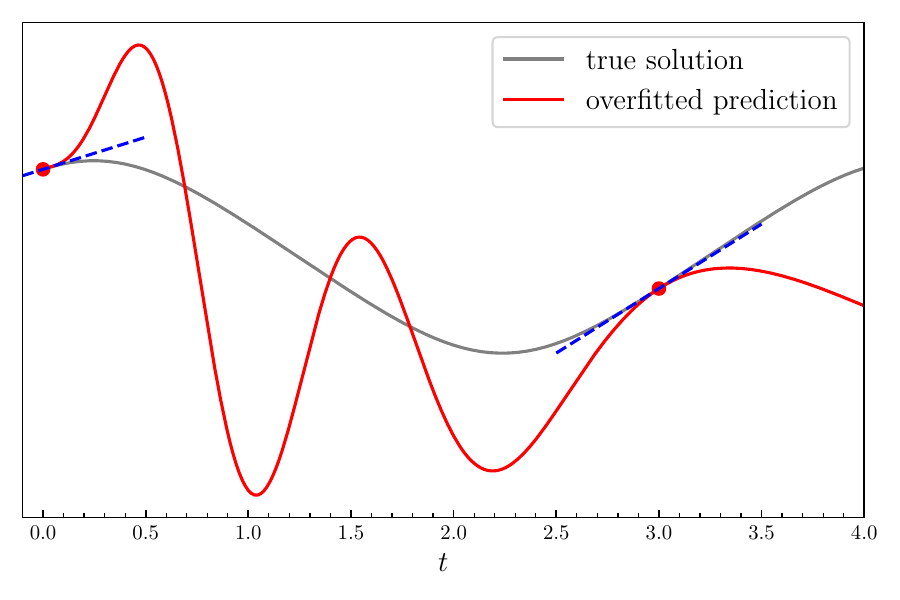}
\caption{\label{fig:overfitting} 
\footnotesize{
Demonstrative example of overfitting for a HNN. In this case the dynamical loss is minimized only for $t=0$ and $t=3$ (red points), in correspondence of which the predicted solution is accurate, while it rapidly deviates from true one otherwise. }
}
\end{figure}

Let us stress that \eqref{eq:qp_hat} are continuous and differentiable in the variable $t$ and then, after the training, they represent an \textit{analytical} solution to the Hamilton equations, providing a significant advantage over numerical integrators, where the solution is only known at predefined discrete time points. However, a natural concern is the reliability of the prediction for times different from the $t_j$ entering $L_\mathrm{dyn}$. As well as regression tasks, also HNNs can suffer from over-fitting by providing an accurate solution for $t=t_j$ but considerably different from the ground truth for $t$ in between $t_j$ and $t_{j+1}$. A demonstrative example of the problem is shown in FIG.~\ref{fig:overfitting}. This problem is expected to be particularly severe as $\Delta t$ increases ($N\to 0$), since the solution is less constrained in between two consecutive points, and vanish in the limit $\Delta t \to 0$ ($N\to \infty$). A possibility to overcome the problem is to sample the interval $[0,T]$ differently at the beginning of each epoch, for instance by adding gaussian noise to a given initial set of times $t_j$, or to choose $N$ large and check a posteriori that the loss function, computed for a set of times different from $t_j$, keeps at the same level as the one minimized during the training. In this work we followed the latter strategy. Of course, there is no chance that the prediction could be accurate for $t$ outside the interval $[0,T]$, since the network has not been instructed on the dynamics in that regime.

Another great advantage of the procedure outlined above is the possibility to extend the definition of the loss function by including terms accounting for conservation of physical quantities. Particularly important is the term imposing the energy conservation, already introduced in \cite{Harvard}, that can be defined as

\begin{flalign}\label{eq:loss_energy}
    L_\mathrm{energy}(\textbf{w})=\dfrac{1}{N} \sum_{j=0}^{N}\big[\mathcal{H}\big(\widehat{\textbf{z}}(t_j,\textbf{w})\big)-E_0\big]^{2},  
\end{flalign}

where $E_0=\mathcal{H}(\textbf{z}_0)$ is the value of the energy to be conserved along the trajectory. We then define the total loss function that has to be minimized as a convex linear combination of $L_\mathrm{dyn}$ and $L_\mathrm{energy}$,

\begin{flalign}\label{eq:total_loss}
    L(\lambda,\textbf{w})=(1-\lambda)L_\mathrm{dyn}(\textbf{w})+\lambda L_\mathrm{energy}(\textbf{w}),   
\end{flalign}

where we have introduced the trade-off parameter $\lambda\in~[0,1]$. The practical advantage of introducing the energy conservation term in the loss function is making the convergence towards the correct solution faster and the training procedure more stable. Moreover, setting $\lambda>0$ serves also as regulator to select a unique solution in those situations where $L_\mathrm{dyn}$ might admit multiple ones. We will show in fact that setting $\lambda >0$ dramatically affects the performance of the neural network in some cases. Similarly one can introduce other terms enforcing the predicted trajectories to conserve physical quantities.

However, the loss function of \eqref{eq:total_loss} is not particularly helpful in comparing the performances of the neural network at varying $\lambda$ since $\lambda$ itself, especially when close to 0 or 1, alters the order of magnitude of $L(\lambda,\textbf{w})$ at a given  $\textbf{w}$. During the training we therefore also monitor individually $L_\mathrm{dyn}(\textbf{w})$ and $L_\mathrm{energy}(\textbf{w})$ in order to have a measure of the goodness of the solution independently from $\lambda$.

Further regularization terms limiting the space of the solutions according to additional knowledge about the system can be introduced as well to improve the accuracy of the network. We will come back to it later in the results' Section since this is case-dependent.

We conclude this Section by observing that we have specialized (\ref{eq:loss_dyn}) for the case of the Hamilton equations, but in general one can write a proper loss function to search for the solution of any system of ordinary differential equations. For instance, the geometry described in the next Section enjoys also the separability of the motion and it allows the derivation of a different, but physically equivalent, set of equations of motion. We will consider this alternative point of view in Appendix \ref{APP:PINN_separability}, with the aim of highlighting differences between the usage of HNN.
The scope of applicability of PINNs is therefore extremely wide and such technology is a valuable tool that can complement standard numerical integrators and, at times, replace them.

In the next Section we will introduce the physical system whose equations of motion are the ones we want to solve by implementing a HNN.

\section{\label{sec:fuzzball}D1-D5 fuzzball geometry}
The physical configuration we consider in this work is the motion of a scalar massless neutral particle inside the geometry of a D1-D5 circular fuzzball. This Section is divided in the following subsections: in \ref{sec:Presentation} we furnish a presentation of the background in which the fuzzball comes out, with the aim of understanding its nature and why it is playing an interesting role in String Theory; in \ref{Formalism} we recall the basics for determining the  Hamiltonian of a particle that is moving inside a generic spacetime; in \ref{sec:Geometry} we focus on the aforementioned case of D1-D5 fuzzball geometry, writing the  Hamiltonian of the particle and presenting some relevant properties of such geometry; in \ref{sec:Planar_Case} and \ref{Non-Planar Case} we distinguish the two cases of interest, that is, the planar and non-planar motion. Details concerning the formulae that will be shown can be found in the Appendices.

\subsection{Presentation of  the D1-D5 fuzzball}\label{sec:Presentation}
As already said, in this paper we study the geodesic motion (i.e. the free fall trajectory) of a massless particle moving inside a specific geometry derived from a low energy limit of String Theory, exploring it as another potential application of the powerful and versatile techniques of ML in this field. To grasp the physical context, we give a brief review of the concept of Black Holes (BHs) and its relevance in modern theoretical physics. These gravitational objects, first discovered within the context of General Relativity, exhibit several peculiar properties. They are surrounded by a surface called the event horizon, beyond which nothing, not even light, can escape - hence the term \quote{black} in their name. Furthermore, at the core of the BH lies a singularity, known as a curvature singularity, where spacetime is no longer well-defined. The cause is the fact that General Relativity becomes unreliable at scales of energy of the singularity and quantum effects should be properly taken into account.
\\

As a matter of fact, BHs are an interesting puzzle for theoretical physicists and this is due to their simple nature: they are described by a handful of parameters - namely their mass, charge and angular momentum - while being the result of the gravitational collapse of very complex physical systems, such as stars. Moreover, no matter the complexity of the quantum state that crosses the event horizon, this only results in a change of those parameters. This is typically referred to as the \textit{information-loss paradox} \cite{Mathur:2009hf} and reveals a curious - albeit not yet understood - relationship between General Relativity and Quantum Mechanics.
Remaining still at the classical level, it was shown that BHs are akin to thermodynamical systems: they obey laws similar to the ones of Thermodynamics \cite{PhysRevD.7.949} and possess an associated entropy-like quantity in the form of the event horizon’s area. 
It is well understood that entropy has a microscopic interpretation in terms of the number of microstates, a description that is woefully absent in General Relativity, since here it is usually said that `black holes have no hair' (meaning, as said before, that they are described only by mass, charge and angular momentum). 

As it is, both the existence of a singularity and of the event horizon call for a theory of quantum gravity and a better understanding of the classical BHs. In the context of String Theory a proposal can be formulated (the \quote{fuzzball proposal} \cite{Lunin:2001fv,Mathur:2005zp,Bena:2022rna}), according to which BHs should be thought of superpositions of quantum states, some of which admit a classical description as smooth and horizonless gravity solutions with the same mass, charge and angular momentum of the BH. In addition, these microstates can emit radiation, preserving quantum information and hence solving the information-loss paradox.
\newline
There are many ways to construct these smooth-horizonless geometries and one of these is to consider particular objects naturally arising in String Theory, the D$p$-branes. Without entering into too much detail, (super) String Theory can be defined in a mathematically consistent way only in 9+1 spacetime dimensions; D$p$-branes extend both along the time direction of spacetime and along $p$ spatial dimensions. They carry mass and charge and, consequently, are subjected to the action of forces. One can combine a number of different branes in order to form a stable gravitational object, with a given mass and charge. For the scope of this work we are interested in the gravitational object - and its associated spacetime metric - obtained by combining D1 and D5 branes and which has a circular profile. 

\subsection{Formalism} \label{Formalism}
Given a spacetime of dimension $d$, it is well known that the scalar product between any two of its vectors is codified by a symmetric $d \times d$ matrix, which is the metric $g_{\mu \nu}$. In this way the squared length of the line element is given by

\begin{flalign}
    \mathrm{d}s^2 = g_{\mu \nu} \dd x^\mu \dd x^\nu,
\end{flalign}

where $x^\mu$ are the spacetime coordinates ($\mu = 0,\dots,d-1$). The free fall of a neutral particle inside a certain spacetime with metric $g_{\mu \nu}$ happens along the geodesics, which are solutions of differential equations known as \textit{geodesic equations}. On the other hand, it is also known that the trajectories in the physical space are such that the action is minimized, which can be done by solving the \textit{Euler-Lagrange equations}

\begin{flalign} \label{E-L Eqns}
\frac{\mathrm{d}}{d\mathrm{s}}\frac{\partial \mathcal{L}}{\partial \dot{x}^\mu}=\frac{\partial \mathcal{L}}{\partial x^\mu},
\end{flalign}

where $s$ is the affine parameter of the geodesic of the particle and $\mathcal{L}$ its lagrangian. From all of this we understand that the geodesic equations are equivalent to the Euler-Lagrange ones and it can be shown that it is possible if the Lagrangian has the following form

\begin{flalign} \label{Lagrangian}
    \mathcal{L}=\frac{1}{2} g_{\mu\nu}\dot{x}^\mu\dot{x}^\nu\quad,\quad \dot{x}^\mu=\frac{\mathrm{d} x^\mu}{\mathrm{d}s}.
\end{flalign}

In order to pass to the  Hamiltonian viewpoint, we have to compute the canonical conjugate momenta

\begin{flalign} \label{eq:Momenta}
P_\mu= \frac{\partial \mathcal{L}}{\partial \dot{x}^\mu}
\end{flalign}

and the  Hamiltonian $\mathcal{H}$ is obtained through a Legendre transformation of the Lagrangian

\begin{flalign}\label{eq:Hamiltonian}
    \mathcal{H} = P_\mu \dot{x}^\mu (P) - \mathcal{L}(x,\dot{x}(P)) = \frac{1}{2} g^{\mu \nu} P_\mu P_\nu.
\end{flalign}

Notice that what we have called here $x$ and $P$ correspond to $\mathbf{q}$ and $\mathbf{p}$ of Section \ref{sec: Hamiltonian_mechanics}, but it is the typical notation in the gravity context. In addition we consider as time the affine parameter $s$ in place of the coordinate time $t$ used in Section~\ref{sec:HNN_Presentation}. This implies no conceptual and practical differences in the ML strategy proposed. The mass shell condition for a particle of mass $\mu_0$ reads \cite{Bianchi:2017sds,Chervonyi:2013eja} $\mathcal{H}= - \frac{1}{2}\mu_0^2$. Since we are interested in massless geodesics, we obtain the following important condition: $\mathcal{H}=0$. Finally, we recall that, if the metric components and thus the Lagrangian do not depend on a specific coordinate $x^\mu$ (which is said to be cyclic), then the corresponding momentum $P_\mu$ is conserved, as can be read from the Euler-Lagrange equations \ref{E-L Eqns}.

\subsection{Geometry}\label{sec:Geometry}
Now we pass to examine the geometry of the D1-D5 circular fuzzball and the properties of the motion of a massless particle inside this. 
The spacetime metric generated by this object is

\begin{flalign} \label{eq:metric}
\dd s^2
=&H^{-1}\left[-(\dd t+\omega_\phi \dd \phi)^2+(\dd z+\omega_\psi \dd \psi)^2\right]
\nonumber\\[6pt]
&+H\Bigg[(\rho^2+a_f^2\cos^2\theta)\left(\frac{\dd\rho^2}{\rho^2+a_f^2}
+\dd\theta^2\right)
\nonumber\\[6pt]
&+\rho^2\cos^2\theta \dd\psi^2+(\rho^2+a_f^2)\sin^2\theta \dd\phi^2\Bigg]
\nonumber\\[6pt]
&+\left(\frac{H_1}{H_5}\right)^{1/2}\dd z_{1,2}^2+\left(\frac{H_5}{H_1}\right)^{1/2}\dd z_{3,4}^2
\end{flalign}

where

\begin{flalign} \label{eq:def_params}
H  &=\sqrt{H_1H_5} \,, \quad H_i =1+\frac{L_i^2}{\rho^2+a_f^2\cos^2\theta}\,,
\nonumber\\[6pt]
\omega_\phi&=\frac{a_fL_1L_5\sin^2\theta}{\rho^2+a_f^2\cos^2\theta}\,, \quad \omega_\psi=\frac{a_fL_1L_5\cos^2\theta}{\rho^2+a_f^2\cos^2\theta}.
\end{flalign}

$a_f$ is the radius of the circular profile, $L_1$ and $L_5$ are the charges of, respectively, D1 and D5 branes \footnote{Notice that $L_1$ and $L_5$ have dimensions of a length, while in measure units with $c=G=\frac{1}{4 \pi \epsilon_0}=1$, the charge has dimension of a squared length. Consequently the real charge is related to $L_i$ via multiplication by a dimensional parameter.}. These are the parameters of the geometry that we are considering and it is also rotating. It has to be said that the complete solution describing the fuzzball does not contain only the metric \eqref{eq:metric}, but also other fields, which are not relevant for the dynamics of the neutral probe we are considering and for this reason we have not written them. This spacetime is 10-dimensional \footnote{For $L_1=L_5=0$ (i.e. $H_1=H_5=1$ and $\omega_\phi = \omega_\psi = 0$) and under a suitable coordinate change the metric \eqref{eq:metric} describes flat spacetime.}, but four out of ten coordinates, $z_I$ with $I=1,\dots,4$, parametrize compact directions and for this reason we can safely set their corresponding momenta $P_{z_I}$ equal to 0 and neglect these coordinates. The coordinates $(\rho,\theta,\phi,\psi)$ are called \textit{oblate spheroidal coordinates} and they are related to the standard cartesian ones $(x_1,x_2,x_3,x_4)$ through the relations

\begin{flalign}\label{eq:spheroidal_coordinates}
    x_1 &= \sqrt{\rho^2 + a_f^2} \sin\theta \cos \phi \,,\\[6pt] 
    x_2 &=  \sqrt{\rho^2 + a_f^2} \sin\theta \sin \phi  \,,\nonumber\\[6pt]
    x_3 &= \rho \cos \theta \cos \psi \,,\nonumber\\[6pt]
    x_4 &= \rho \cos \theta \sin \psi.\nonumber
\end{flalign}

Notice that the metric components do not depend on $t$, $\phi$, $\psi$ and $z$, consequently the associated momenta $P_t = - E$, $P_\phi = J_\phi$, $P_\psi=J_\psi$ and $P_z$ are conserved. Following \eqref{Lagrangian}, \eqref{eq:Momenta} and \eqref{eq:Hamiltonian}, we can arrive at the expression of the  Hamiltonian of a particle moving inside this geometry, which is

\begin{flalign}\label{eq:Ham_tot}
    \mathcal{H} = & \frac{1}{2 \left(\rho^2 +a_f^2\cos^2\theta\right)H}\Bigg[P_\rho^2 (\rho^2+a_f^2) 
    \nonumber\\[6pt]
    &  + P_\theta^2 + \frac{(J_\psi a_f - P_z L_1 L_5)^2}{\rho^2} + \frac{J_\psi^2}{\cos^2\theta} 
    \nonumber\\[6pt]
    &  + \frac{J^2_\phi}{\sin^2\theta} - \frac{(J_\phi a_f - E L_1 L_5)^2}{\rho^2 + a_f^2} 
    \nonumber\\[6pt]
    &- (E^2-P_z^2)(\rho^2+a_f^2+L_1^2+L_5^2)
    \nonumber\\[6pt]
    & + (E^2-P_z^2) a_f^2 \sin^2\theta \Bigg].
\end{flalign} 

From this, we can compute the derivatives, as required by Hamilton equations, that should then be used inside the loss function as explained in the previous Section \ref{sec:HNN}. We will not report these derivatives for the most general case, since they are not important for our study; we write in Appendix \ref{APP:hamilton} the derivatives that are related to the \textit{planar} and a specific \textit{non planar} motions, as we will soon explain.\\
Before proceeding with this, we have to pay attention to a particular feature of this geometry and that will constitute another aspect to probe the goodness of the HNN, that is the \textit{separability} of the dynamics along the radial $\rho$ and the angular $\theta$ directions. Indeed, from the previous expression of the Hamiltonian, since it has to be 0 for the massless particle, we can neglect the overall factor and, by examining the terms within the square brackets, we notice that, by introducing the \textit{Carter constant} $K^2$ \cite{Chandrasekhar:1984siy}, we can distinguish the motion along the direction parametrized by $\rho$ from the one along the direction parametrized by $\theta$. In particular

\begin{flalign}\label{eq:QR}
P_\rho^2 = Q_R(\rho)&=(E^2-P_z^2) \left(1+\frac{L_1^2+L_5^2}{\rho^2+a_f^2}\right)+  
\nonumber\\[6pt]
&+\frac{(J_\phi a_f {-} E L_1L_5)^2}{(\rho^2{+}a_f^2)^2}{-}\frac{(J_\psi a_f {-} P_z L_1L_5)^2}{\rho^2(\rho^2{+}a_f^2)}
\nonumber\\[6pt]
&-\frac{K^2}{\rho^2{+}a_f^2}
\end{flalign}
and
\begin{flalign}\label{eq:QA}
P_\theta^2{=}&Q_A(\theta){=}K^2{-}\frac{J_\psi^2}{\cos^2\theta}{-}\frac{J_\phi^2}{\sin^2\theta}{-}(E^2{-}P_z^2) a_f^2\sin^2\theta .
\end{flalign}

Thanks to this, it is possible to integrate these two equations, getting expressions in terms of elliptic integrals. We are not going to do this in the most general case, since we are interested in the planar and a specific non planar motions. In the former situation we report also the procedure that shows how to integrate these equations, getting the ground truth that will constitute the reference point with respect to which we compare the results coming from the HNN and the standard numerical integrator.\\

A common peculiarity of BHs, D-branes and
other compact gravitating objects is the presence of \quote{photon-spheres} or light-rings (light-halos for rotating objects), that are stable or unstable bound orbits of particles separating the asymptotically flat region from the horizon (or the inner region for fuzzballs) \cite{Bianchi:2020yzr,Bianchi:2021yqs, Bianchi:2022wku}. In the following we will denote them as \textit{critical geodesics}. The simultaneous vanishing of the radial momentum $P_\rho^2(\rho) \equiv Q_R(\rho)$ and its derivative
\begin{equation}\label{critcond}
Q_R(\rho_c,J_c)=Q'_R(\rho_c,J_c)=0
\end{equation}
individuate the critical radius $\rho_c$ and the corresponding critical angular momentum $J_c$.
The radial effective potential $-Q_R(\rho)$ depends also on all the other parameters of the geometry, hence fixing all of them we get a particular shape corresponding to which $\rho_c$ can correspond both to a maximum or a minimum. In the former case the critical geodesic is unstable, in the latter is stable. It is worth noticing that this identification is possible only if the geometry is separable. Determining the critical geodesics for non separable geometries is something not clear and HNNs could help in this direction.\\ 
We pass now to examine the planar and non-planar cases separately, determining also the critical geodesics.

\begin{figure}[t]
  \centering
  \includegraphics[width=\columnwidth]{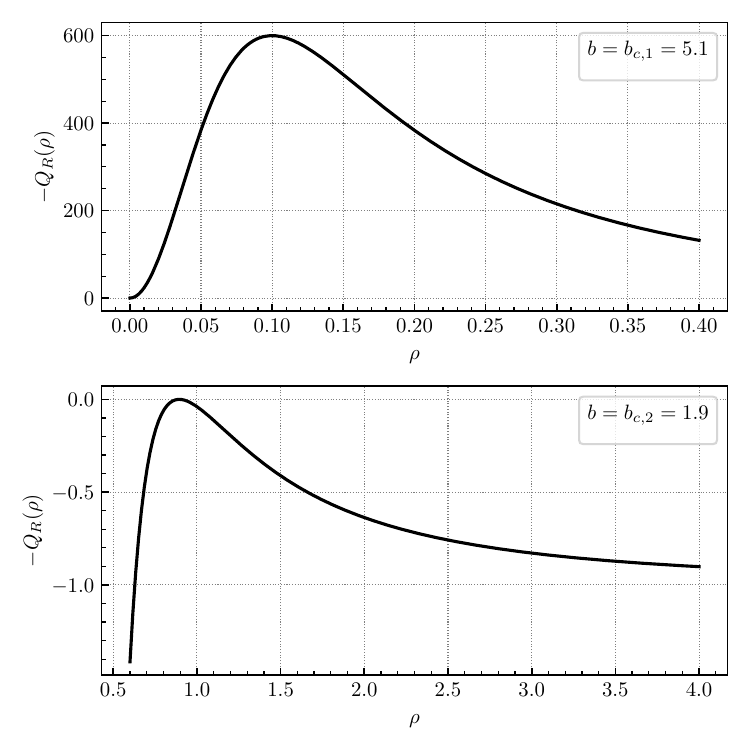}
  \caption{
  \footnotesize{
  D1-D5 circular fuzzball radial effective potential in the equatorial plane $\theta=\pi/2$ for $L_1 = L_5 = 1$ and $a_f=0.1$. \emph{Top-panel}:  at $b_c = b_{c,1} = 5.1$, the stable photon-sphere corresponds to $\rho_c = \rho_{c,1}=0$. \emph{Bottom-panel}:  
  at $b_c = b_{c,2} = 1.9$, the unstable photon-sphere corresponds to $\rho_c = \rho_{c,2}=\sqrt{0.8}= 0.8944$.
  }}
  \label{fig:plotQ}
\end{figure}

\subsection{Planar case} \label{sec:Planar_Case}
Due to the axial (non fully spherical) symmetry, the geodesics are non planar in general. Some simplifications arise for geodesics when the motion exists in a fixed plane. In order to find this, we should solve the Euler-Lagrange equation for $\theta$ coordinate (which is a second order differential equation) imposing as initial conditions $\theta(s_0) = \theta_0$ (the value of the angle corresponding to which we have the plane) and $\dot{\theta}(s_0)=0$, where $s_0$ is the value of the affine parameter at the beginning of the motion. In this way we obtain a Cauchy problem and if the only solution is $\theta(s)=\theta_0$, then the motion takes place only along the plane corresponding to $\theta = \theta_0$.\\
For the geometry we are studying, this happens if $\theta_0=~0,\pi/2$ \footnote{Actually these are specific cases of a more general class of geodesics, denoted as \textit{shear free} \cite{Bianchi:2022qph}.}, which correspond to the two equatorial planes. Since $\theta$ is constant, then $P_\theta=0$. Furthermore, in order to avoid divergences in \eqref{eq:QA} for $\theta=\pi/2$, we are forced to take $J_\psi = 0$, conversely $J_\phi=0$ for $\theta=0$. Furthermore, we set also $P_z=0$, which simplifies the treatment without altering the final results. In the following we will consider only the case with $\theta_0 = \pi/2$.\\
The  Hamiltonian \eqref{eq:Ham_tot} becomes

\begin{flalign}\label{eq:Ham_plan}
    & \mathcal{H} = \frac{1}{2 \rho^2 H(\theta=\pi/2)}\Bigg[P_\rho^2 (\rho^2+a_f^2) + J^2_\phi +  
    \nonumber\\[6pt]
    & - \frac{(J_\phi a_f - E L_1 L_5)^2}{\rho^2 + a_f^2} - E^2 (\rho^2+a_f^2+L_1^2+L_5^2)+ E^2 a_f^2 \Bigg].
\end{flalign} 

The derivatives of it are reported in Appendix \ref{APP:hamilton}.\\

For what concerns the separability, the equation \eqref{eq:QA} with $P_\theta = 0, \theta=\pi/2, J_\psi=0, P_z=0$ can be solved for the Carter constant

\begin{equation}
 K^2 = J_\phi^2+E^2 a_f^2,
\end{equation}

which plugged into the radial equation \eqref{eq:QR}, gives

\begin{flalign}\label{eq:Prho_Plan}
    P_\rho^2 & = E^2 \left(1+\frac{L_1^2+L_5^2}{\rho^2+a_f^2}\right) 
    \nonumber\\[6pt]
    & + \frac{(J_\phi a_f - E L_1 L_5)^2}{(\rho^2+a_f^2)^2} - \frac{J_\phi^2+a_f^2 E^2}{\rho^2+a_f^2}.
\end{flalign}

We introduce now the impact parameter $b$
\begin{flalign} \label{impact param}
    b = \frac{J_\phi}{E}
\end{flalign}

In this way these equations can be expressed in terms of such parameter. Using the expressions of the momenta $P_i$ in terms of the derivatives of the coordinates, $\dot{x}^i$, from \eqref{eq:Momenta} and inverting all these relations, we can find an equation for the derivative of $\rho$ and $\phi$ with respect to the coordinate time $t$. At the end we get

\begin{widetext}
\begin{flalign}\label{eq:diffeq}
\frac{\mathrm{d} \rho(t)}{\mathrm{d} t} = \frac{\dot{\rho}}{\dot{t}} = &\pm \frac{(\rho^2+a_f^2)\sqrt{L_1^2 (L_5^2+\rho^2)+\rho^2 (L_5^2+\rho^2-b^2)+ a_f^2 (L_1^2+L_5^2+\rho^2) - 2 a_f b L_1 L_5}}{(L_1^2+\rho^2)(L_5^2+\rho^2)+a_f^2(L_1^2+L_5^2+\rho^2)-a_f b L_1 L_5},
\nonumber\\[6pt]
\frac{\mathrm{d} \phi(t)}{\mathrm{d} t} = \frac{\dot{\phi}}{\dot{t}} = &\frac{b \rho^2 + a_f L_1 L_5}{(L_1^2+\rho^2)(L_5^2+\rho^2)+a_f^2(L_1^2+L_5^2+\rho^2)-a_f b L_1 L_5}.
\end{flalign}
\end{widetext}

Details can be found in Appendix \ref{APP:separability}. In the first equation of \eqref{eq:diffeq}, the negative sign refers to the phase in which the particle is moving closer to the central object, the positive sign refers to the opposite situation.\\
We can find the critical geodesics by imposing the vanishing of \eqref{eq:Prho_Plan} and its radial derivative. Introducing the impact parameter $b$, we have two equations in the variables $\rho,b$ and we can solve them in terms of the parameters $L_1,L_5,a_f$. Keeping in mind that $\rho$ and $b$ are non negative, then we obtain the final results

\begin{flalign}\label{eq:stable_ph}
\rho_{c,1} & = 0\,,
\nonumber\\[6pt]
b_{c,1} & = \frac{a_f^2 (L_1^2+L_5^2)+L_1^2 L_5^2}{2 a_f L_1 L_5}\,,
\end{flalign}

and

\begin{flalign}\label{eq:unstable_ph}
\rho_{c,2} & = \sqrt{L_1 L_5 - a_f (L_1+L_5)}\,,
\nonumber\\[6pt]
b_{c,2} &= L_1+L_5-a_f \,.
\end{flalign}

FIG.~\ref{fig:plotQ} shows the plot of the effective radial potential $-P^2_\rho=-Q_R(\rho)$ for these two cases by setting $L_1=L_5=1$ and $a_f=0.1$. As it can be seen, $\rho_{c,1}$ describes a stable photon-sphere (top-panel), while $\rho_{c,2}$ describes an unstable one (bottom-panel). By plugging $\theta = \pi/2$ and $\rho=0$ in \eqref{eq:spheroidal_coordinates} we obtain

\begin{flalign}
 x_1 &= a_f \cos \phi\,,
 \\[4pt]
 x_2 &=  a_f \sin \phi\,,
 \nonumber\\[4pt]
 x_3 &= x_4 = 0\,.\nonumber
\end{flalign}

Hence when $\rho = 0$ we get on the plane $(x_1,x_2)$ a circle of radius $a_f$ and it is the stable geodesic \footnote{In the $\theta = 0$ plane described by the coordinates $(x_3,x_4)$ (since $x_1=x_2=0$), the fuzzball instead corresponds to a point located at $x_3=x_4=0$.}. In the following of the paper we will indicate $(x_1,x_2)$ as the usual $(x,y)$ and only focus on the orbit with radius $\rho_{c} = \sqrt{L_1 L_5 - a_f (L_1+L_5)}$ and impact parameter equal to $b_c = L_1+L_5-a_f$. From the above discussion, it is clear that, when $b<b_c$ the particle, coming from infinity, arrives at the compact object and then it bounces, while for $b>b_c$ the particle reaches a minimum value of the radius and then it moves away again. The former case will be denoted as \textit{sub-critical}, while the latter as \textit{over-critical}.\\
Another way of formulating this distinction from the first equation of \eqref{eq:diffeq} is the following:
\begin{itemize}
    \item when $b<b_c$, the radicand does not admit real zeros in $\rho$. It implies that $\mathrm{d} \rho / \mathrm{d} t$ never vanishes, meaning that the distance always decreases (with minus sign) or always increases (with plus sign). If the particle comes from infinity, then it reaches $\rho = 0$ and then it bounces;
    \item when $b=b_c$, the radicand has two real and positive zeros that are coincident and they correspond to $\rho_c$;
    \item when $b>b_c$, the radicand has two distinct real and positive zeros, hence we have two critical radii $\rho_{-}$ and $\rho_{+}$, with $\rho_{-} < \rho_{+}$. If the particle comes from infinity, it reaches first $\rho_{+}$ and after this it goes away again.
\end{itemize}
It has to be noticed that, in this case, we can rewrite the equatorial geodesics in terms of elliptic integrals and it is done from \eqref{eq:diffeq}. This constitutes the ground truth in the comparison of the results of the HNN and the numerical integrators. Details of the determination of the ground truth can be found in Appendix \ref{APP:groundtruth}. 
In the following we will set, in the planar case, $L_1 = L_5$ without loss of generality, simplifying a lot the involved equations.

\subsection{Non-planar case}\label{Non-Planar Case}
Now we want to analyze the motion that takes place not only on the equatorial plane. In particular we set
\begin{equation}
    P_z = J_\psi = 0
\end{equation}
and we fix $\psi = 0$. It corresponds to a family of 3-dimensional geodesics that take place in the cartesian space through the relations outlined in \eqref{eq:spheroidal_coordinates}, that is

\begin{flalign}
\label{eq:change_coord_3d}
x &= \sqrt{\rho^2 + a_f^2} \sin\theta \cos \phi\,,
\\[6pt]
y &= \sqrt{\rho^2 + a_f^2} \sin \theta \sin \phi \,,
\nonumber\\[6pt]
z &= \rho \cos \theta\,,\nonumber
\end{flalign}

where we have denoted $x_3$ as $z$. The general  Hamiltonian \eqref{eq:Ham_tot} now reads

\begin{flalign}\label{eq:Ham_Non-Plan}
    \mathcal{H} & = \frac{1}{2 \left(\rho^2 +a_f^2\cos^2\theta\right)H}\Bigg[P_\rho^2 (\rho^2+a_f^2)  
    \nonumber\\[6pt]
    & + P_\theta^2 + \frac{J^2_\phi}{\sin^2\theta} - \frac{(J_\phi a_f - E L_1 L_5)^2}{\rho^2 + a_f^2} 
   \nonumber\\[6pt]
    &  - E^2(\rho^2+a_f^2+L_1^2+L_5^2-a_f^2 \sin^2\theta) \Bigg].
\end{flalign} 

The derivatives with respect to the variables that define the Hamilton equations are  reported in Appendix \ref{APP:hamilton}.\\

We discuss now the determination of the critical geodesics, that is the only case we study with the HNN since it is the most physically relevant one for observational reasons. The momenta \eqref{eq:QR} and \eqref{eq:QA} now become

\begin{flalign}
P_\rho^2  = E^2 \left(1+\frac{L_1^2+L_5^2}{\rho^2+a_f^2}\right) + \frac{(J_\phi a_f - E L_1L_5)^2}{(\rho^2+a_f^2)^2}-\frac{K^2}{\rho^2+a_f^2} \label{eq:QR_Non Plan} 
\end{flalign}
and
\begin{flalign} \label{eq:QA_Non_Plan} 
P_\theta^2  = K^2-\frac{J_\phi^2}{\sin^2\theta}- E^2 a_f^2 \sin^2\theta,
\end{flalign}

respectively.  For $P_\theta\propto \dot{\theta}\neq 0$ the angular dynamics is non-trivial. Let us rewrite \eqref{eq:QR_Non Plan} as follows,

\begin{flalign}
\frac{(\rho^2+a_f^2)^2P_\rho^2}{E^2}&=\mathcal{R}(\rho)=\rho^4+A\rho^2+B\,,
\nonumber\\[6pt]
A&=2a_f^2+L_1^2+L_5^2-b^2\,,
\nonumber\\[6pt]
B&=a_f^2(\zeta^2-b^2+a_f^2+L_1^2+L_5^2)\,,
\end{flalign}

where

\begin{flalign}
\zeta=b_\phi-\frac{L_1L_5}{a_f},\quad b_\phi=\frac{J_\phi}{E},\quad b=\frac{K}{E}\,.
\end{flalign}

The critical conditions are $\mathcal{R}(\rho_c)=\mathcal{R}'(\rho_c)=0$ and these correspond to two equations in terms of the variables $\rho_c, \zeta_c$ and $b_c$. We can solve them in terms of $\zeta_c$ and $b_c$, which become functions of $\rho_c$, that is

\begin{flalign}\label{eq:critreg}
\zeta_c^2=&\frac{(\rho_c^2+a_f^2)^2}{a_f^2},
\nonumber\\[6pt]
b_c^2=&2\rho_c^2+2a_f^2+L_1^2+L_5^2.
\end{flalign}

In order to fix $\rho_c$ we must work on the condition $P^2_\theta \geq 0$. Details of all steps that have to be done in order to arrive at the determination of $\rho_c$ and the examination of the corresponding critical geodesics are reported in Appendix \ref{APP:nonplangeo}. Indeed, as it is shown there, these critical geodesics generically wrap around spheroidal zones. There is only one case in which the critical geodesics wrap around an oblate spheroid and we will study only this last scenario. The critical radius has the following expression

\begin{equation} \label{eq:crit_radius}
\rho_c=\sqrt{L_1L_5-a_f^2}, \qquad  0<a_f\leq \sqrt{L_1L_5}.
\end{equation}

These geodesics are \textit{counterrotating} with respect to the fuzzball, i.e. they rotate in the opposite sense with respect to the one of the central object. From \eqref{eq:crit_radius} we can obtain the expressions of $\zeta_c$ and $b_c$, which are

\begin{align} \label{eq:final_params_nonplan}
& \zeta_c = -\frac{L_1 L_5}{a_f} \leftrightarrow b_{\phi,c}=0, 
\nonumber\\[6pt]
& \qquad b_c = L_1+L_5.
\end{align}

The reason why we have the negative sign of \eqref{eq:critreg} for $\zeta_c$ is explained in Appendix \ref{APP:nonplangeo}.

\section{\label{results:planar}Results for planar geodesics}

In this Section we present in detail the implementation of the HNNs, the strategies and the numerical results for the planar geodesics in the $\theta=\pi/2$ plane and in the three scenarios, over-critical ($b>b_c$), critical ($b=b_c$) and sub-critical ($b<b_c$), illustrated in Section~\ref{sec:Planar_Case}. The Hamilton equations that describe the geodesics are given by the following system of coupled differential equations,
\begin{align}
   &\dot \rho = \pdv{\mathcal{H}}{P_\rho},
   \nonumber\\[6pt]
   &\dot P_\rho  =-\pdv{\mathcal{H}}{\rho},
   \nonumber\\[6pt]
   &\dot \phi = \pdv{\mathcal{H}}{J_\phi}.
\end{align}
The derivatives on the right-hand side are explicitly written in Appendix\ref{APP:hamilton}. The integration of the equations obtained from the separability of the motion \eqref{eq:diffeq}, which are different but physically equivalent, will be presented in Appendix~\ref{APP:PINN_separability}.

For all the three cases,  we fix the geometry parameters to $a_f=0.1$, $L_1=L_5=L=1$ and $E=1$. With this choice, the critical values of the impact parameter \eqref{eq:unstable_ph} and of the critical radius are
\begin{equation}\label{eq:planar_crit_params_values}
  b_\mathrm{c}=1.9, \qquad \rho_c = \sqrt{0.8} = 0.8944.
\end{equation}
In the remainder of this Section we outline the general strategy, while the case-dependent details and the associated results will be presented separately in the following three subsections.

The neural networks are explicitly designed to map the input time $s$ to the three-dimesional output $(O_\rho(s),O_{P_\rho}(s),O_\phi(s))$, from which the solutions are built according to 
\begin{align}
& \widehat{\rho}(s) = \rho_0  +f(s)O_\rho(s),
\nonumber\\[6pt]
& \widehat{P_\rho}(s) = P_{\rho,0}+f(s)O_{P_\rho}(s),
\nonumber\\[6pt]
& \widehat{\phi}(s) = \phi_0 +f(s)O_{\phi}(s).
\end{align}

 Let us stress that the role of the time $t$ in Section~\ref{sec: Hamiltonian_mechanics} is now played by the affine parameter $s$ and this  does not entail any difference either in the strategy or in the implementation of the neural networks. Indeed, Section~\ref{sec: Hamiltonian_mechanics} can be re-read just replacing $t$ with $s$.
 
Our implementation of the neural networks\footnote{A demonstrative notebook is provided in \cite{notebook}. The complete code is available upon request.} relies on the libraries Keras \cite{keras} and TensorFlow \cite{tensorflow}. The trainings are performed by employing the Adam optimizer of \cite{Adam}.  In all the cases the weights are initialized by drawing random numbers from a normal distribution with center $0$ and width $0.05$.

The loss function we minimize is the combination given in \eqref{eq:total_loss} and during the training we monitor individually also $L_\mathrm{dyn}$ and $L_\mathrm{energy}$, which are defined  respectively in \eqref{eq:loss_dyn} and \eqref{eq:loss_energy}. Unless otherwise specified, we set $\gamma_\rho=~\gamma_{P_\rho}=~\gamma_\phi=1$ inside the loss function \eqref{eq:loss_dyn}.
As usually happens when training a neural network, the loss function fluctuates continuously over the epochs. Therefore, we save the updated weights at the end of each epoch and, at the end of the training, we consider the set of weights corresponding to the lowest value reached by the loss function as the optimal set for the trajectory prediction.

If not otherwise specified, the learning rate $\eta$ is defined as function of the number of the epoch according to the scheduler function
\begin{flalign}\label{eq:learning_rate_scheduler}
\eta(\mathrm{epoch}) =  \eta_f + \frac{\eta_i-\eta_f}{1+\exp\big[-(\eta_c-\mathrm{epoch})/\sigma_\eta \big]},
\end{flalign}
with $\eta_i=8\times 10^{-4}$, $\eta_f=\eta_i/10$, $\eta_c=10^5$ and $\sigma_\eta=20\times10^3$. It can be easily seen that \eqref{eq:learning_rate_scheduler} is a decreasing function such that $\eta\sim \eta_i$ for $\mathrm{epoch}\ll \eta_c$ and $\eta\sim\eta_f$ for $\mathrm{epoch}\to \infty$. Setting a decreasing learning rate during the training is a standard technique to improve the convergence towards the minimum of the loss function and the choice of the aforementioned parameters has been done after a careful investigation.

In addition, for each case we perform several trainings by changing $\lambda$ and the neural network architecture and judge the best choice according to the lowest loss function. We will show an example of this study for the over-critical case and for the other cases (including the non-planar motion in Section~\ref{results:non_planar}) we will only quote the configuration that according to our analysis is the one that leads to the best performance.

As explained in Section~\ref{sec:Planar_Case} from \eqref{eq:spheroidal_coordinates}, the solution in the $x-y$ Cartesian plane (recall that now $\theta=\pi/2$) is given by

\begin{flalign}\label{eq:cartesian_map}
x =\sqrt{\rho^2+a_f^2}\cos \phi \,,
\nonumber\\[6pt]
y =\sqrt{\rho^2+a_f^2}\sin \phi \,.
\end{flalign}

For the three cases we set the  initial value of the radius $\rho_0$ to 10. The initial values of the variables $P_{\rho}$ and $\phi$ depend instead on the choice of the value of the impact parameter $b$. On the Cartesian plane, the impact parameter $b$ is identified with $y_0$ and therefore, the initial value of the angular variable is given by inverting the second equation of 
 \eqref{eq:cartesian_map}, that is

\begin{equation}\label{eq:plan_phi0}
    \phi_0 = \arcsin{\left(\dfrac{b}{\sqrt{\rho_0^2+a_{f}^{2}}}\right)}.
\end{equation}

The value of $P_{\rho,0}$ can be obtained by solving $\mathcal{H}(0)=0$ in (\ref{eq:Ham_plan}).

In each case we set the interval of discrete times in correspondence of which the loss function \eqref{eq:total_loss} has to be minimized by splitting the interval $\mathcal{T}_\mathrm{train}=[0,T]$ in $N$ equally spaced points. We keep this set fixed during the entire training process. We then construct a second interval of discrete times, denoted by $\mathcal{T}_\mathrm{val}$, by adding 5 additional equally spaced points between two consecutive times in $\mathcal{T}_\mathrm{train}$ and excluding the times in $\mathcal{T}_\mathrm{train}$. The interval $\mathcal{T}_\mathrm{val}$ thus contains $5\times(N-1)$ points which are not used during the training. For all the trainings performed in this work we have explicitly checked that, at the end of the training, the loss function evaluated for the set $\mathcal{T}_\mathrm{val}$ is compatible with that actually minimized within 5\%. This is a numerical check that the over-fitting of the solution has not occurred. Finally, we show in the plots the results predicted for the joint interval $\mathcal{T}_\mathrm{train}\cup\mathcal{T}_\mathrm{val}$ which contains $N+5\times(N-1)$ points in total.

In order to assess the performance of the neural networks we compare our results with those obtained by other numerical integrators. In particular we consider the first-order semi-implicit Euler integrator \eqref{eq:Euler}, that, as well known, preserves the symplectic structure of  Hamiltonian systems by conserving the energy up to a shift term proportional to the discretization step, and the more advanced Runge-Kutta method of order 5(4), \texttt{RK45} in short, which is not symplectic though. To implement the latter we use the SciPy package \texttt{solve\_ivp}. The package (see the documentation at \cite{solve_ivp}) allows conveniently to set the precision of the solution $y(t)$ in terms of absolute and relative tolerances, determined by the parameters $a_\mathrm{tol}$ and $r_\mathrm{tol}$ respectively, by keeping the local error estimates smaller than $a_\mathrm{tol}+ r_\mathrm{tol} |y(t)|$.  We have set in all the cases the relative and absolute tolerances to the minimum possible values $r_\mathrm{tol}=10^{-15}$ and $a_\mathrm{tol}=10^{-15}$, respectively.

\begin{figure}[t!]
\includegraphics[width=\columnwidth]{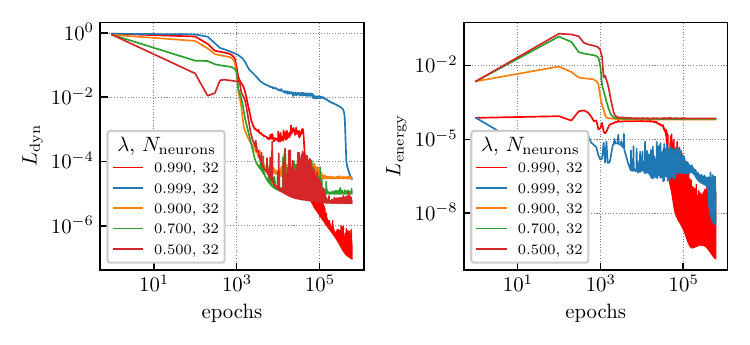}
\includegraphics[width=\columnwidth]{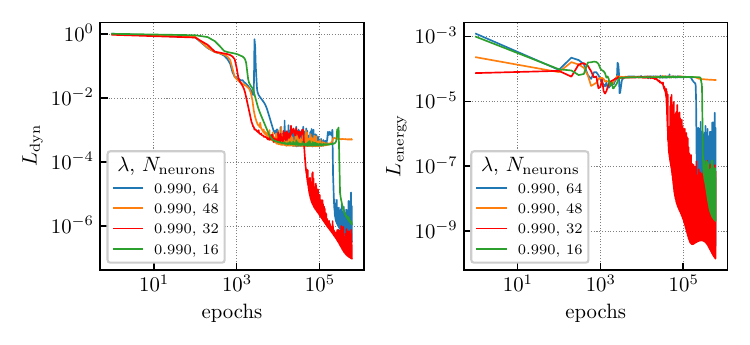}
\caption{
\label{fig:hamilton_overcritical_tuning} 
\footnotesize{
Tuning of the parameters for the over-critical case with $b=1.91$. In all the tests the neural network has 2 hidden layers and $\tanh(x)$ as activation function. \emph{Top-panels}: $L_\mathrm{dyn}$ (left) and $L_\mathrm{energy}$ (right) at fixed number of 32 neurons and changing $\lambda$ from 0.5 to 0.99. \emph{Bottom-panels}: the same as the top-panel but at fixed $\lambda=0.99$ and varying the number of neurons from 16 to 64. The four plots show that the best performance, in terms of minimization of the loss function, is given by the choice $\lambda=0.99$ and $N_\mathrm{neurons}=32$, corresponding to the red color.
}}
\end{figure}

\begin{figure}[t]
    \centering
    \includegraphics[width=\columnwidth]{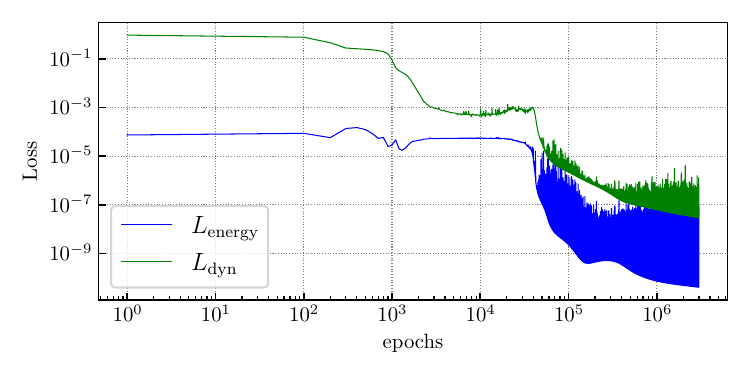}
    \caption{
    \label{fig:Hamilton_overcritical_loss}
    \footnotesize{
    Loss functions for the over-critical case with $b=1.91$. The training has been performed with $\lambda=0.99$, $N_\mathrm{neurons}=32$, 2 hidden layers and $\tanh(x)$ as activation function.
    }
    }
\end{figure}

We evaluate the numerical solution obtained from \texttt{RK45} in correspondence of the interval $\mathcal{T}_\mathrm{train}\cup\mathcal{T}_\mathrm{val}$ as well. The numerical solution provided by the Euler method is instead strongly dependent on the discretization step $\Delta s$, which regulates the update of solution in \eqref{eq:Euler}. We consider two time steps, $\Delta s=T/N$ which corresponds to the same step of $\mathcal{T}_\mathrm{train}$ and $\Delta s=T/(100N)$, which is instead 100 times finer. In the discussion of the results we will refer to these different solutions by Euler with $1\times N$ and $100\times N$ points respectively.

As explained in Appendix~\ref{APP:groundtruth}, we can compute exactly $\phi$ as function of $\rho$ and then, by plugging these values in \eqref{eq:cartesian_map}, we get the exact trajectory in the Cartesian plane. We call this trajectory \quote{ground truth} as customary in Machine Learning studies. The comparison with the ground truth only allows for a qualitative check of the correctness of the predicted geodesics. In order to have a quantitative measure of the error committed by the neural network and the numerical integrators, we can compare the exact function $\phi(\widehat{\rho})$ with $\widehat{\phi}(\widehat{\rho})$ and $\phi(\tilde{\rho})$ with $\tilde{\phi}(\tilde{\rho})$. The same comparison can be carried out also for $P_\rho$, whose true value can be found as function of $\rho$ by solving 
$\mathcal{H}(s)=0$ as explicitly done in \eqref{eq:Prho_exact_overcriical}-\eqref{eq:Prho_exact_subcritical}. 

We pass now to examine separately the three regimes.

\subsection{\label{sec:planar_overcritical}Over-critical case}
In this subsection we discuss the results for the over-critical case and set $b=1.91$. The initial conditions are therefore given by

\begin{flalign}\label{eq:init_val_over}
& \phi_0 = 0.1922, 
\nonumber\\[6pt]
& \mathcal{H}(0) = 0,
\nonumber\\[6pt]
& P_{\rho,0} = -0.9917.
\end{flalign}

\begin{figure}
\includegraphics[width=\columnwidth]{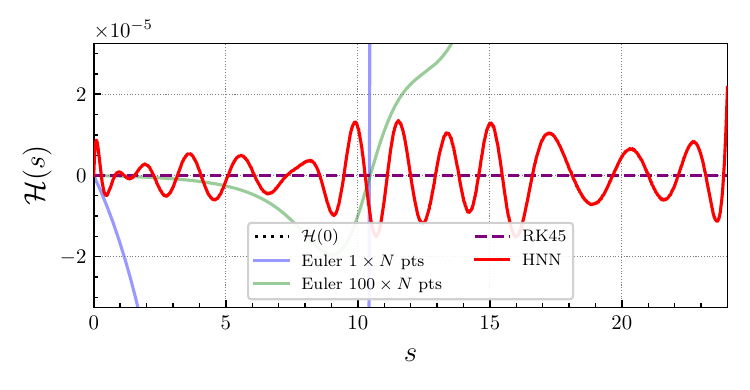}
\caption{
\label{fig:hamilton_overcritical_H}
\footnotesize{
Energy conservation over the time in the over-critical case with $b=1.91$  and $N=400$ for different methods. The conserved value of the energy $\mathcal{H}(0)=0$ is represented by the dotted black line.  
}
}
\end{figure}

To find the optimal setup, we have repeated the training several times by varying the trade-off parameter $\lambda$ and the number  of neurons $N_\mathrm{neurons}$  while keeping fixed the number of hidden layers to 2 and the activation function $\sigma(x)$ to the hyperbolic tangent
\begin{equation}\label{eq:act_fun}
    \sigma(x) = \tanh(x)
\end{equation}
For $\lambda $ we have considered the values 0.5, 0.7, 0.9, 0.99 and 0.999 while for $N_\mathrm{neurons}$  we have considered the values 16, 32, 48 and 64. 
The interval of time $\mathcal{T}_\mathrm{train}$ is given by splitting the interval $[0,24]$ in $N=400$ points. The neural network is trained with $600\times 10^3$ epochs for each combination of parameters. The loss functions $L_\mathrm{dyn}$ and $L_\mathrm{energy}$ are displayed in FIG.~\ref{fig:hamilton_overcritical_tuning}, where the top-panel shows the dependence on $\lambda$ at fixed $N_\mathrm{neurons}=32$, while the bottom-panel shows the dependence on $N_\mathrm{neurons}$ at fixed $\lambda=0.99$. The study shows a strong dependence upon both $\lambda$ and $N_\mathrm{neurons}$. In particular, at fixed $N_\mathrm{neurons}=32$, the loss functions get stuck in a plateau for $\lambda\le 0.9$ highlighting that, indeed, the addition of the conservation term $L_\mathrm{energy}$ favors a faster and better minimization of the total loss function. We have observed no significant improvement by passing from 2 to 3 hidden layers. 

The study shows that the best performance is achieved by the combination $\lambda=0.99$ and $N_\mathrm{neurons}=32$. We have then trained the neural network with such combination up to $3\times 10^6$ epochs and we have used it to predict the geodesic. The corresponding loss functions are shown in FIG.~\ref{fig:Hamilton_overcritical_loss}.

FIG.~\ref{fig:hamilton_overcritical_H} shows the function $\mathcal{H}(s)$ for the predicted geodesic (labeled by \quote{HNN} and represented by the solid red line), for the Runge-Kutta method (purple dashed line) and for the Euler method with respectively $N$ (blue) and $100\times N$ (green) points. The plot shows that the HNN conserves the energy very well on average with fluctuations between $-2\times 10^{-5}$ and $2\times 10^{-5}$. The Euler methods with both $N$ and $100\times N$ points exhibit larger violation of the energy that increases with the time $s$. The \texttt{RK45} integrator is remarkably good with violations from $\mathcal{H}(0)=0$ of order $10^{-9}$.

\begin{figure}
\includegraphics[width=\columnwidth]{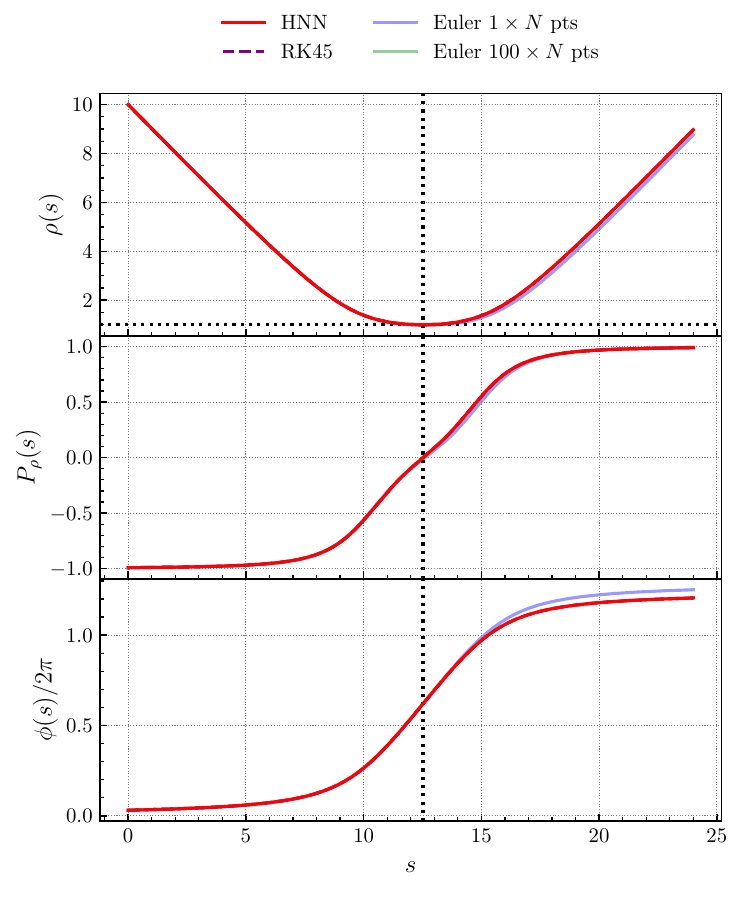}
\caption{
\label{fig:Hamilton_overcritical_variables}
\footnotesize{
From top to bottom: $\rho(s)$, $P_\rho(s)$ and $\phi(s)/2\pi$ for different methods in the over-critical case with $b=1.91$ and $N=400$. The vertical dotted line in the three panels represents $\widehat{s}_+=12.521$, that is the time at which the minimum of $\rho$ is reached and with respect to which the motion is symmetric. The motions to the left and to the right of $\widehat{s}_+$ are respectively incoming and outgoing. The horizontal dotted line in the top-panel represents $\rho_+=1.000138$.
}
}
\end{figure}

\begin{figure}
\includegraphics[width=\columnwidth]{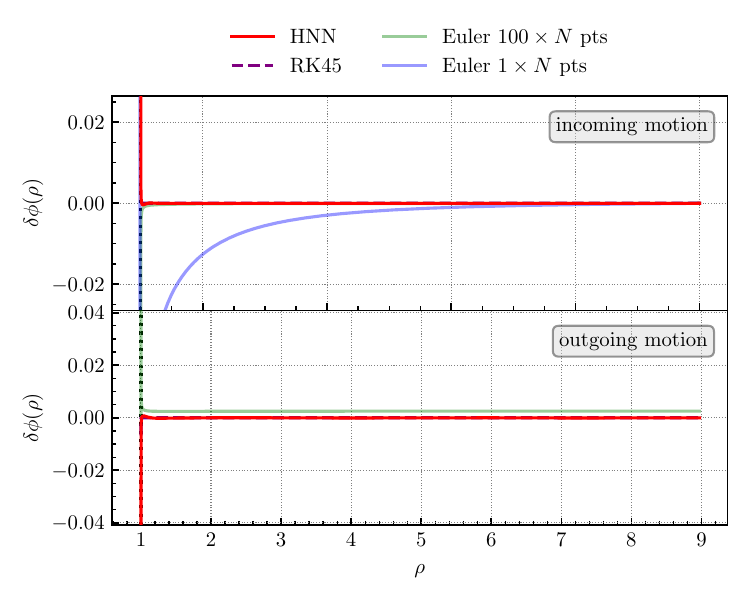}
\caption{
\label{fig:Hamilton_overcritical_phi_error}
\footnotesize{Error on the function $\phi(\rho)$ for different methods and for the incoming motion (\emph{top-panel}}) and the outgoing motion (\emph{bottom-panel)} in the over-critical case with $b=1.91$. The vertical dotted line represents $\rho_+=1.000138.$
}
\end{figure}

\begin{figure}
\includegraphics[width=\columnwidth]{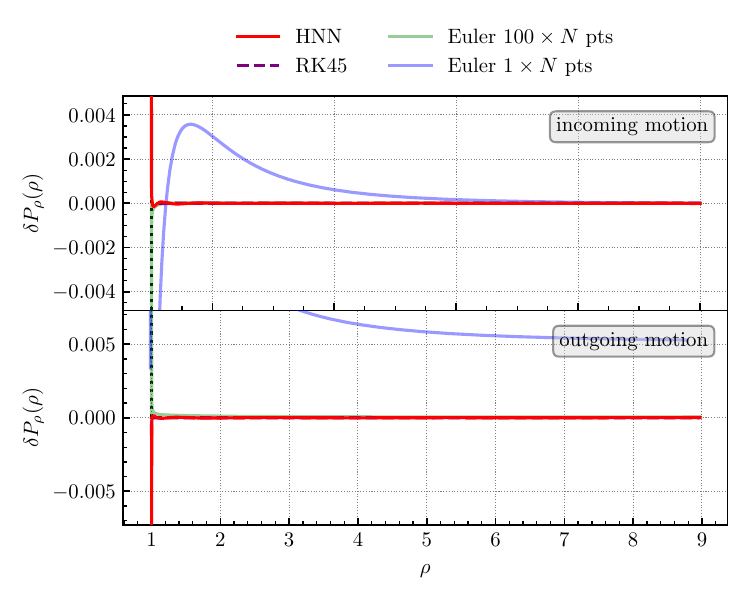}
\caption{
\label{fig:Hamilton_overcritical_Prho_error}
\footnotesize{Error on the variable $P_\rho(\rho)$  for different methods and for the incoming motion (\emph{top-panel}}) and the outgoing motion (\emph{bottom-panel)} in the over-critical case with $b=1.91$.  The vertical dotted line represents $\rho_+=1.000138.$
}
\end{figure}

Before showing the results for $\rho(s)$, $P_\rho(s)$ and $\phi(s)$, let us discuss their expectations based on physical arguments of Section~\ref{sec:Planar_Case}. At the beginning of the motion, the particle approaches the fuzzball (we refer to this phase as \quote{incoming motion}), and consequently $\rho$ decreases, until it reaches the minimum value $\rho_+=1.000138$, which is the largest root of the radicand in \eqref{eq:diffeq}. Afterward, the particle moves away from the fuzzball (we refer to this phase as \quote{outgoing motion}), and $\rho$ starts increasing again. We denote by $s_+$ the time at which the variable $\rho$ attains its minimum $\rho_+$. Due to the symmetry of the system under $\rho\mapsto -\rho$, the function $\rho(s)$ must be symmetric with respect to $s_+$.   On the other hand, $P_\rho$ starts from the negative value $P_{\rho, 0}$, given in \eqref{eq:init_val_over}, and increases over time until it reaches zero at $s_+$, eventually approaching a plateau at infinite time. Similarly, $\phi$ starts from the initial value $\phi_0$ written in \eqref{eq:init_val_over}, then increases and asymptotically reaches a plateau at infinite time.

This qualitative description of the motion is numerically confirmed by the functions $\rho(s)$, $P_\rho(s)$ and $\phi(s)/2\pi$ that are shown in  FIG.~\ref{fig:Hamilton_overcritical_variables} for different methods. The three dynamical variables behave qualitatively as expected and there is no qualitative difference between the prediction of the HNN, \texttt{RK45} and the Euler method with $100\times N$ points, while Euler with $N$ points visibly deviates from the other. From the numerical solution of the HNN we find that the minimum value of $\rho$ is  $\widehat{\rho}_+=1.000153$, 0.002\% away of $\rho_+$ , at time  $\widehat{s}_+=12.521$.

In order to have a quantitative measure of the quality of the solutions predicted by the HNN we consider the quantities
\begin{flalign}\label{eq:Prho_error}
\delta P_\rho(\hat\rho)&=\widehat{P_\rho}(\hat{\rho})-P_\rho(\hat{\rho}),
\end{flalign}
and
\begin{flalign}\label{eq:phi_error}
\delta \phi(\hat\rho)&=\widehat{\phi}(\hat{\rho})-\phi(\hat{\rho}).
\end{flalign}

\begin{figure}[t]
\includegraphics[width=\columnwidth]{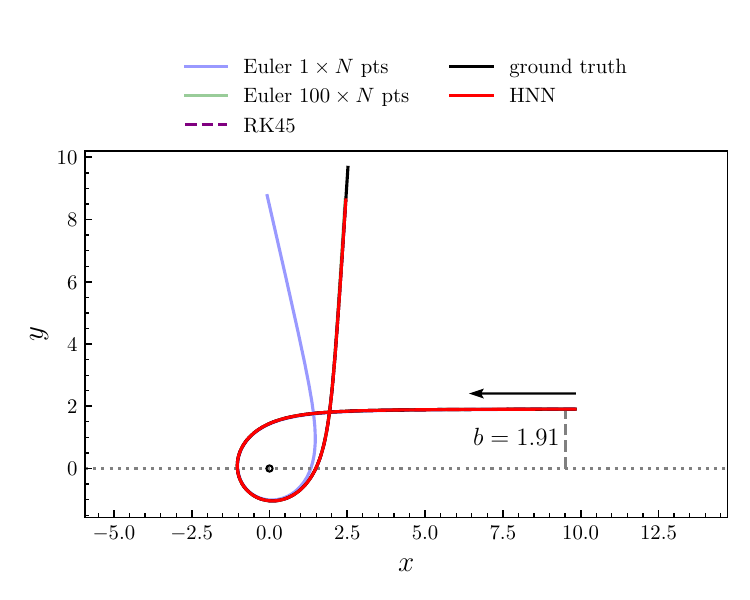}
\caption{
\label{fig:Hamilton_overcritical_cartesian}
\footnotesize{
Geodesics in the Cartesian plane obtained from different methods in the over-critical case with $b=1.91$ and $N=400$. The ground truth is represented by the solid black line, while the small circle at the center (0,0) has radius $a_f$ and represents the fuzzball. The black arrow indicates the incoming direction of the particle.
}
}
\end{figure}

Equivalent quantities can be considered also for the solutions obtained from numerical integrators. As anticipated before, the functions $P_\rho(\rho)$ and $\phi(\rho)$ can be computed exactly and therefore \eqref{eq:Prho_error} and \eqref{eq:phi_error} are good indicators to quantify the error made by a given method. For this reason we will refer to these quantities as \quote{errors} henceforth. The functions  $\phi(\rho)$ and $P_\rho(\rho)$ are obtained as explained in Appendix~\ref{APP:groundtruth}.

The errors of the HNN and the numerical integrators are shown in FIG.~\ref{fig:Hamilton_overcritical_phi_error} and FIG.~\ref{fig:Hamilton_overcritical_Prho_error} for the variable  $\phi$ and $P_\rho$ respectively. Both the figures show the error in the incoming motion (top-panel) and outgoing motion (bottom-panel). As it can be seen, the error of the HNN on the variable $\phi$ oscillates uniformly around 0 with an order of magnitude of $10^{-4}$ and with a spike occurring for $\rho$ close to $\rho_+$. An analogous picture is found for the variable $P_\rho$ but with an order of magnitude of $10^{-5}$. The small scale of the errors (compare with $P_\rho(\rho)$ and $\phi(\rho)$ in FIG.~\ref{fig:phi_exact} and FIG.~\ref{fig:Prho_exact} in Appendix \ref{APP:groundtruth}) is a reassuring evidence that the Hamilton equations have been solved by the HNN with a high accuracy.  The Euler method with both $N$ and $100\times N$ points has larger errors than HNN. \texttt{RK45} is instead the most precise, with errors that fluctuate on the scale $10^{-9}$ and $10^{-10}$ for $\phi$ and $P_\rho$ respectively.

\begin{figure}[t]
\includegraphics[width=\columnwidth]{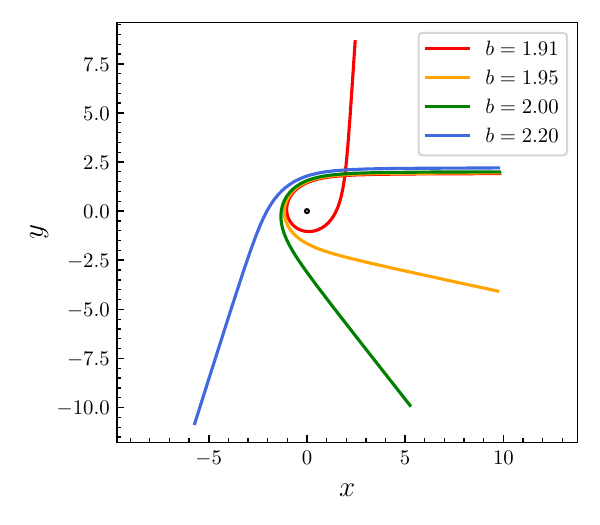}
\caption{
\label{fig:Hamilton_overcritical_multib}
\footnotesize{Geodesics in the Cartesian plane predicted by the HNN for different values of the impact parameter $b$.}
}
\end{figure}

We eventually project the motion obtained for the different methods  to the Cartesian plane by using \eqref{eq:cartesian_map} and compare to the ground truth. The geodesics in the Cartesian plane are shown in FIG.~\ref{fig:Hamilton_overcritical_cartesian}. The HNN and \texttt{RK45} have no visible differences with respect to the ground truth, showing that they are both efficient integrators of the Hamilton equation of motions. Concerning the Euler method, for the same number of points as those used to train the neural network the trajectory shows a clear deviation from the ground truth over long timescales and a reliable result is obtained only by considering a number of points (time step) larger (smaller) by a factor 100. 

We conclude this subsection by employing the HNN to compute the geodesics for impact parameters $b=1.95$, $2.0$ and $2.2$. The three cases have been obtained by considering the same setup as $b=1.91$ discussed above, that is $\lambda=0.99$, $N_\mathrm{neurons}=32$, $N=400$, $\mathcal{T}_\mathrm{train}=[0,24]$ and $3\times10^6$ epochs. For all the three cases we have computed the errors and checked that the quality of the solution is at the same level as the case with $b=1.91$. In FIG.~\ref{fig:Hamilton_overcritical_multib} we show the final geodesics predicted by the HNN in the Cartesian plane.  As it can be seen from the plot, the scattering angle at which the particle moves away from the fuzzball increases as the impact parameter increases and this is expected by theory (in fact, as the impact parameter increases, the particle is less affected by the gravitational field of the fuzzball, and therefore its trajectory is less bended). We leave to a future work the intriguing possibility to treat the impact parameter $b$ not as fixed algorithmic parameter but as an additional variable to feed to the neural network in input together with the time.

\subsection{\label{sec:planar_critical}Critical case}
In this subsection we present the critical case obtained by setting $b=b_c=1.9$. The initial conditions now read

\begin{flalign}\label{eq:init_val_critical}
& \phi_0 = 0.1912, 
\nonumber\\[6pt]
& \mathcal{H}(0) = 0,
\nonumber\\[6pt]
& P_{\rho,0} = -0.9919.
\end{flalign}

\begin{figure}[t]
    \centering
    \includegraphics[width=\columnwidth]{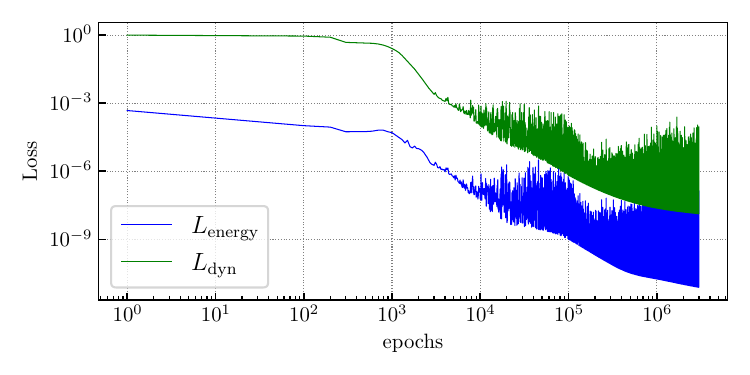}
    \caption{
    \label{fig:Hamilton_critical_loss}
    \footnotesize{
    Loss functions for the critical case with $b=b_c=1.9$. The training has been performed with $\lambda=0.99$, $N_\mathrm{neurons}=48$, 2 hidden layers and $\tanh(x)$ as activation function.
    }
    }
\end{figure}

\begin{figure}[t]
    \centering
    \includegraphics[width=\columnwidth]{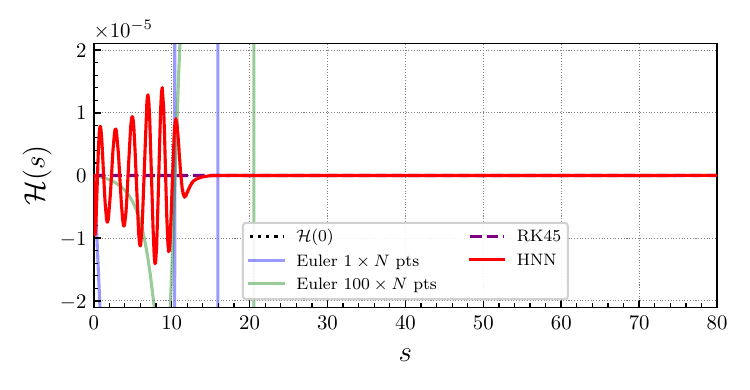}
    \caption{
    \label{fig:Hamilton_critical_H}
    \footnotesize{
Energy conservation over the time in the critical case with $b=b_c=1.9$  and $N=800$ for different methods. The conserved value of the energy $\mathcal{H}(0)=0$ is represented by the dotted black line.  
    }
    }
\end{figure}

For this case we have considered the time interval $\mathcal{T}_\mathrm{train}=[0,80]$ sampled with $N=800$ points. The reason behind such a long time interval is to probe the stability of the trajectory on the critical radius circumference over long timescales. We have repeated the tuning procedure discussed in the over-critical case and found that the optimal setup is given by $\lambda=0.99$ and $N_\mathrm{neurons}=48$. The number of hidden layers is 2 and the activation function is given by \eqref{eq:act_fun}. The neural network is trained with $3\times 10^6$ epochs and the corresponding loss functions $L_\mathrm{dyn}$ and $L_\mathrm{energy}$ are shown in FIG.~\ref{fig:Hamilton_critical_loss}.

\begin{figure}
    \centering
    \includegraphics[width=\columnwidth]{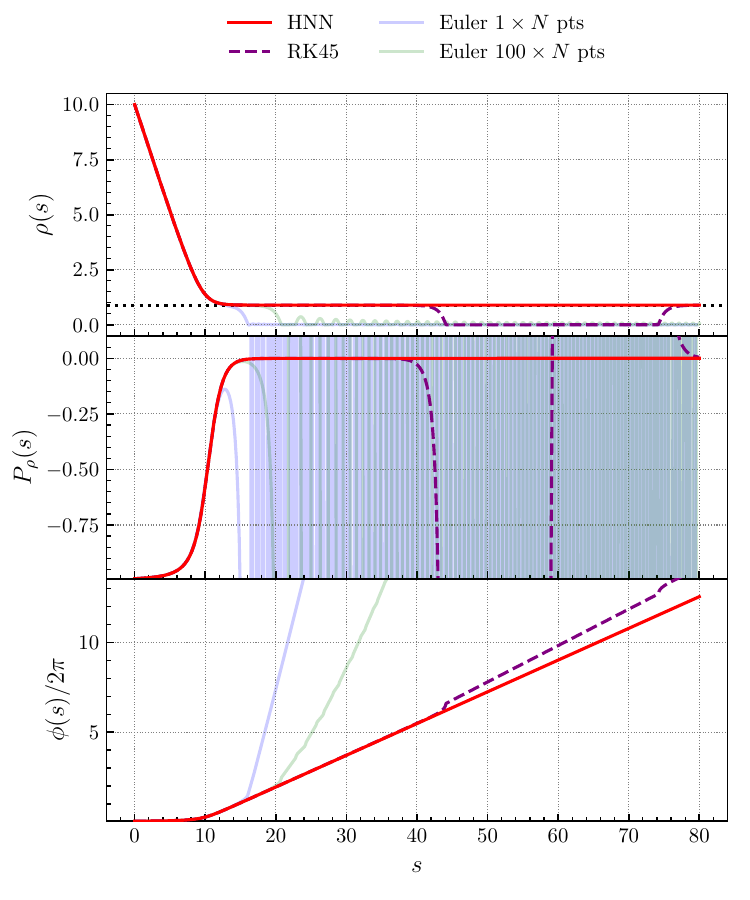}
    \caption{
    \label{fig:Hamilton_critical_variables}
    \footnotesize{
    From top to bottom: $\rho(s)$, $P_\rho(s)$ and $\phi(s)/2\pi$ for different methods in the critical case with $b=b_c=1.9$ and $N=800$. The horizontal black dotted line in the top panel represents the critical radius $\rho_c=0.8944$, on which the trajectory must settle. The trajectory predicted by the HNN is the only one among the different methods to correctly reproduce this feature.
    }
    }   
\end{figure}

In FIG.~\ref{fig:Hamilton_critical_H} we show the function $\mathcal{H}(s)$ for the different methods. The energy is very well conserved on average by the HNN with oscillations of order $10^{-5}$ for $s$ smaller than $\simeq 14$ and of order $10^{-9}$ for larger times. The Euler method with $N$ and $100\times N$ points does not conserve the energy from the very beginning. Particular attention should be paid to the \texttt{RK45}, which conserves the energy on a scale of $10^{-13}$. This information seems to indicate the \texttt{RK45} method as the best for trajectory prediction. However, as we will see shortly, exact energy conservation by a non-symplectic method does not automatically imply an exact solution of the equations of motion.

\begin{figure}
    \centering
    \includegraphics[width=\columnwidth]{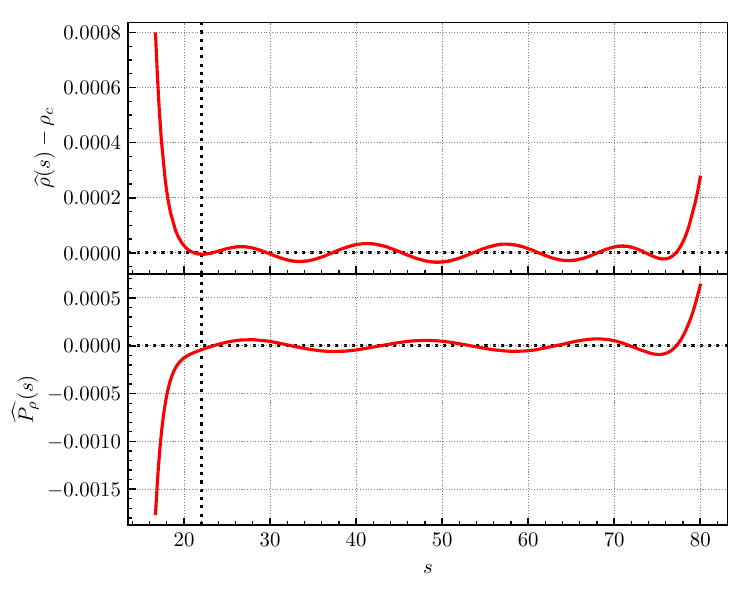}
    \caption{
    \label{fig:Hamilton_critical_differences}
    \footnotesize{Stability of the trajectory predicted by the HNN over the long timescale.  The vertical dotted line is drawn at $s\simeq 22$, the time at which approximately the trajectory reaches the critical circumference.   \emph{Top-panel}: difference between $\widehat{\rho}(s)$ and $\rho_c=0.8944$. \emph{Bottom-panel}:  predicted $\widehat{P_\rho}(s)$.
    }
    }
\end{figure}

Concerning the physics of the motion, in this case we only have an incoming phase in which the radial coordinate $\rho(s)$ decreases from $\rho_0$ up to reach asymptotically the value $\rho_c=0.8944$. 
 Correspondingly, the radial momentum $P_\rho(s)$ increases from the negative value $P_{\rho,0}$ given in \eqref{eq:init_val_critical}  up to freezing at 0. The function $\phi(s)$ is instead expected to have a linear grow as the particle stabilizes rotating on the critical geodesic.

FIG.~\ref{fig:Hamilton_critical_variables} shows the functions $\rho(s)$, $P_\rho(s)$ and $\phi(s)/2\pi$ for HNN and the numerical integrators. As it can be seen, the HNN is the only method that correctly reproduces the expected dynamics of the motion. The Euler method with both $N$ and $100\times N$ points starts failing around $s\simeq 15$ and exhibit a badly oscillating behavior. The  \texttt{RK45},  despite exactly conserving energy, fails for $s>44$. In particular $\rho(s)$ for both Euler and \texttt{RK45} approaches touches 0, which corresponds to a situation in which the particle trajectory has collapsed onto the sphere of radius $a_f$ and it is physically impossible for this impact parameter. The variable $\phi(s)$ predicted by the HNN exhibits the correct linear grow for times larger  than $s\simeq 22$, which we can consider as the time at which the trajectory stabilizes on the critical circumference. The angle $\phi(s)$ is plotted normalized by $2\pi$ in order to give an idea of the number of windings of the trajectory around the fuzzball. From time $s = 22$ to time $s = 80$, the trajectory completes approximately 10 full windings and remains stable throughout.

Since numerical integrators clearly fail, it does not make sense to compare the error of the HNN with the error of the integrators. Rather, to assess the stability of the solution predicted by the neural network, we show in the FIG.~\ref{fig:Hamilton_critical_differences} the quantities  $\widehat{\rho}(s)-\rho_c$ and $\widehat{P_\rho}(s)$ for large times. Both quantities oscillate around 0 on a remarkably small scale. A small deviation from 0 is observed for $s\simeq 76$. We will investigate the cause of this effect in the future, but it is possible that it could simply be reduced by increasing the number of epochs. As can be seen in the FIG.~\ref{fig:Hamilton_critical_loss}, in fact, the loss functions have not yet reached their minimum.

At this point, we emphasize that although the \texttt{RK45} method is very accurate and the Euler method is symplectic, the HNN is the only one that correctly integrates the equations of motion, outperforming both methods in terms of efficiency and accuracy. The HNN ability to predict a stable trajectory over long timescales is due to the introduction of the energy conservation term, which penalizes solutions that violate this constraint. The ability to enforce constraints within the Machine Learning framework proves to be an extremely beneficial factor in this case, enabling accurate results where standard methods fail.

\begin{figure}
    \centering
    \includegraphics[width=\columnwidth]{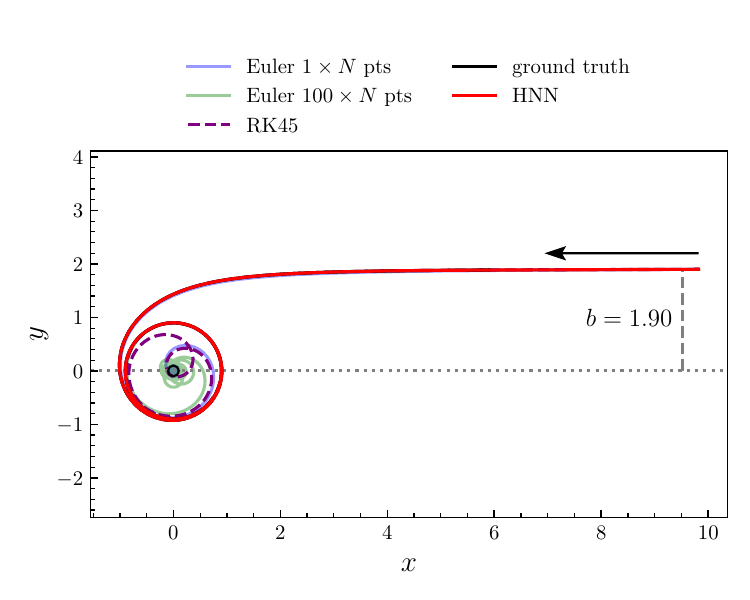}
    \caption{
    \label{fig:Hamilton_critical_cartesian}
    \footnotesize{
Geodesic in the Cartesian plane obtained from different methods in the critical case with $b=b_c=1.9$ and $N=800$. The ground truth is represented by the solid black line, while the small circle at center (0,0) has radius $a_f$ and it represents the fuzzball. The black arrow indicates the incoming direction of the particle.
    }
    }
\end{figure}

We then project the  motion to the Cartesian plane by means of (\ref{eq:cartesian_map}) and the result is shown in FIG.~\ref{fig:Hamilton_critical_cartesian} together with the ground truth. As it can be appreciated, the geodesic predicted by the HNN remains stable along the circumference of radius $\sqrt{\rho_c^2+a_f^2}$, while all the other methods predict a geodesic that collapses onto the fuzzball.

\subsection{Sub-critical case} \label{subsec:sub-critical}

In this subsection we present the results for the last regime, that is the sub-critical one, and we study the situation with $b=1.89$. The initial conditions are

\begin{flalign}\label{eq:init_val_subcritical}
& \phi_0 = 0.1901, 
\nonumber\\[6pt]
& \mathcal{H}(0) = 0,
\nonumber\\[6pt]
& P_{\rho,0} = -0.9921.
\end{flalign}

This case is particularly challenging from the HNN perspective, as the underlying physics involves an impact of the particle against the fuzzball, hence the radius $\rho$ decreases from $\rho_0$ down to zero (this can be also deduced from the first of \eqref{eq:diffeq}, since the radicand never vanishes). As a result, the conjugate momentum $P_\rho$ increases in absolute value until it reaches the value
\begin{equation} \label{minimal_prho_sub_critical}
    \overline{P_{\rho}} \equiv P_\rho(\rho=0)= -80.1249,
\end{equation}
where we have used $P_\rho(\rho)$ in \eqref{eq:Prho_exact_subcritical}. The problem arises from the fact that the decrease from $P_{\rho,0}$ to $\overline{P_\rho}$ is very rapid in a small interval of values of $\rho$, which are close to 0, and this is shown in the right-panel of FIG.~\ref{fig:Prho_exact}. The same behavior is observed in the time domain, as it can be  seen beforehand in the central panel of FIG.~\ref{fig:Hamilton_subcritical_variables}. From this perspective, the differential equation governing the dynamics of $P_\rho$ is a stiff equation and, as well known, stiff equations can be a challenge to numerical methods, HNN included. Indeed, we did not manage to obtain accurate solutions by using the same strategy adopted in the past two subsections. Therefore, in the following we propose two alternative approaches to overcome this difficulty and get a reliable prediction for the geodesic.

\subsubsection{\label{sec:splitting}Approach 1: splitting the time interval}

In the first approach we split the time interval over which we want to compute the solution in such a way to separate the stiff regime of $P_\rho$ from that in which its decrease is smoother. This can be easily achieved by dividing the total time interval into consecutive sub-intervals and we will refer to this as \quote{time-splitting approach}. For each sub-interval, a neural network is trained from scratch using as initial conditions the values of the predicted variables at the final time of the previous interval. We consider the following three time intervals, $\mathcal{T}_\mathrm{train}^{(1)}=[0,14]$, for which the initial conditions are given in    \eqref{eq:init_val_subcritical}, $\mathcal{T}_\mathrm{train}^{(2)}=[14,14.8]$, for which the initial conditions are $\rho_0=0.6552$, $P_{\rho,0}=-0.8982$ and $\phi_0=5.8452$, and finally
$\mathcal{T}_\mathrm{train}^{(3)}=[14.8,15.22]$, for which the initial conditions are $\rho_0=0.3014$, $P_{\rho,0}=-7.0688$ and $\phi_0=7.2790$. All the three intervals are divided into $800$ points for a total number of $N=2400$ points. Notice that, at practical level, the time interval to feed to the neural network always starts at time 0, that is $\mathcal{T}_\mathrm{train}^{(2)}$ and $\mathcal{T}_\mathrm{train}^{(3)}$ actually are $[0,0.8]$ and $[0,0.42]$. Setting as initial conditions the values predicted at the end of the previous interval is what guarantees that the predictions, on the sub-intervals, glued together eventually correspond to the solution in the overall interval $\mathcal{T}_\mathrm{train}=[0,15.22]$. 

\begin{figure}
    \centering
    \includegraphics[width=\columnwidth]{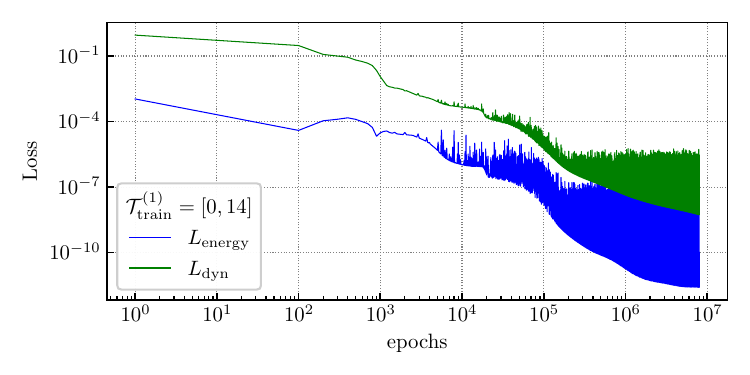}
    \includegraphics[width=\columnwidth]{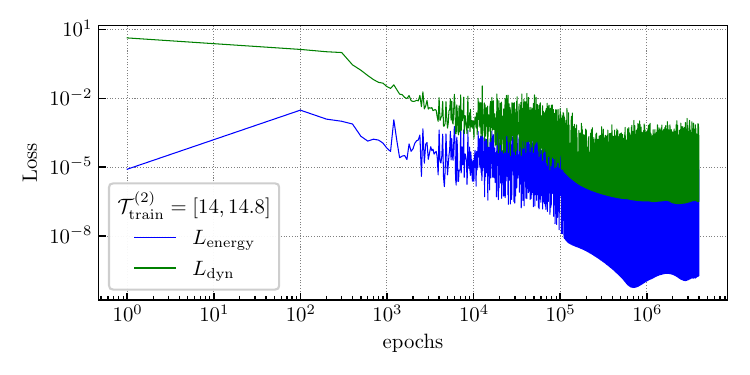}
    \includegraphics[width=\columnwidth]{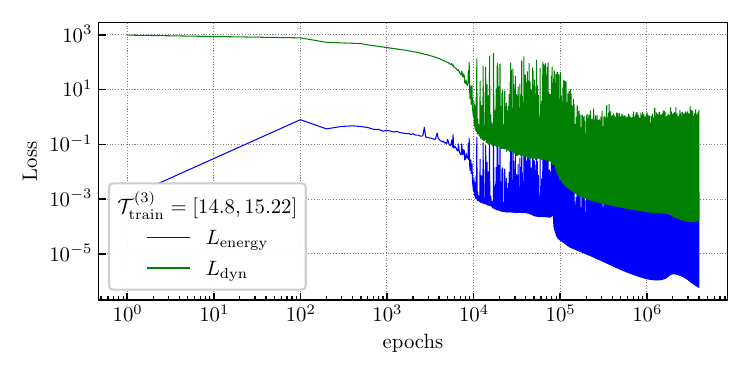}
    \caption{
    \label{fig:Hamilton_subcritical_losses}
    \footnotesize{
    Loss functions for the subcritical case with $b=1.89$. From top to bottom the losses refer to the training intervals $\mathcal{T}_\mathrm{train}=[0,14]$, $[14,14.8]$ and $[14.8,15.22]$.
    }
    }
\end{figure}

FIG.~\ref{fig:Hamilton_subcritical_losses} shows the loss functions for the three time intervals. The numbers of epochs, respectively, are $8 \times 10^6$, $4\times 10^6$ and $4\times 10^6$. The neural networks in all three cases consist of 3 hidden layers with 64 neurons per each. $\lambda=0.99$ again and the activation function always is $\tanh{(x)}$.

\begin{figure}
\includegraphics[width=\columnwidth]{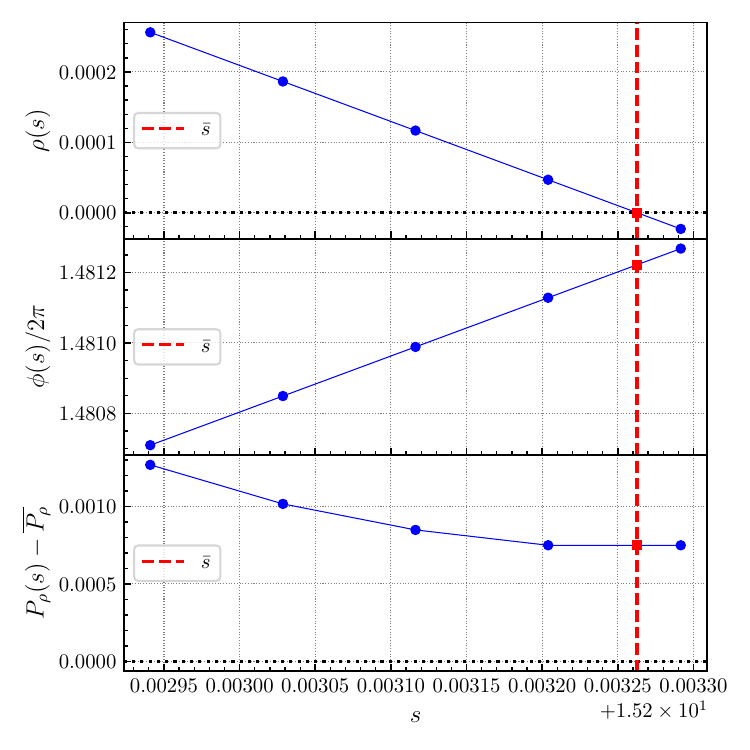}
\caption{
\label{fig:prediction_interpolation}
\footnotesize{
From top to bottom: interpolation of $\rho(s)$, $\phi(s)/2\pi$ and $P_\rho(s)-\overline{P_\rho}$ at the time $\bar{s}$. The plot refers to the HNN with the time-splitting approach for which $\bar{s}=15.2032$ (vertical dashed line). In the top and bottom panel we highlight the value 0 with a horizontal dotted line.
}
}
\end{figure}

The final time $s=15.22$ has been chosen such that $\rho$ reaches and slightly exceeds the value 0. We call $\bar{s}$ the  time at which $\rho(\bar{s})=0$. The motion in the interval $[0,\bar{s}]$ corresponds to the incoming phase, in which the particle approaches the fuzzball up to bounce from it, and, thanks to the symmetry $\rho\mapsto -\rho$ of the system, the outgoing phase can be found by reflecting the incoming motion with respect to $\bar{s}$.

\begin{figure}[t]
    \centering
    \includegraphics[width=\columnwidth]{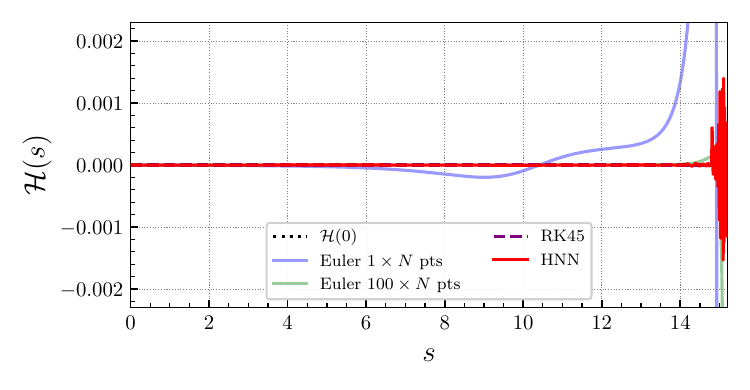}
    \caption{
    \label{fig:Hamilton_subcritical_H}
    \footnotesize{
Energy conservation for the sub-critical case with $b=1.89$  for the different methods and by following the time-splitting approach. The conserved value of the energy $\mathcal{H}(0)=0$ is represented by the dotted black line.}
    }
\end{figure}

FIG.~\ref{fig:prediction_interpolation} shows the procedure we have employed to determine $\bar{s}$ in the case of HNN. In the top-panel we determine $\bar{s}$ by calculating the time at which the curve $\rho(s)$ intersects the line $\rho=0$ (assuming a linear ansatz due to the small range of times we are focusing on). The resulting time is $\bar{s}=15.2032$ and it is marked by the red vertical dashed line. The fact that the mathematical solution to the system predicts $\rho <0$ for $s>\bar{s}$ is irrelevant to the physics since the geodesics in the Cartesian plane (see \eqref{eq:cartesian_map}) depends on $\rho^2$.  We then interpolate $\phi(s)/2\pi$ and $P_\rho(s)$ at the time $\bar{s}$ (middle- and bottom-panels) and discard all the points for which $s>\bar{s}$.  The procedure is repeated independently also for the numerical integrators.

The conservation of the energy is shown in FIG.~\ref{fig:Hamilton_subcritical_H} for the different methods. The HNN conserves very well the energy over the time and an increasing of the oscillations up to $\pm 0.001$ is observed as $s\to \bar{s}$. The Euler method has larger deviations both for $N$ and $100\times N$, points while the \texttt{RK45} method conserves the energy much better compared to the others.

\begin{figure}
\includegraphics[width=\columnwidth]{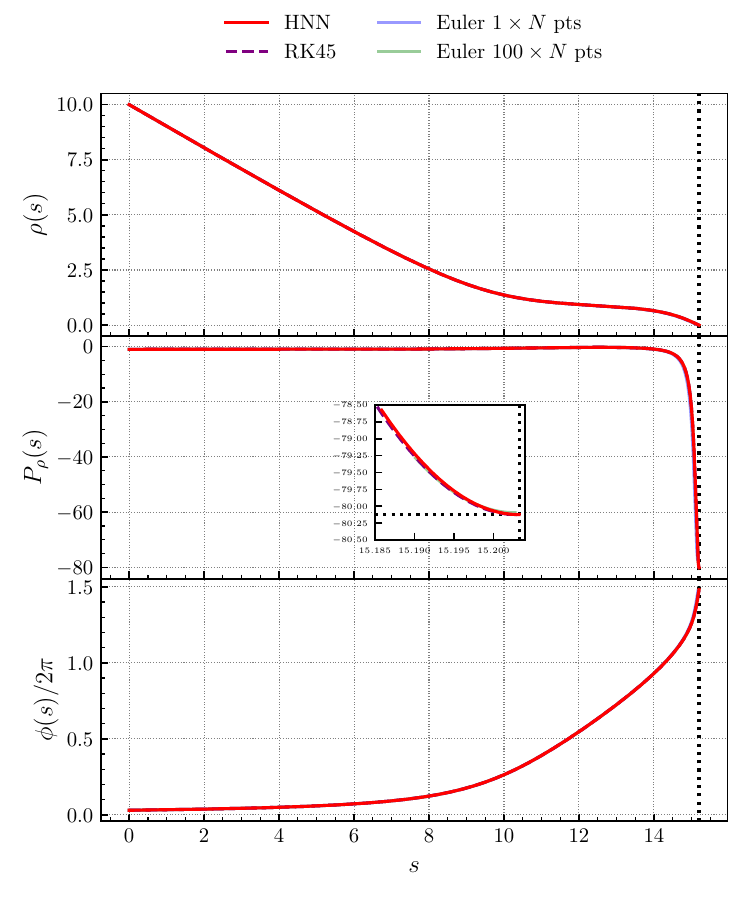}
\caption{
\label{fig:Hamilton_subcritical_variables}
\footnotesize{
From top to bottom: $\rho(s)$, $P_\rho(s)$ and $\phi(s)/2\pi$ for the different methods in the sub-critical case with $b=1.89$ and $N=2400$ obtained with the time-splitting approach. The vertical dotted line in the three panels represents $\bar{s}$ for the HNN. In the middle-panel we show also a zoom in the region $s\in[15.185,\bar{s}]$.}
}
\end{figure}

\begin{figure}
\includegraphics[width=\columnwidth]{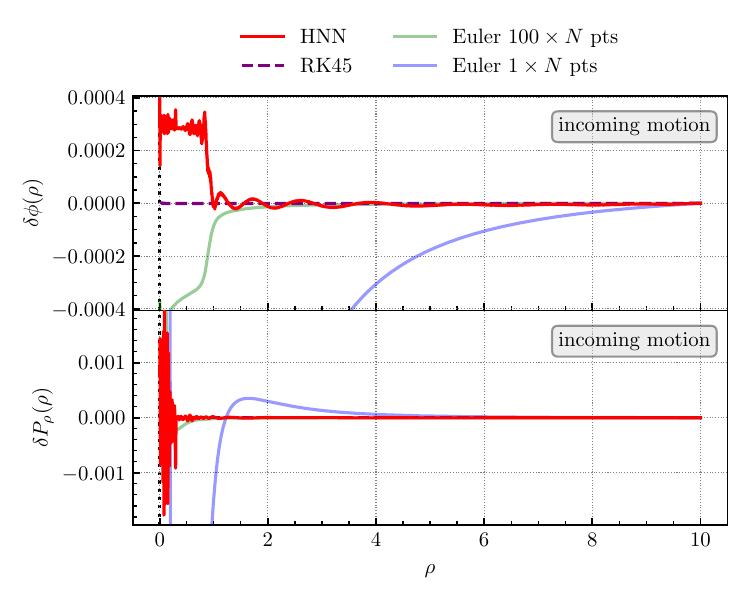}
\caption{
\label{fig:Hamilton_subcritical_errors}
\footnotesize{Errors on the variables $\phi(\rho)$ and $P_\rho(\rho)$ in the sub-critical case with $b=1.89$ obtained from the time-splitting approach. 
}
}
\end{figure}

The dynamical variables $\rho(s)$, $P_\rho(s)$ and $\phi(s)/2\pi$ obtained from the different methods are shown in FIG.~\ref{fig:Hamilton_subcritical_variables}. No visible differences are observed among them. 

To quantify the goodness of the solutions, we still rely on the errors as defined in \eqref{eq:phi_error} and \eqref{eq:Prho_error}, where now for the exact $P_\rho(\rho)$ we use \eqref{eq:Prho_exact_subcritical}. The errors are plotted in FIG.~\ref{fig:Hamilton_subcritical_errors} for the different methods in the incoming motion (the outgoing motion would be identical, since we obtain it from reflection of the first one). The errors made by the HNN tend to increase for both the variables as $\rho \to 0$, but remain very small compared to the scale of the variables in question, showing that the HNN is able to solve the system with high accuracy. The Euler method with both $N$ and $100\times N$ points show larger deviations, while \texttt{RK45} is very accurate.

\begin{figure}
    \centering
    \includegraphics[width=\columnwidth]{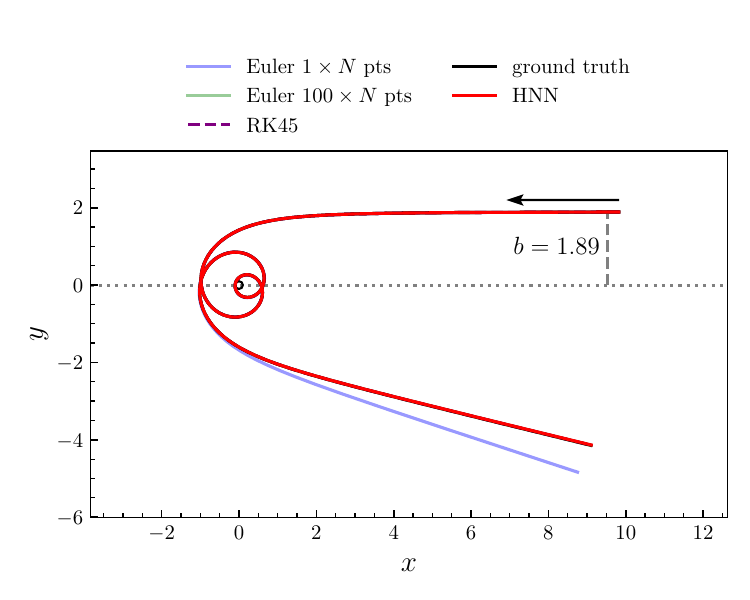}
    \caption{
    \label{fig:Hamilton_subcritical_cartesian}
    \footnotesize{
Geodesic in the Cartesian plane obtained from the different methods in the sub-critical case with $b=1.89$ and by applying the time-splitting approach. The ground truth is represented by the solid black line, while the small circle at center (0,0) has radius $a_f$ and it represents the fuzzball. The black arrow indicates the incoming direction of the particle.
    }
    }
\end{figure}

FIG.~\ref{fig:Hamilton_subcritical_cartesian} shows the complete (incoming and ougoing phases) geodesic in the Cartesian plane obtained from the different methods. HNN, Euler with $100\times N$ points and \texttt{RK45} are qualitatively overlapping with the ground truth, while Euler with $N$ points (blue line) deviates significantly from it during the outgoing phase.

\subsubsection{Approach 2: preconditioning of the HNN output}

The second approach we propose consists in providing an offset to the network's output associated with the problematic variable, in this case $P_\rho$, relieving the network from having to learn a particularly stiff behavior. In order to understand better this concept, let us consider the following redefinition 
\begin{flalign}\label{eq:z_hat_precond}
    \widehat{P_{\rho}}(s,\textbf{w}) = P_{\rho,0}+f(s)\big[O_{P_\rho}(s,    \textbf{w})+h(s)\big],
\end{flalign}
where we have introduced the function $h(s)$, representing the new offset with respect to which the output $O_{P_\rho}$ has to be trained to solve the Hamilton equation for $P_\rho(s)$. The function $h(s)$ is chosen arbitrarily and does not depend on the network weights. The idea is to incorporate into $h(s)$ the complexity of the solution $P_\rho(s)$, in such a way that the residual function $O_{P_\rho}(s,\textbf{w})$, that has to be learned by the HNN, is close to zero or has a much smoother behavior. Let us emphasize that minimizing the loss function leads to finding the solution for $P_{\rho}(s)$ regardless of the definition of $h(s)$. If $h(s)$ is completely different from the expected solution, the offset will be automatically corrected by the output $O_{P_\rho}(s,\textbf{w})$. From this perspective, the strategy remains model-independent. We will refer to this approach as \quote{preconditioning} of the HNN output.

In order to capture the rapid decrease of $P_{\rho}(s)$ near $\overline{s}$, we have considered the following Breit-Wigner-type parametrization for $h(s)$
\begin{flalign} \label{eq:fit_ansatz}
    h(s)= \dfrac{\big(\overline{P_\rho} -P_{\rho,0}\big)\,\Gamma^{2}}{\Gamma^{2}+(s-\bar{s})^2}.
\end{flalign}

\begin{figure}[t]
    \centering
    \includegraphics[width=\columnwidth]{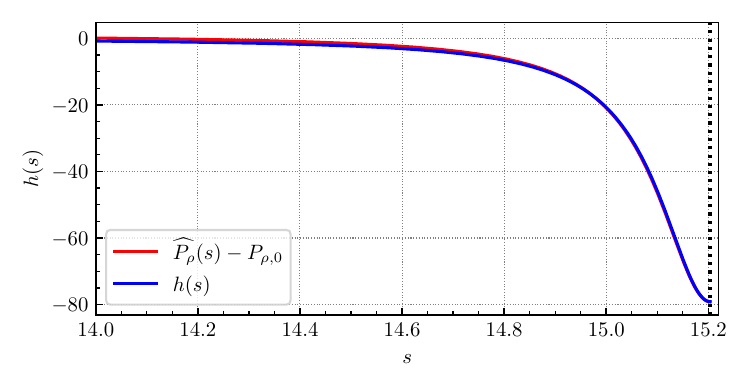}
    \caption{ Result of the fit (blue line) compared to the fitted quantity (red line). In this case $\widehat{P_\rho}(s)$ is the prediction of the HNN obtained from the time-splitting approach in Section~\ref{sec:splitting}.
    \label{fig:h}
    \footnotesize{}
    }
\end{figure}

With the aim of minimizing the amount of prior information, we have considered the accurate prediction $\widehat{P_\rho}(s)$ obtained from the time-splitting procedure of the previous subsection, then $\bar s =15.2032$ and fitted the quantity  $\widehat{P_\rho}(s)-P_{\rho,0}$ in the time range $s\in[14,\bar s]$ with the fit ansatz \eqref{eq:fit_ansatz}, getting $\Gamma = 0.1213$. In the fitting procedure we have neglected the effect of the function $f(s)$ since we are interested in parameterizing the solution at $s\gg 1$, for which $f(s)\simeq 1$. The result of the fit is drawn in blue in FIG.~\ref{fig:h}. As it can be seen, the so-defined $h(s)$ function provides and excellent parametrization for $\widehat{P_\rho}(s)-P_{\rho,0}$ (in red) for times close to $\bar s$. Let us say that in general one can rely on other source of information, if available, such as the result of the numerical integrators, or one can design ansätze by leveraging theoretical knowledge of the system.

With this setup, we have trained a neural network made of 3 hidden layers and 64 neurons per each for $30\times 10^6$ epochs.  $\eta_i$, appearing in \eqref{eq:learning_rate_scheduler}, is set to $4\times 10^{-4}$. The time interval is $\mathcal{T}_\mathrm{train}=[0,15.22]$ and it is divided in $N=800$ points. 
In addition we have set $\gamma_{P_\rho}=0.01$ (recall the definition of the dynamics loss function in \eqref{eq:loss_dyn}) in order to give more importance to the dynamical equations for $\rho(s)$ and $\phi(s)$, since the one for $P_\rho(s)$ is basically already solved.

\begin{figure}[t]
    \centering
    \includegraphics[width=\columnwidth]{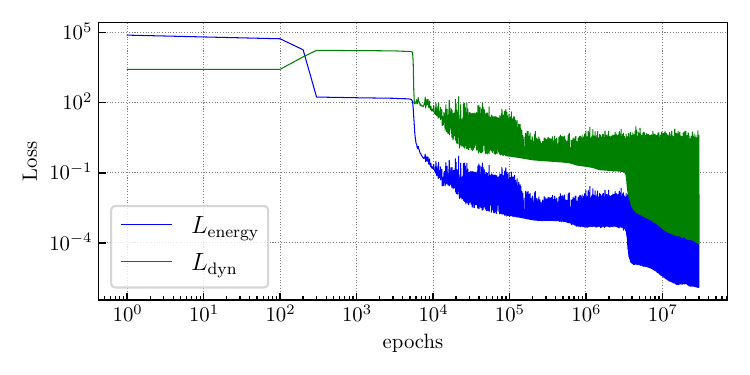}
    \caption{
    \label{fig:prec_loss}
    \footnotesize{Loss functions for the sub-critical case with $b=1.89$ by using the preconditioning approach.}
    }
\end{figure}

The loss functions are shown in FIG.~\ref{fig:prec_loss} and are at a similar level as those corresponding to the third interval in the time-splitting approach, which are displayed in the bottom-panel of FIG.~\ref{fig:Hamilton_subcritical_losses}.

\begin{figure}
    \centering
    \includegraphics[width=\columnwidth]{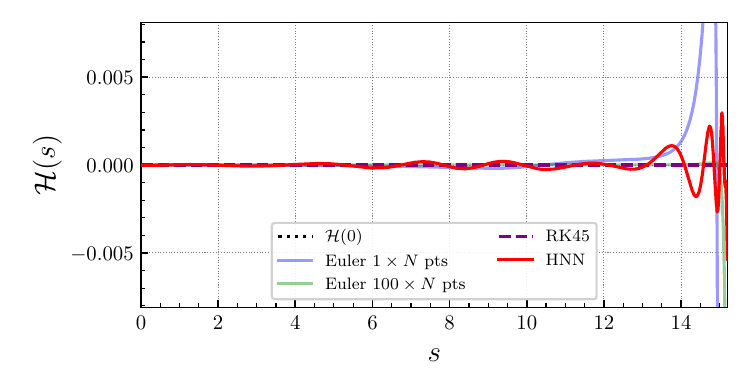}
    \caption{
    \label{fig:prec_H}
    \footnotesize{Energy conservation for the sub-critical case with $b=1.89$  for the different methods and by following the preconditioning approach. The conserved value of the energy $\mathcal{H}(0)=0$ is represented by the dotted black line.}
    }
\end{figure}

The results from the numerical integrators are taken from the previous subsection and we will not comment them here again. For the HNN predictions we have repeated the interpolation procedure illustrated in FIG.~\ref{fig:prediction_interpolation} to determine $\bar s$ (which still remains at the value $15.2032$) and consequently to discard the points for which $s>\bar s$. This procedure is also needed to obtain the outgoing trajectory by reflection of the incoming one. 

As shown in FIG.~\ref{fig:prec_H}, the energy is well conserved on average and, compared to the time-splitting method (see FIG.~\ref{fig:Hamilton_subcritical_H}), the deviations from $\mathcal{H}(0)=0$ are slightly larger and reach 0.005 in absolute value for $s\to \bar {s}$. It should be noted that both $L_\mathrm{dyn}$ and $L_\mathrm{energy}$ have not yet reached a plateau, meaning that the accuracy of the solution, and then the conservation of the energy, can be further improved by training the network for more epochs.

\begin{figure}
    \centering
    \includegraphics[width=\columnwidth]{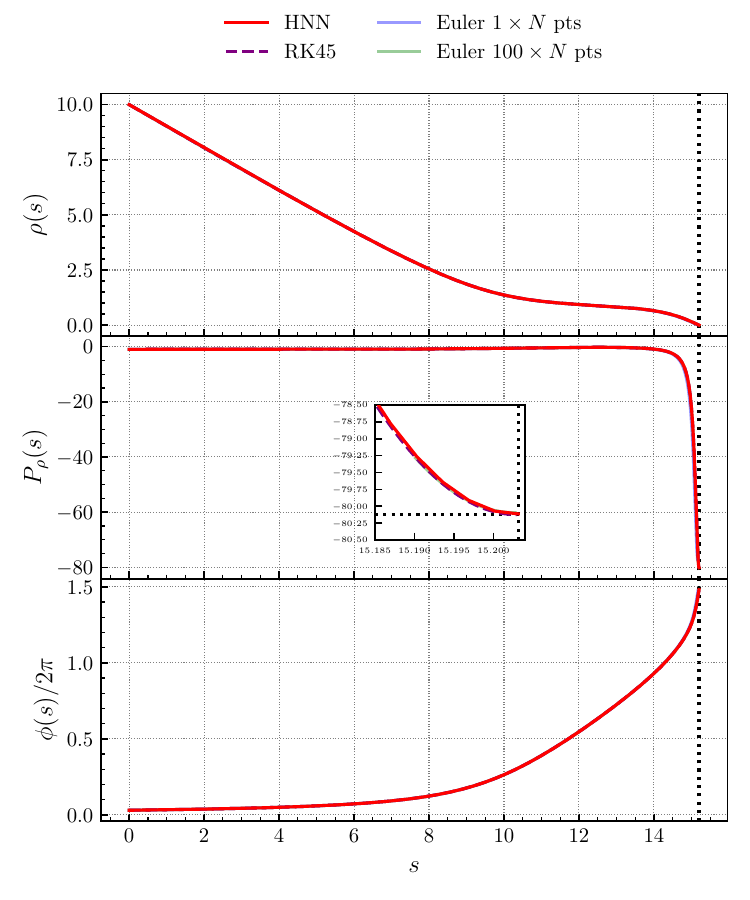}
    \caption{
    \label{fig:prec_variables}
    \footnotesize{From top to bottom: $\rho(s)$, $P_\rho(s)$ and $\phi(s)/2\pi$ for the different methods in the sub-critical case with $b=1.89$ and $N=2400$ obtained with the preconditioning approach. The vertical dotted line in the three panels represents $\bar{s}$ for the HNN. In the middle-panel we show also a zoom in the region $s\in[15.185,\bar{s}]$.}
    }
\end{figure}

The predicted variables $\rho(s)$, $P_\rho(s)$ and $\phi(s)/2\pi$ are shown in FIG.~\ref{fig:prec_variables} from top to bottom. Analogously to what observed in the time-splitting approach (FIG.~\ref{fig:Hamilton_subcritical_variables}), there is no visually appreciable difference between the different methods.

\begin{figure}
    \centering
    \includegraphics[width=\columnwidth]{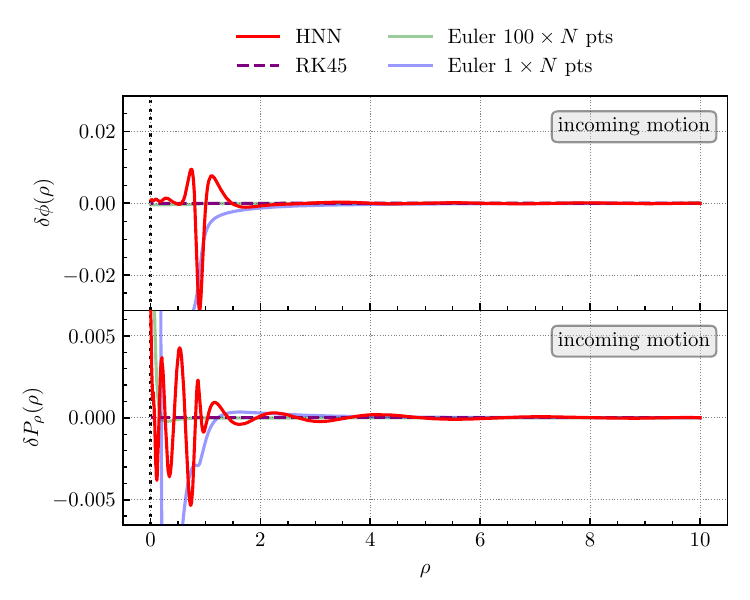}
    \caption{
    \label{fig:prec_errors}
    \footnotesize{Errors on the variables $\phi(\rho)$ and $P_\rho(\rho)$ in the sub-critical case with $b=1.89$ obtained from the preconditioning approach. }
    }
\end{figure}

In order to make a quantitative statement, we plot the errors $\delta \phi(\rho)$ and $\delta P_\rho( \rho)$ in, respectively, the top- and bottom-panel of FIG.~\ref{fig:prec_errors}. For $\phi(\rho)$ we observe errors that are about 100 times larger than than the ones shown in the top panel of FIG.~\ref{fig:Hamilton_subcritical_errors}, while for $P_\rho$ we see that the error increases by about a factor of 5. The worse performance compared to before can be improved by increasing the number of epochs, but we emphasize that the error scale remains very small.

\begin{figure}
    \centering
    \includegraphics[width=\columnwidth]{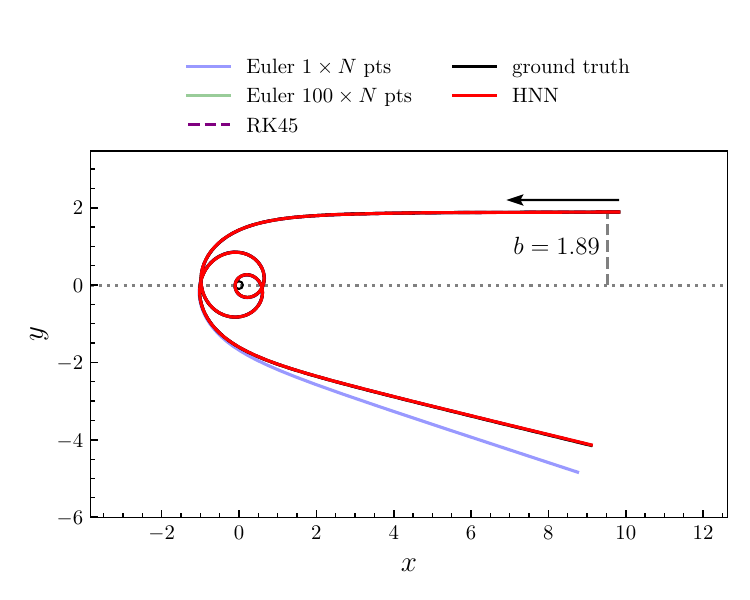}
    \caption{
    \label{fig:prec_cartesian}
    \footnotesize{Geodesic in the Cartesian plane obtained from different methods in the sub-critical case with $b=1.89$ and by applying the preconditioning approach. The ground truth is represented by the solid black line, while the small circle at the center (0,0) has radius $a_f$ and represents the fuzzball. The black arrow indicates the incoming direction of the particle.}
    }
\end{figure}

Finally, in FIG.~\ref{fig:prec_cartesian} we show the predicted geodesic in the Cartesian plane compared to the other methods and to the ground truth. As we can see, the solution predicted by the HNN is still very accurate and in agreement with the ground truth. The strategy proposed here effectively allows incorporating information about the solutions during the training phase while preserving model independence and facilitating the minimization process of the loss function. We leave the exploration of further approaches for future work. With this we conclude the results concerning the planar geodesics.

\section{\label{results:non_planar}Results for non-planar geodesics}

In this Section we present the results for the non-planar case which is discussed in Section~\ref{Non-Planar Case} and whose dynamics is determined by the following five ($J_\phi$ is not considered since it is constant) Hamilton equations

\begin{align}\label{eq:Hamilton_nonplanar}
   \dot \rho = & \pdv{\mathcal{H}}{P_\rho}, \quad \dot P_\rho  =-\pdv{\mathcal{H}}{\rho}, \quad \dot \phi = \pdv{\mathcal{H}}{J_\phi},
   \nonumber\\[6pt]
   &\dot \theta = \pdv{\mathcal{H}}{P_\theta}, \quad \dot P_\theta = -\pdv{\mathcal{H}}{\theta}. 
\end{align}

The derivatives on the right-hand side are explicitly written in Appendix \ref{app:non-planar} and one can immediately appreciate the extreme complexity of such set of equations. Indeed, the non-planar case presents a much greater challenge compared to the planar one, due to the higher number of coupled dynamical variables and the increased complexity of the dynamics. 

As already hinted in Section \ref{Non-Planar Case}, our analysis of non-planar geodesics has been limited to the critical ones, as they are crucial from both an observational and theoretical perspective. Their determination relies on the separability of motion. However, when separability does not hold, determining critical geodesics becomes challenging and we leave a detailed investigation of this case to future work.

The geometry parameters that we fix in this analysis are $L_1=0.5$, $L_5=1$, $a_f=0.2$, $E=1$ (from Appendix \ref{APP:nonplangeo} it is clear that $L_1$ and $L_5$ cannot be equal). With this choice, the critical parameters $\zeta_c$, $b_c$ and $\rho_c$ assume the values

\begin{flalign}\label{eq:crit_params_values}
  &\zeta_c = -2.5,
  \\[4pt]
  & b_c = 1.5,
\\[4pt]
  & \rho_c = \sqrt{0.46} = 0.6782,
\end{flalign}
while $b_{\phi,c}=0$ (see Appendix~\ref{APP:nonplangeo}).\\

In order to accommodate the Hamilton equations \eqref{eq:Hamilton_nonplanar}, we need a neural network with 5 output nodes, from which the solutions are assembled according to

\begin{flalign}
& \widehat{\rho}(s) = \rho_0  +f(s)O_\rho(s),
\nonumber\\[6pt]
& \widehat{P_\rho}(s) = P_{\rho,0}+f(s)O_{P_\rho}(s),
\nonumber\\[6pt]
& \widehat{\phi}(s) = \phi_0 +f(s)O_{\phi}(s),
\nonumber\\[6pt]
& \widehat{P_\theta}(s) = P_{\theta,0} +f(s)O_{P_\theta}(s),
\nonumber\\[6pt]
& \widehat{\theta}(s) = \theta_0 +f(s)O_{\theta}(s).    
\end{flalign}

We have chosen the following initial conditions

\begin{equation}\label{eq:init_cond_nonplan1}
  \rho_0 = 3, \qquad \theta_0 = 0.9, \qquad \phi_0 = 1.
\end{equation}

The value $P_{\theta,0}$ is obtained from \eqref{eq:QA_Non_Plan}, in particular we have (recall that $E=1$)
\begin{equation}
    \frac{P^2_{\theta,0}}{E^2} = P^2_{\theta,0} = b^2_c - a_f^2 \sin^2 \theta_0
\end{equation}

from which
\begin{flalign}\label{eq:init_cond_nonplan2}
    P_{\theta,0} = 1.4918.
\end{flalign}

The positive sign of the solution determines the initial direction of the trajectory. Finally, $P_{\rho,0}$ is obtained by requiring that the  Hamiltonian in \eqref{eq:Ham_Non-Plan} vanishes at $s=0$. Using \eqref{eq:crit_params_values}, \eqref{eq:init_cond_nonplan1} and \eqref{eq:init_cond_nonplan2}, we gain
\begin{equation}
P_{\rho,0} = -0.9447.
\end{equation}

Again, we will compare the prediction of the HNN with the \texttt{RK45} method and with the symplectic first-order Euler method. However, in this case, we are unable to derive a ground truth for the geodesic in the Cartesian space and, therefore, we cannot quantify the errors introduced by the different integration methods. To assess the quality of a solution, we can only rely on the expected theoretical behavior of the variables. 

Since we have chosen the parameters to select the critical trajectory, $\rho(s)$ is expected to decrease from $\rho_0$  until it reaches and settles at the value $\rho_c$.  Consequently, $P_\rho(s)$ is expected to increase from the negative $P_{\rho,0}$ to 0 and remain at that value. Concerning $\theta(s)$ and $\phi(s)$, we expect that, when the particle enters the critical regime with $\rho(s)\simeq \rho_c$ and it wraps around the oblate spheroid \eqref{eq:oblate_spheroid}, they have a  linear grow in time. Consequently, $P_\theta(s)$ in the critical regime should have an oscillating trend with constant period. 

For completeness, we project the oblate spheroid \eqref{eq:oblate_spheroid} onto the three planes $x-y$ ($\theta = \pi/2$), $x-z$ ($\phi = 0$) and $y-z$ ($\phi = \pi/2$). To do this we rely on \eqref{eq:change_coord_3d}, which maps the spherical coordinates $\rho$, $\theta$ and $\phi$ to the Cartesian ones $x$, $y$ and $z$. It turns out that the profile of the geodesic in the plane $x-y$ is given by the circumference

\begin{flalign}
\label{eq:ellipse_xy}
& x^2 + y^2 = \rho^2_c+a_f^2 \qquad x-y \quad \text{plane}\,,
\end{flalign}

and, in the other two planes, by the two following ellipses 

\begin{flalign}
\label{eq:ellipse_xz}
& \frac{x^2}{\rho^2_c+a_f^2} + \frac{z^2}{\rho^2_c} = 1 \qquad x-z \quad \text{plane}\,,
\\[6pt]
\label{eq:ellipse_yz}
& \frac{y^2}{\rho^2_c+a_f^2} + \frac{z^2}{\rho^2_c} = 1 \qquad y-z \quad \text{plane}.
\end{flalign}

This concludes our a priori knowledge of the geodesic, based on physical principles.

\begin{figure}
    \centering
    \includegraphics[width=\columnwidth]{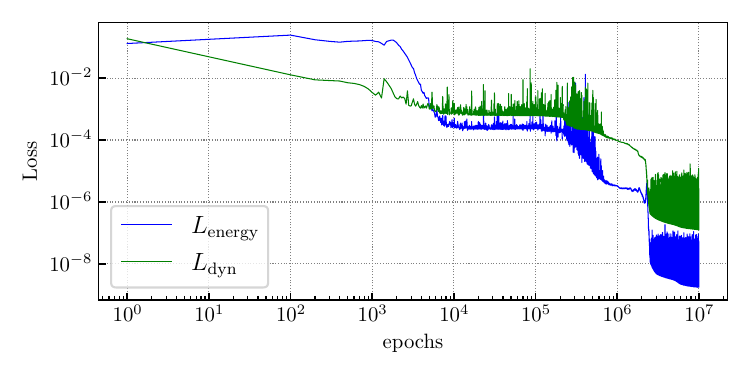}
    \caption{
    \label{fig:nonplanar_loss}
    \footnotesize{
    Loss functions for the non-planar case.}
    }
\end{figure}

\begin{figure}
    \centering
    \includegraphics[width=\columnwidth]{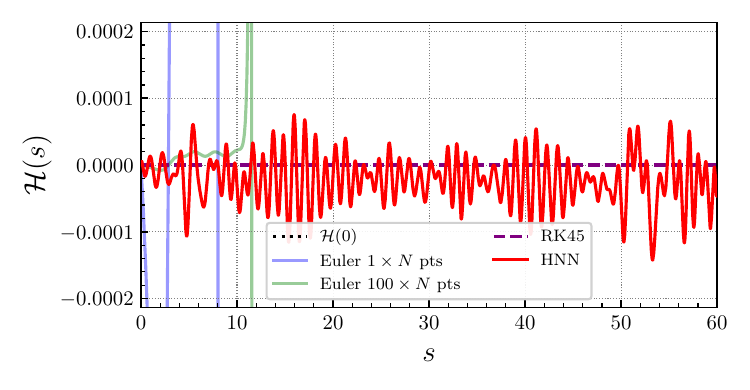}
    \caption{
    \label{fig:nonplanar_H}
    \footnotesize{
    Energy conservation over the time for the non-planar case. The conserved value of the energy $\mathcal{H}(0)=0$ is represented by the dotted black line. 
    }
    }
\end{figure}

We now discuss the training phase of the neural network. We have thoroughly investigated various parameter choices and architectures. Below, we present the strategy that, according to our analysis, has proven to be the most efficient. As training time interval we have set the range $\mathcal{T}_\mathrm{train}=[0,60]$ equally sampled with $N=2400$ points. Also in this case, after the training, we have validated the loss function on the interval $\mathcal{T}_\mathrm{val}$ to check the no over-fitting has occurred.
 
The first difference with respect to the planar case is the choice of the neural network architecture. As anticipated, the variable $P_\theta(s)$ is expected to have a periodic trend, in contrast to the monotonic trend of the others. For this reason we have found convenient to consider a two-block architecture with a common input node given by the time. The first block consists in 2 hidden layers with 64 neurons per each and the output $O_{P_\theta}$. For this first block we have used a sinusoidal activation function since, as remarked in Ref.~\cite{Harvard}, using trigonometric functions is particularly convenient when the output is expected to have a periodic behavior. The second block consists in 2 hidden layers with 128 neurons per each, hyperbolic-tanget activation function and the remaining four outputs  $O_\rho$, $O_{P_\rho}$, $O_\phi$ and $O_\theta$.

The second difference is the addition of a penalty term to the loss function  \eqref{eq:total_loss}.  This contribution is introduced to penalize the  solutions such that $P_\rho(s)>0$ and takes the form

\begin{flalign}
    L_{\text{penalty}}(\textbf{w}) = \sum_{s \in \mathcal{S}_p} \bigg[\max(\widehat{P_\rho}(s,\mathbf{w}),0)\bigg]^2 
\end{flalign}

where $\mathcal{S}_p=\{8,20,30,40,50,60\}$. The total loss function actually minimized is therefore

\begin{flalign}
    L(\lambda,\textbf{w}) = (1-\lambda)L_\mathrm{dyn}(\textbf{w})+\lambda L_\mathrm{energy}(\textbf{w})+ L_\mathrm{penalty}(\textbf{w}).
\end{flalign}

The penalty term is activated only for $\widehat{P_\rho}(s,\textbf{w})>0$, while it does not contribute to the loss function when, correcltly, $\widehat{P_\rho}(s,\textbf{w})<0$. We have observed that without this additional term, the HNN tends to predict unstable trajectories. Let us stress that we are using an information from the physics that can be deduced without any prior knowledge of the actual solution. This is a clear example of how various types of information about the solutions can be fruitfully incorporated into the loss function to restrict the solution space and improve the network's predictability. The training is performed by using the Adam optimizer and we monitor individually both $L_\mathrm{dyn}$ and $L_\mathrm{energy}$.

\begin{figure}
    \centering
    \includegraphics[width=\columnwidth]{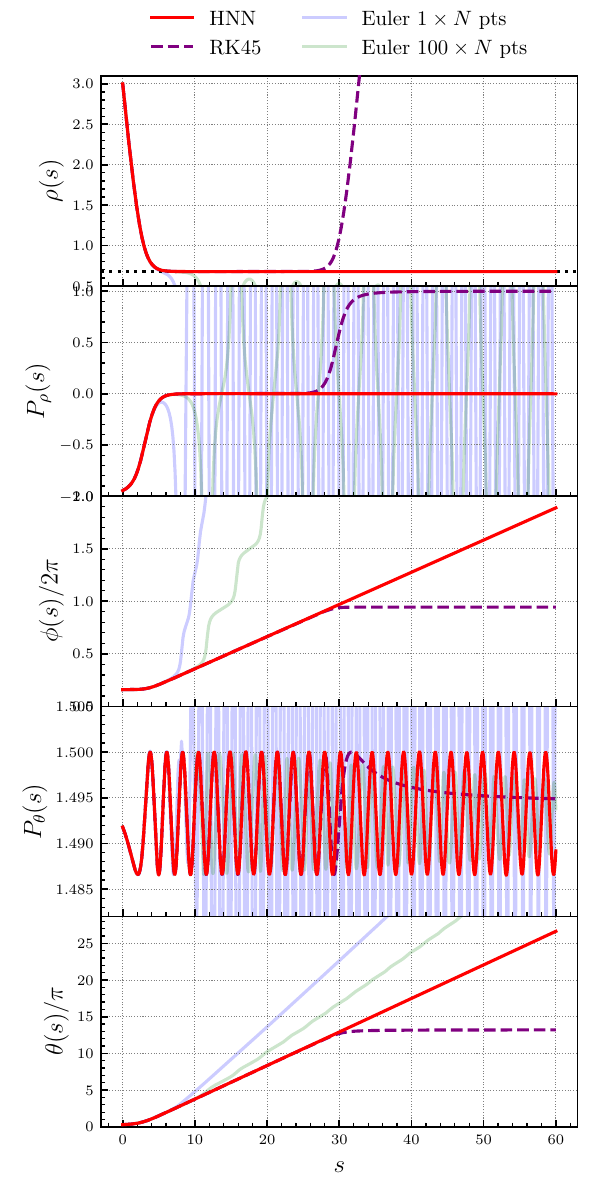}
    \caption{
    \label{fig:nonplanar_variables}
    \footnotesize{
 From top to bottom: $\rho(s)$, $P_\rho(s)$, $\phi(s)/2\pi$, $P_\theta(s)$ and $\theta(s)/\pi$ for the different methods in the non-planar case. The horizontal black dotted line in the top panel represents the critical radius $\rho_c=0.6782$, on which the trajectory must settle. The trajectory predicted by the HNN is the only one among the different methods to correctly reproduce this feature.    
    }
    }
\end{figure}

\begin{figure}
    \centering
    \includegraphics[width=\columnwidth]{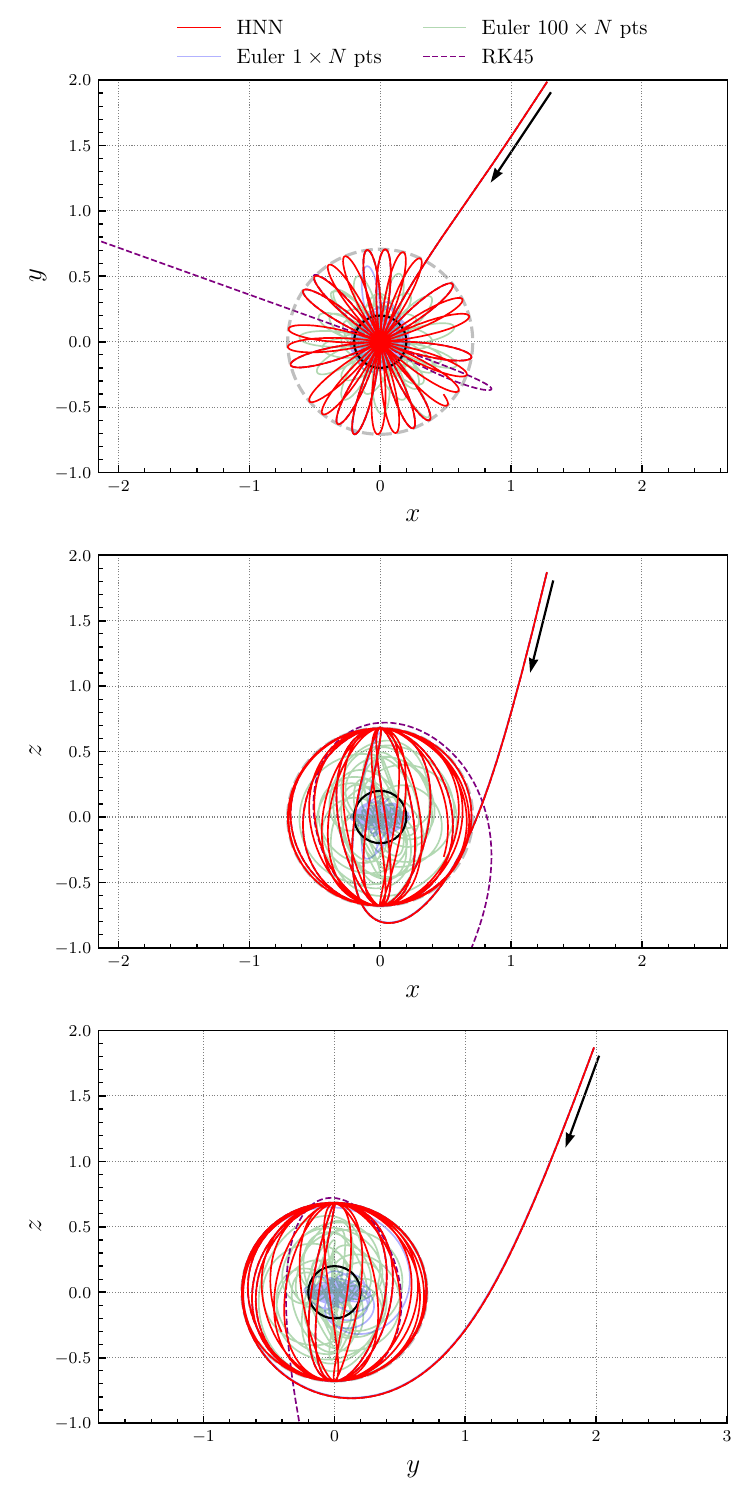}
    \caption{
    \label{fig:nonplanar_cartesian}
    \footnotesize{
    Geodesics in the non-planar case for the different methods, projected onto the, respectively from top to bottom, $x-y$, $x-z$ and $y-z$ Cartesian plane.  The  solid black line represents the fuzzball surface. The dashed light-gray  lines represent the circumference and ellipses defined in  \eqref{eq:ellipse_xy}-\eqref{eq:ellipse_yz}. The black arrow indicates the incoming direction of the particle. 
    }
    }
\end{figure}

Concerning the remaining algorithmic parameters, the ones entering $\eta(\mathrm{epoch})$ in \eqref{eq:learning_rate_scheduler} are $\eta_i=8\times 10^{-8}$, $\eta_f=\eta_i/50$, $\eta_c=500\times 10^3$ and $\sigma_\eta=100\times 10^3$.  In this case, it turns out that the small value $\lambda=0.1$ gives better performances compared to $\lambda = 0.99$ used in all the cases of the planar study. We interpret this result in light of the fact that the dynamics are much more complex and, therefore, should be prioritized over energy conservation during the minimization process.

Having fixed the setup, we have trained the neural network with $10\times 10^6$ epochs and the loss functions are shown in FIG.~\ref{fig:nonplanar_loss}. The plot shows a good performance which is comparable to those obtained in the planar case. 

In FIG.~\ref{fig:nonplanar_H} we show the energy $\mathcal{H}(s)$ for the different methods. The HNN conserves the energy on average with small fluctuations between $-0.0001$ and $+0.0001$. The Euler methods starts failing in conserving the energy already at small times. The \texttt{RK45} method appears to conserve energy with great accuracy, but, as we have already seen in the planar critical case in Section~\ref{sec:planar_critical}, this does not necessarily mean that its numerical solution is the correct one. 

We realize this by looking at the predicted $\widehat{\rho}(s)$, $\widehat{P_\rho}(s)$, $\widehat{\phi}(s)/2 \pi$, $\widehat{P_\theta}(s)$ and $\widehat{\theta}(s)/\pi$, which are shown, form top to bottom in the given order, in FIG.~\ref{fig:nonplanar_variables}. As it can be seen, the different methods yield very different solutions and the one predicted by the HNN is the only one which is in fully agreement with the theoretical expectations discussed above. The critical regime for which $\rho(s)\simeq \rho_c$ is reached around $s=6$ and it is maintained stable for all subsequent times. The prediction of $P_\theta(s)$ as an oscillating function is remarkably good and such a result is due to the use of a trigonometric activation function in the architecture block responsible for the output associated with $P_{\theta}$. The Euler method with both $N$ and $100\times N$ points fails in remaining in critical regime and even predicts $\rho(s)\simeq 0$, which corresponds to a particle which has collapsed onto the fuzzball.  The \texttt{RK45} method is in agreement with the HNN up to $s\simeq 28$ and after this $\rho(s)$ starts increasing, while the angular variables flatten, showing a particle that moves away from the critical trajectory as if it were in a situation analogous to the over-critical case presented in Section~\ref{sec:planar_overcritical}.

We now convert the spherical coordinates to the Cartesian ones by using \eqref{eq:change_coord_3d}. In FIG.~\ref{fig:nonplanar_cartesian} we show the trajectories from the different methods on the $x-y$, $x-z$ and $y-z$ planes, respectively from top to bottom. We also display the curves (colored in gray) given in \eqref{eq:ellipse_xy}, \eqref{eq:ellipse_xz}  and \eqref{eq:ellipse_yz}, which represent the profile of the oblate spheroid. We can appreciate how the geodesic predicted by the HNN, once entered in the critical regime,  exactly describes the profile given by the aforementioned curves in the three Cartesian planes. With this observation we can conclude that the HNN is the only method that, in this challenging situation due to the instability of the critical solution,  solves the Hamilton equations with high accuracy and allows for a solid and stable numerical study of the critical non-planar geodesic of a massless particle immersed in the D1-D5 circular fuzzball. This is the main result of this paper and paves the way for future studies on more complex geometries.

\begin{figure}
    \centering
    \includegraphics[width=\columnwidth]{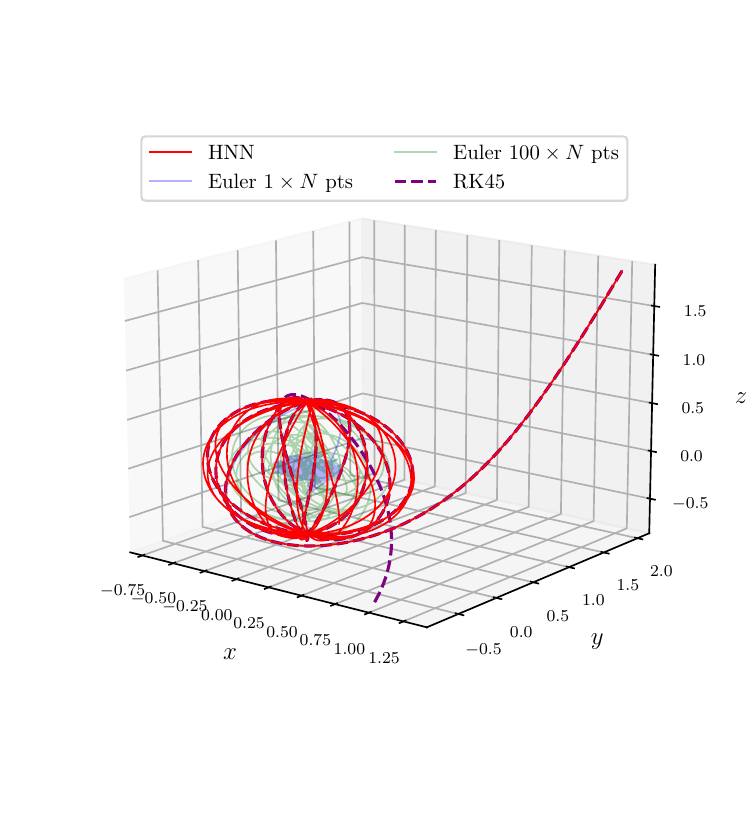}
    \caption{
    \label{fig:nonplanar_3D}
    \footnotesize{ Geodesic in the 3D Cartesian space for different methods in the non-planar geometry.
    }
    }
\end{figure}

We conclude this Section by showing the geodesics in the 3D Cartesian space in FIG.~\ref{fig:nonplanar_3D}. While it does not add anything new to the previous discussion, it  still looks beautiful and, frankly, quite promising.

\section{\label{conclusions}Conclusions and outlooks}

Inspired by \cite{Harvard}, we studied the massless and neutral geodesics in a peculiar, smooth, and horizonless fuzzball geometry as an innovative application of Hamiltonian Neural Networks (HNNs). For this physical system, extensively studied in \cite{Bianchi:2017sds, Bianchi:2022qph, DiRusso:2024hmd}, it is possible to obtain an explicit ground truth, which has allowed for a comparative study of the performance of HNNs and other numerical integrators, specifically the first-order semi-implicit Euler method and the \texttt{RK45} method. The former is symplectic, while the latter is not but allows for setting bounds on the relative and absolute errors on the solution. We have analyzed sub-critical, critical and over-critical planar geodesics, as well as non-planar critical geodesics. The critical ones that are non trivial are unstable, making it challenging to obtain accurate results using standard numerical integrators, as they are known to accumulate errors over time.

Our analysis has shown that, despite the indisputable complexity of the equations learned, that can be appreciated in Appendix~\ref{APP:hamilton}, HNNs were able to accurately reproduce the geodesics across all the mentioned cases, even over large time scales, whereas the considered numerical integrators failed, as expected, when computing unstable trajectories. This result highlights another significant aspect that emerged from our analysis: the performance of HNNs, due to the fact that the solution is obtained by simultaneously and independently considering all time steps, does not depend on the intrinsic stability or instability of the dynamics described by the Hamilton equations and it remains at the same level for the different cases. In contrast, the performance of numerical integrators is highly case-dependent. In order to complete the comparison, we emphasize that HNNs are model-independent and produce an analytic time-dependent output, whereas numerical integrators predict the solution at a finite and discrete set of time steps. Additionally, HNNs allow for easy incorporation of constraints, symmetries and data-driven information to enhance the solution’s predictability as done, for example, in the non-planar case. The set of theorems and statements, collectively known as the Universal Approximation Theorem \cite{hornik1989multilayer,Goodfellow-et-al-2016}, further guarantees that neural networks, as function approximators, are always improvable in terms of accuracy. All these advantages make HNNs not only an alternative, but a preferable tool in terms of flexibility and robustness compared to standard numerical integrators. This remains true even in cases where the \texttt{RK45} method appears to produce more precise results. In fact, as observed in the critical cases, the non - symplectic nature of \texttt{RK45} can lead to solutions that exactly conserve energy but are nonetheless incorrect. On the other hand, a symplectic method like the Euler approach often requires an infinitesimally small time step - and consequently, an immense number of iterations—to produce accurate solutions.

The results and evidence gathered in this paper are clearly generalizable to other physical systems, since the concept of stability and instability extends far beyond that of a particle moving in a fuzzball geometry. Our work thus contributes to the already flourishing field of ML and PINN applications across various domains of physics and science more broadly.

However, in our view, a major downside of ML remains the fact that implementing a neural network and tuning its parameters to find the optimal setup is a significantly more laborious and time-consuming process compared to applying standard methods, which are often readily available as default packages in common programming languages. Nevertheless, this issue can be mitigated by the numerous libraries and dedicated tools that have made neural network implementation accessible and straightforward, even on standard desktop notebooks.

For the authors, this work represents the initial step in a long-term project, aimed at systematically studying geometries emerging in String Theory through HNNs and PINNs. For instance, as a future outlook a detailed study of geodesic motion in the so-called D1-D5-P (three-charge) fuzzball would be highly interesting. This is a nontrivial generalization of the present D1-D5 (two-charge) fuzzball with an additional Kaluza-Klein (KK) charge. The main complication arises from the fact that, in general, this geometry does not allow for a separation of radial and angular dynamics via the introduction of a Carter-like constant; separation is possible only on the equatorial planes \cite{Bianchi:2018kzy}. After the detailed experience with the separable two-charge case, we plan to apply the HNN to this more complex physical system, which, due to its considerable intricacy, has never been studied in terms of either geodesic motion or wave propagation.
 
In this case, there is no ground truth for comparison and a challenging aspect we intend to address is finding a way to assess the quality of the prediction. This issue, as discussed above, also applies to numerical integrators. For HNNs, we can first rely on the loss function itself, which provides a quantitative measure of how well the predicted solution satisfies the equations of motion. Secondly, multiple optimal network setups may exist and studying the stability of the predicted solution under variations in algorithmic parameters - such as the number of time steps, the length of the time interval, the choice of activation functions and the network architecture - serves as both a qualitative and quantitative indicator of accuracy. We plan to further explore and refine this type of analysis in our next work.

 Finally, studying the geodesic motion in this new scenario will help to identify the photon spheres, which serve as a polar star in analyzing the Quasi-Normal Mode (QNM) spectrum \cite{Bianchi:2023rlt,Bianchi:2023sfs,Cipriani:2024ygw}, an investigation entrusted to a future stage of project.

\begin{acknowledgments}
We are grateful to N.~Tantalo for his illuminating comments on a preliminary version of the manuscript. We warmly thank M.~Bianchi for the discussion that led to the beginning of this work and for useful comments. We also want to thank the PRIN \quote{String Theory as a Bridge between Gauge Theory and Quantum Gravity} for having created a stimulating environment. A. C. would like to thank the Department of Theoretical Physics at CERN for kind hospitality during the final stages of the present work.

\end{acknowledgments}

\FloatBarrier
\appendix
\section{\label{APP:hamilton} Hamilton equations}
In this Appendix we report the derivatives of the Hamiltonian in the planar and non-planar cases that are used inside the loss functions. Due to the huge length of these relations, symbolic writing will be used and that will be specified.

\subsection{Planar Case}
From the  Hamiltonian \eqref{eq:Ham_plan}, the derivatives to be considered are
\begin{widetext}
\begin{flalign}
 - \dot{P}_\rho  =\frac{\partial \mathcal{H}}{\partial \rho} &= \frac{E^2 \Bigg[-H_1 \left(-4 H_5 \omega _{\phi} 
  \left(\left(a_f^2 + \rho^2 \right) \, \omega_{\phi,\rho} - \rho \, \omega _{\phi}\right) \, + \, \omega_{\phi}^2 \, H_{5,\rho} \, \left(a_f^2+\rho^2\right) \, + \, H_5^2 H_{1,\rho} \, \left(a_f^2+\rho^2\right)^2\right)\Bigg]}{4 H_1^{3/2} \, H_5^{3/2} \, \left(a_f^2+\rho^2\right)^2} +
  \nonumber\\[6pt]
& + \frac{- E^2 \left(H_5 \, \omega_{\phi}^2 \, H_{1,\rho} \left(a_f^2+\rho^2\right) \, + \, H_5 
  H_1^2 \, H_{5,\rho} \left(-\left(a_f^2+\rho^2\right)^2\right)\right)}{4 H_1^{3/2} \, H_5^{3/2} \, \left(a_f^2+\rho^2\right)^2}  +
  \nonumber\\[6pt]
& +\frac{E \, J_{\phi} \left(H_1 \left(2 \, H_5 \left(\left(a_f^2+\rho^2\right) \omega_{\phi,\rho} 
  \, - \, 2\rho \, \omega_{\phi}\right) - \omega_{\phi} \, H_{5,\rho} \, \left(a_f^2+\rho^2\right)\right) \, - \, H_5 \, \omega_{\phi} \, H_{1,\rho} \left(a_f^2+\rho^2\right)\right)}{2 H_1^{3/2} \, H_5^{3/2} \, \left(a_f^2+\rho^2\right)^2}  +
  \nonumber\\[6pt]
& +\frac{J_{\phi}^2 \, \left(-\left(H_5 H_{1,\rho} \, \left(a_f^2+\rho^2\right)\right) \, - \, H_1
   \left(H_{5,\rho} \left(a_f^2 \, + \, \rho^2\right) \, + \, 4 H_5 \rho \right)\right)}{4 H_1^{3/2} \, H_5^{3/2} \, \left(a_f^2 + \rho^2\right)^2} +
   \nonumber\\[6pt]
& +\frac{P_{\rho}^2 \left(H_5 (-\chi ) \, H_{1,\rho} \, \left(a_f^2+\rho^2\right) \, - \, H_1 
  \left(\chi \, H_{5,\rho} \, \left(a_f^2 + \rho^2\right)+ 4 \, H_5 \left(\chi_{\rho} \left(a_f^2 \, + \, \rho^2\right) \,- \, \rho \, \chi \right)\right)\right)}{4 H_1^{3/2} \, H_5^{3/2} \chi^3} \,,
  \\[6pt]
\dot{\rho} = \frac{\partial \mathcal{H}}{\partial P_\rho} &= \frac{P_{\rho } \left(a_f^2 \, + \, \rho ^2\right)}{\sqrt{H_1} \, \sqrt{H_5} \, \chi ^2} \,,
\\[6pt]
- \dot{J_\phi} = \frac{\partial \mathcal{H}}{\partial \phi} &= 0  \label{cons J} \,,
\\[6pt]
\dot{\phi}= \frac{\partial \mathcal{H}}{\partial J_\phi} &= \frac{E \, \omega_{\phi} \, + \, J_{\phi}}{\sqrt{H_1} \, \sqrt{H_5} \, \left(a_f^2+\rho ^2\right)}\,,
\end{flalign}    
\end{widetext}
where we have adopted the following conventions
\begin{itemize}
    \item  $\chi^2 \equiv \chi^2(\rho,\theta) = \rho^2 + a_f^2 \cos^2 \theta$;
    \item  all the quantities that depend on $\theta$ (such as $H_1, H_5, \chi, \omega_\phi$) are evaluated at $\theta = \pi/2$;
    \item $f_{,x} \equiv \partial f / \partial x$ for any quantity $f$, except for $\chi_\rho = \partial \chi / \partial \rho$. These derivatives are evaluated in $\theta = \pi/2$ as well.
\end{itemize}
Notice that \eqref{cons J} is due to the fact that $\phi$ is cyclic and consequently the corresponding conjugate momentum is conserved. The functions $H_1$ and $H_5$ have already been defined in  \eqref{eq:def_params}.

\subsection{\label{app:non-planar}Non-Planar Case}
From the  Hamiltonian \eqref{eq:Ham_Non-Plan}, the derivatives to be considered are

\begin{widetext}
\begin{flalign}
-\dot{P}_\rho  & =  \frac{\partial \mathcal{H}}{\partial \rho} = \frac{E^2 \left(H_1 \left(H_5^2 \left(c_{\theta }^2 \, - \, 1\right) H_{1,\rho} \left(-\left(a_f^2 + \rho^2\right)^2\right) \, - \, 4  H_5 \omega_{\phi } \left(\left(a_f^2 + \rho^2\right) \omega_{\phi,\rho} \, - \, \rho  \omega_{\phi} \right) \, + \, \omega_{\phi }^2 H_{5,\rho} \left(a_f^2+\rho^2\right)\right)\right)}{4 H_1^{3/2} \, H_5^{3/2} \left(c_{\theta}^2-1\right) \left(a_f^2 \, + \, \rho^2\right)^2} + 
\nonumber\\[6pt] 
& + \frac{E^2 \left(H_5 H_1^2 \left(c_{\theta }^2 \, - \, 1\right) H_{5,\rho} \left(-\left(a_f^2+\rho^2\right)^2\right) \, + \, H_5 \omega_{\phi}^2 H_{1,\rho} \left(a_f^2 \, + \,\rho^2\right)\right)}{4 H_1^{3/2} \, H_5^{3/2} \left(c_{\theta}^2-1\right) \left(a_f^2 \, + \, \rho^2\right)^2}+
\nonumber\\[6pt] 
& + \frac{P_{\theta}^2 \left(-H_5 \left(\chi \, H_{1,\rho} \, + \, 4 H_1 \chi_{\rho} \right)\, - \, H_1 \, \chi \, H_{5,\rho}\right)}{4 H_1^{3/2} H_5^{3/2} \chi^3} +
\nonumber\\[6pt] 
& + \frac{P_{\rho}^2 \left(H_5 (-\chi) H_{1,\rho} \left(a_f^2 \, + \, \rho^2\right) \, - \, H_1 \left(\chi \, H_{5,\rho} \left(a_f^2 \, + \, \rho^2\right)+4 H_5 \left(\chi_{\rho} \left(a_f^2 \, + \, \rho^2\right) \, - \, \rho \, \chi \right)\right)\right)}{4 H_1^{3/2} H_5^{3/2} \chi^3} +
\nonumber\\[6pt] 
& + \frac{E J_{\phi} \left(H_5 \, \omega_{\phi} \, H_{1,\rho} \left(a_f^2 \, + \, \rho^2\right) \, + \, H_1 \left(\omega_{\phi} \, H_{5,\rho} \left(a_f^2 \, + \, \rho^2\right) \, - \, 2 H_5 \left(\left(a_f^2 \, + \, \rho^2\right) \omega _{\phi,\rho} \, - \, 2 \rho \, \omega_{\phi}\right)\right)\right)}{2 H_1^{3/2} \, H_5^{3/2} \left(c_{\theta }^2\, - \, 1\right) \left(a_f^2 \, + \, \rho^2\right)^2} +
\nonumber\\[6pt] 
& + \frac{J_{\phi}^2 \left(H_5 \, H_{1,\rho} \left(a_f^2 \, + \, \rho^2\right) \, + \, H_1 \left(H_{5,\rho} \left(a_f^2 \, + \, \rho^2 \right) \, + \, 4 H_5 \rho \right)\right)}{4 H_1^{3/2} H_5^{3/2} \left(c_{\theta}^2\, - \, 1\right) \left(a_f^2 \, + \, \rho^2\right)^2} \,, \\[6pt]
\dot{\rho} &= \frac{\partial \mathcal{H}}{\partial P_\rho} = \frac{P_{\rho} \left(a_f^2 \, + \, \rho^2\right)}{\sqrt{H_1} \, \sqrt{H_5} \chi^2} \,, \qquad
\\[6pt]
- \dot{J}_\phi &=\frac{\partial \mathcal{H}}{\partial \phi} = 0\,,\qquad
\\[6pt]
\dot{\phi} & = \frac{\partial \mathcal{H}}{\partial J_\phi} = \frac{\csc^2(\theta) \left(E \, \omega_{\phi} \, + \, J_{\phi}\right)}{\sqrt{H_1} \, \sqrt{H_5} \left(a_f^2 \, + \, \rho^2\right)}\,,\qquad
\\[6pt]
\dot{\theta}& =\frac{\partial \mathcal{H}}{\partial P_\theta} = \frac{P_{\theta }}{\sqrt{H_1} \sqrt{H_5} \chi ^2}\,,\qquad
\end{flalign}

\begin{flalign}
-\dot{P}_\theta  &= \frac{\partial\mathcal{H}}{\partial \theta} = -\frac{E^2 \left(H_1 \left(H_5^2 \, s_{\theta}^3 H_{1,\theta} \left(a_f^2 \, + \, \rho^2\right) \, + \, 4 H_5 \omega_{\phi} \left(c_{\theta} \, \omega _{\phi} \, - \, s_{\theta} \, \omega_{\phi,\theta} \right) \, + \, s_{\theta} \, \omega_{\phi}^2 \, H_{5,\theta}\right) \, + \, H_1^2 H_5 s_{\theta}^3 \, H_{5,\theta} \left(a_f^2 \, + \, \rho^2\right)\right)}{4 H_1^{3/2} \, H_5^{3/2} \, s_{\theta}^3 \left(a_f^2 \, + \,\rho^2\right)} 
\nonumber\\[6pt] 
& - \frac{E^2 \left(H_5 s_{\theta} \, \omega_{\phi}^2 \, H_{1,\theta}\right)}{4 H_1^{3/2} \, H_5^{3/2} \, s_{\theta}^3 \left(a_f^2 \, + \,\rho^2\right)} -\frac{E \, J_{\phi} \left(H_1 \left(H_5 \left(4 \, c_{\theta} \, \omega_{\phi} \, - \, 2 s_{\theta} \, \omega_{\phi,\theta} \right) \, + \, s_{\theta} \, \omega_{\phi} \, H_{5,\theta} \right) \, + \, H_5 \, s_{\theta} \, \omega_{\phi} \, H_{1,\theta}\right)}{2 H_1^{3/2} \, H_5^{3/2} \, s_{\theta}^3 \left(a_f^2 \, + \, \rho^2\right)}
\nonumber\\[6pt]
& + \frac{P_{\theta}^2 \left(-H_5 \left(\chi \, H_{1,\theta} \, + \, 4 H_1 \, \chi_{\theta} \, \right) \, - \, H_1 \, \chi \, H_{5,\theta}\right)}{4 H_1^{3/2} \, H_5^{3/2} \, \chi^3} \, -\, \frac{P_{\rho}^2 \left(a_f^2 \, + \, \rho^2\right) \left(H_5 \left(\chi \, H_{1,\theta} \, + \, 4 H_1 \, \chi_{\theta}\right) \, + \, H_1 \, \chi \, H_{5,\theta} \right)}{4 H_1^{3/2} \, H_5^{3/2} \chi^3} 
\nonumber\\[6pt]
& -\frac{J_{\phi}^2 \left(H_1 \left(s_{\theta} \, H_{5,\theta} \, + \, 4 H_5 c_{\theta}\right) \, + \, H_5 s_{\theta} \, H_{1,\theta}\right)}{4 H_1^{3/2} \, H_5^{3/2} \, s_{\theta}^3 \left(a_f^2 \, + \, \rho ^2\right)} \,,
\end{flalign}
\end{widetext}

where the conventions are the same as for the planar case with the only difference that now the quantities are no longer evaluated anymore in $\theta = \pi/2$, but they are functions of $\theta$. Furthermore, we have introduced the symbols $c_\theta = \cos \theta, s_\theta = \sin \theta$. 

\section{\label{APP:separability} Separability}
In this Appendix we report the details that allow to determine the equations \eqref{eq:diffeq} of the separable case.\\
We start by computing the conjugate momenta using \eqref{eq:Momenta}, knowing the Lagrangian in the more general case (meaning keeping in count also the coordinates $z_I$, with $I=1,\dots,4$, that have been neglected throughout all the work). We get the following results
\begin{widetext}
\begin{equation} \label{momenta}
P_t=-\frac{\dot{t}+\omega_\phi \dot{\phi}}{H}=-E\quad,\quad P_z=\frac{\dot{z}+\omega_\psi \dot{\psi}}{H}\quad,\quad P_\rho=\frac{H(\rho^2+a_f^2\cos^2\theta)}{\rho^2+a_f^2}\dot{\rho}\quad,\quad P_\theta=H(\rho^2+a_f^2\cos^2\theta)\dot{\theta} 
\end{equation}
$$P_\phi=-\frac{\omega_\phi}{H}\dot{t}+\Bigg[H(\rho^2+a_f^2)\sin^2\theta-\frac{\omega_\phi^2}{H}\Bigg]\dot{\phi}=J_\phi \quad,\quad P_\psi=\frac{\omega_\psi}{H}\dot{z}+\left(H\rho^2\cos^2\theta+\frac{\omega_\psi^2}{H}\right)\dot{\psi}=J_\psi$$
$$P_{1,2}=\left(\frac{H_1}{H_5}\right)^{1/2}\dot{z}_{1,2}\quad,\quad P_{3,4}=\left(\frac{H_5}{H_1}\right)^{1/2}\dot{z}_{3,4}$$
where the only non conserved momenta are $P_\rho$ and $P_\theta$. By inverting the previous relations, we obtain
\begin{equation}\label{dotxinP}
\dot{t}=H E-\frac{\omega_\phi(J_\phi +E\omega_\phi)}{H(\rho^2+a_f^2)\sin^2\theta} \quad,\quad \dot{z}=H P_z-\frac{\omega_\psi(J_\psi-P_z\omega_\psi)}{H\rho^2\cos^2\theta} \quad,\quad \dot{\rho}=\frac{\rho^2+a_f^2} {H(\rho^2+a_f^2\cos^2\theta)}P_\rho
\end{equation}
$$\dot{\theta}=\frac{1}{H(\rho^2+a_f^2\cos^2\theta)}P_\theta \quad,\quad \dot{\phi}=\frac{J_\phi+E \omega_\phi}{H(\rho^2+a_f^2)\sin^2\theta} \quad,\quad \dot{\psi}=\frac{J_\psi-P_z \omega_\psi}{ H\rho^2\cos^2\theta}$$
$$\dot{z}_{1,2}=\left(\frac{H_5}{H_1}\right)^{1/2}P_{1,2}\quad,\quad \dot{z}_{3,4}=\left(\frac{H_1}{H_5}\right)^{1/2} P_{3,4}$$
\end{widetext}
In order to find \eqref{eq:diffeq}, we have to divide $\dot{\rho}$ by $\dot{t}$ and $\dot{\phi}$ by $\dot{t}$, where $\omega_{\phi}$ and $H$ are given in \eqref{eq:def_params}, $J_\phi = b E$ (from \eqref{impact param}) and $P^2_\rho$ is written in \eqref{eq:Prho_Plan}. By doing all these substitutions, the equations \eqref{eq:diffeq} are obtained.

\section{\label{APP:groundtruth} Derivation of the ground truth for the planar case}
\begin{figure*}[t]
\includegraphics[width=2 \columnwidth]{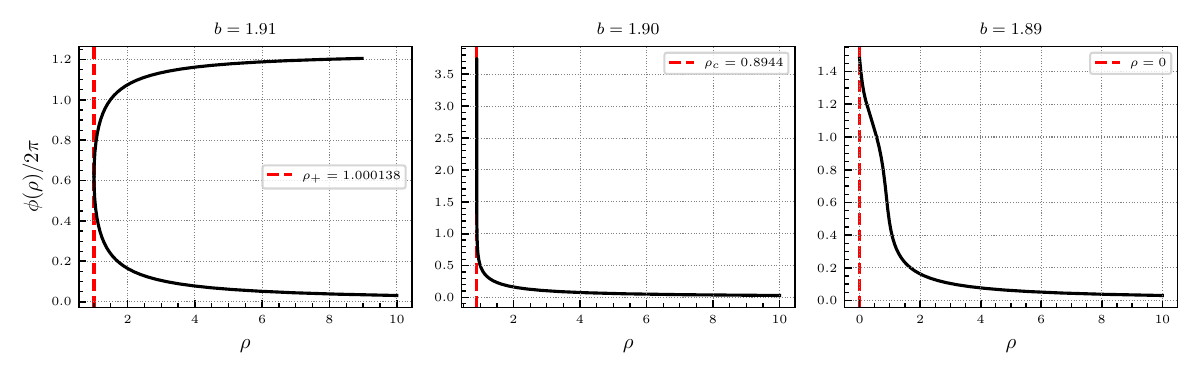}
\caption{
\label{fig:phi_exact}
Exact functions $\phi(\rho)/2\pi$ for the planar geometry. \emph{Left-panel}: over-critical case with $b=1.91$. The vertical dashed  line represents $\rho_+=1.000138$, which is the largest root of the radicand in \eqref{eq:diffeq}. \emph{Central-panel}: critical case with $b=b_c=1.9$. The vertical dashed  line is the critical radius $\rho_c=0.8944$, obtained from \eqref{eq:unstable_ph}. \emph{Right-panel}: sub-critical case with $b=1.89$. The vertical dashed line is drawn for $\rho=0$, below which the motion has no physical sense.
}
\end{figure*}

As hinted in Section \ref{sec:Planar_Case}, the planar case can be studied by exactly integrating the equatorial geodesics.\\
The starting point is \eqref{eq:diffeq}, from which we get
\begin{widetext}
\begin{equation} \label{eq:groundtruth}
\phi(\rho)-\phi_0=\pm \int_{\rho_0}^{\rho}\dd\rho'\frac{ \left(L_1 L_5 a_f+b \rho'^{2}\right)}{\left(\rho'^2+a_f^2\right) \sqrt{\rho'^4+\rho'^2
   \left(a_f^2-b^2+L_1^2+L_5^2\right)-2 b L_1 L_5 a_f+\left(L_1^2+L_5^2\right) a_f^2+L_1^2 L_5^2}},
\end{equation}
\end{widetext}
which holds for all the three regimes listed in Section \ref{sec:Planar_Case} (sub-critical, critical and over-critical). By numerically solving this integral, we can find the angle $\phi$ as a function of $\rho$ and, by using \eqref{eq:cartesian_map}, we can build the trajectory that we promote to ground truth in the comparison with the results obtained from the NN and the other numerical integrators. We have used extended-precision arithmetic to compute exactly the integral up to 32 digits. In the main text we have set the geometry parameters to $a_f=0.1$, $L_1=L_5=L=1$ and $E=1$. The functions $\phi(\rho)$, obtained by integrating \eqref{eq:groundtruth}, are shown in FIG.~\ref{fig:phi_exact} for $b=1.91$ (over-critical regime), $b=b_c=1.9$ (critical regime) and $b=1.89$ (sub-critical regime).

\begin{figure*}[t]
\includegraphics[width= 2 \columnwidth]{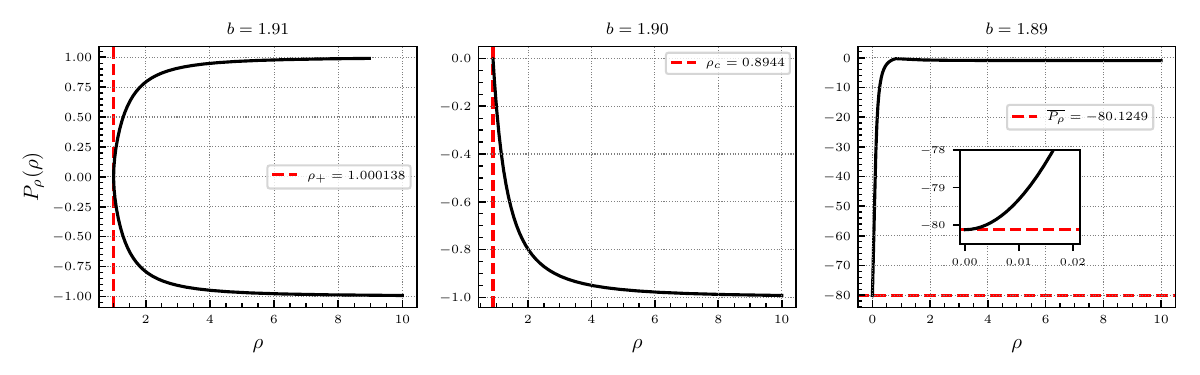}
\caption{
\label{fig:Prho_exact}
Exact functions $P_\rho(\rho)$ for the planar geometry. \emph{Left-panel}: over-critical case with $b=1.91$. The vertical dashed  line represents $\rho_+=1.000138$, which is the largest root of the radicand in \eqref{eq:diffeq}. \emph{Central-panel}: critical case with $b=b_c=1.9$. The vertical dashed  line is the critical radius $\rho_c=0.8944$, obtained from \eqref{eq:unstable_ph}. \emph{Right-panel}: sub-critical case with $b=1.89$. The horizontal dashed line represents $\overline{P_\rho}\equiv P_\rho(0)=-80.1249$, computed from \eqref{eq:Prho_exact_subcritical}. The small frame is a zoom in the region $\rho\in[0,0.02]$ and shows as $P_\rho(\rho)$ has a vanishing derivative for $\rho=0$.
}
\end{figure*}

In the over-critical regime,  the function $\phi(\rho)$ can be written in a closed way, which we now show for completeness. Indeed, when $b > b_c$ the shape of the potential $-Q_R$ is similar to the one in the bottom-panel of FIG.~\ref{fig:plotQ} but shifted upwards and then intersecting the $\rho$-axis twice. This implies that the radicand inside the integral \eqref{eq:groundtruth}, proportional to $Q^2_R$, admits two distinct zeros. In terms of the variable $y$, defined in such a way that $\rho=a_f\sqrt{y}$, the integral becomes
\begin{flalign}\label{eqgeo1}
\phi-\phi_0=\frac{1}{2 a_f^2}\int_{\rho_0^2/a_f^2}^{\rho^2/a_f^2}\dd y\frac{(L_1L_5+a_f b y)}{(1+y)\sqrt{y(y-y_+)(y-y_-)}},
\end{flalign}
where $y_\pm=\rho_\pm^2/a_f^2$. In particular
\begin{widetext}
\begin{flalign}
y_{\pm}=\frac{b^2-a_f^2-L_1^2-L_5^2\pm\sqrt{\left(a_f+b-L_1-L_5\right)
   \left(-a_f+b+L_1-L_5\right) \left(-a_f+b-L_1+L_5\right)
   \left(a_f+b+L_1+L_5\right)}}{2
   a_f^2}.
\end{flalign}
At the end of the day \eqref{eqgeo1} can be written as
\begin{align}
\phi-\phi_0=&\frac{1}{\rho _- a_f}\Big[L_1 L_5 \left(\Pi \left(-\frac{a_f^2}{\rho _-^2};\sin ^{-1}\left(\frac{\rho _-}{\rho }\right)|\frac{\rho _+^2}{\rho _-^2}\right)-\Pi
   \left(-\frac{a_f^2}{\rho _-^2};\sin ^{-1}\left(\frac{\rho _-}{\rho _0}\right)|\frac{\rho _+^2}{\rho _-^2}\right)-F\left(\sin
   ^{-1}\left(\frac{\rho _-}{\rho }\right)|\frac{\rho _+^2}{\rho _-^2}\right)+\right.\nn\\
   &\left.+F\left(\sin ^{-1}\left(\frac{\rho _-}{\rho _0}\right)|\frac{\rho
   _+^2}{\rho _-^2}\right)\right)-b a_f \left(\Pi \left(-\frac{a_f^2}{\rho _-^2};\sin ^{-1}\left(\frac{\rho _-}{\rho }\right)|\frac{\rho
   _+^2}{\rho _-^2}\right)-\Pi \left(-\frac{a_f^2}{\rho _-^2};\sin ^{-1}\left(\frac{\rho _-}{\rho _0}\right)|\frac{\rho _+^2}{\rho
   _-^2}\right)\right)\Big],
\end{align}
\end{widetext}
where $\Pi(n,\phi|m)$ is the incomplete elliptic integral of the third kind and $F(\phi|m)$ is the incomplete elliptic integral of the first kind.

The exact function $P_\rho(\rho)$ is easily obtained by solving the constraint $\mathcal{H}(\rho)=0$, with $\mathcal{H}$ the  Hamiltonian given in \eqref{eq:Ham_plan} at fixed choice of the parameters. For their values reported at the beginning of this section we find
\begin{flalign}\label{eq:Prho_exact_overcriical}
  b=1.91\qquad   P_\rho (\rho) &= \pm \sqrt{\frac{6380 - 16381 \rho^2 + 10000 \rho^4}{1 + 200 \rho^2 + 10000 \rho^4}}\,,
\\[6pt] \label{eq:Prho_exact_critical}
b=1.90\qquad P_\rho (\rho) &= -\frac{20 (-4 + 5 \rho^2)}{1 + 100 \rho^2}\,,
\\[6pt] \label{eq:Prho_exact_subcritical}
     b=1.89\qquad P_{\rho}(\rho)&=-\sqrt{\dfrac{6420-15621\rho^{2}+10000\rho^{4}}{1+200\rho^{2}+10000\rho^{4}}}.
\end{flalign}
In the first equation the minus sign holds for the incoming phase, while the plus sign is valid for the outgoing phase. The three functions are plotted in FIG.~\ref{fig:Prho_exact}

\section{\label{APP:nonplangeo} Examination of the geometry of non-planar critical geodesics}
In this Appendix we investigate in more details what has been said in Section \ref{Non-Planar Case} for the nature of the non-planar critical geodesics, showing that in general they wrap around a spheroidal zone.\\ 
We solved the critical conditions $\mathcal{R}(\rho_c) = \mathcal{R}'(\rho_c) = 0$ in terms of $\zeta_c$ and $b_c$, getting \eqref{eq:critreg}. In order to find the critical radius $\rho_c$, we must work on the condition $P_\theta^2\geq 0$. This can written as
\begin{equation} \label{pos Ptheta}
\beta^2-\frac{\beta_\phi^2}{\sin^2\theta}-\sin^2\theta\geq0 ,\qquad \text{with} \qquad \beta=\frac{b}{a_f},\beta_{\phi}=\frac{b_{\phi}}{a_f}.
\end{equation}
If we introduce $\sin^2\theta=\frac{1}{2}(1-X)$, where $X=\cos(2\theta)$, the solutions of the associated equation are
\begin{equation}
X_\pm=1-\beta^2\pm\sqrt{\beta^4-4\beta_\phi^2}.
\end{equation}
It is very easy to show that the discriminant $\beta^4-4\beta_\phi^2$ is always positive in the critical regime \eqref{eq:critreg}. From the definition of $X$, we can write down the following chain of inequalities
\be
-1\leq X_-\leq X_+\leq 1.
\ee
The condition $X_-\geq-1$ is always false while $X_+\leq 1$ is always verified, so that the first non trivial condition is $X_+\geq-1$, which yields
\begin{equation}
\beta^2-2-\sqrt{\beta^4-4\beta_\phi^2}\leq0 \leftrightarrow \sqrt{b^4-4 b^2_\phi a^2_f} \geq b^2-2 a_f^2.
\end{equation}
Since we want to study the critical geodesics, we should evaluate this relation in the critical regime \eqref{eq:critreg}. In doing so, we have to distinguish two cases depending on the sign of the right hand side. It turns out that if $b^2_c< 2 a_f^2$, then from \eqref{eq:critreg} $2 \rho_c^2 + L_1^2+L_5^2 < 0$ and it is never satisfied, so the only case to consider is when $b^2_c > 2 a_f^2$. This brings to the following condition
\begin{equation}
    b^2_c \geq b^2_{\phi,c}+a_f^2.
\end{equation}
When we substitute the expressions \eqref{eq:critreg}, we have to distinguish two cases according to the sign of $\zeta_c$, with the aim of solving with respect to $\rho_c$
\begin{widetext}
$$a_f\zeta_c=\rho_c^2+a_f^2\Longrightarrow 0\leq\rho_c\leq\sqrt{a_f|L_1-L_5|-L_1L_5}\quad\text{for}\quad a_f\geq\frac{L_1L_5}{|L_1-L_5|}$$
\begin{equation}
    a_f\zeta_c=-(\rho_c^2+a_f^2)\Longrightarrow 
\begin{cases} \label{counter geo}
\sqrt{L_1L_5-a_f(L_1+L_5)}\leq\rho_c\leq \sqrt{L_1L_5+a_f(L_1+L_5)}, & 0<a_f\leq\frac{L_1L_5}{L_1+L_5} \\
0\leq\rho_c\leq \sqrt{a_f(L_1+L_5)-L_1 L_5}, & a_f>\frac{L_1L_5}{L_1+L_5}
\end{cases}
\end{equation}
\end{widetext}
The former case corresponds to a \textit{corotating} geodesic, the latter to a \textit{counterrotating} geodesic.\\
The inequality \eqref{pos Ptheta} has as solution $X_- \leq X \leq X_+$. Since $X_- < -1$ is always valid, then the first inequality is always satisfied. Hence it is left to solve $\cos(2 \theta) \leq X_+$, whose solution is (keeping in mind that $\theta \in [0,\pi]$)
\begin{flalign}\label{eq:interval_theta}
&\theta_{\text{bound}} \leq \theta \leq \pi-\theta_{\text{bound}} 
\nonumber\\[6pt]
&\text{with} \qquad \theta_{\text{bound}} = \frac{\arccos(X_+)}{2}.
\end{flalign}
Since $\phi$ varies from 0 to 2$\pi$, the geodesic describes a spheroidal zone. In order to understand what it is, we recall that the change of coordinates \eqref{eq:change_coord_3d} is a parametrization of an ellipsoid, at $\rho$ fixed, whose cartesian equation is
\begin{equation}\label{eq:oblate_spheroid}
    \frac{x^2}{\rho^2+a_f^2} + \frac{y^2}{\rho^2+a_f^2} + \frac{z^2}{\rho^2} = 1.
\end{equation}
By indicating in this context $a^2 = \rho^2+a_f^2, \, c^2 = \rho^2+a_f^2, \, d^2 = \rho^2$, then, if two of these quantities are equal, the ellipsoid is said to be a spheroid (or ellipsoid of revolution) and since $a=c>d$, then it is an oblate spheroid. As $\theta$ varies inside \eqref{eq:interval_theta}, then the surface is the one bounded by two planes that are parallel to the $x-y$ plane and intersect the oblate spheroid at $\theta=\theta_{\text{bound}}$ and $\theta = \pi-\theta_{\text{bound}}$. This region is called spheroid zone.\\ 
In the text we considered the case with $X_+ = 1$, corresponding to $\theta_{\text{bound}}=0$ and hence $\theta$ varies within its interval of definition, $\theta \in [0,\pi]$. For this reason such trajectory wraps around the complete oblate spheroid \eqref{eq:oblate_spheroid} with $\rho = \rho_c$. Furthermore, the condition $X_+ = 1$ brings to $\beta_{\phi,c} = 0 \leftrightarrow b_{\phi,c} = 0$ and so from the definition \eqref{eq:critreg} we gain $\zeta_c = - \frac{L_1 L_5}{a_f}$. This explains the first condition of \eqref{eq:final_params_nonplan} and hence the fact that this geodesic is counterrotating. The critical radius can be found from $X_+ = 1$ expressed in terms of $\rho_c$ and then solving it with respect to $\rho_c$ or also from the first equation of \eqref{eq:critreg} by substituting $\zeta^2_c$ with $(L_1 L_5/a_f)^2$ and solving with respect to $\rho_c$ (which is, by definition, non negative). At the end we get
\begin{equation}
\rho_c=\sqrt{L_1L_5-a_f^2},\quad 0<a_f\leq \sqrt{L_1L_5}
\end{equation}
which is exactly \eqref{eq:crit_radius}. Notice that this condition has to satisfy \eqref{counter geo}. It turns out that $\sqrt{L1 L5} > \frac{L1 L5}{L1+L5}$ for every $L_1,L_5 >0$ and it can be shown that if $a_f$ satisfies the first or second condition of \eqref{counter geo}, then the corresponding critical radius $\rho_c$ falls within the associated interval.

\section{\label{APP:PINN_separability} Results of the NN exploiting separability of the motion}

In this Appendix we show the results obtained using the NN by integrating the equations coming from separability, i.e. \eqref{eq:diffeq}. This exercise goes beyond the scope of this paper 
and therefore we will keep the discussion of the results short. We suggest to read first Section~\ref{results:planar}, where we have introduced most of the ingredients that will be reused in this Appendix. We do not present a comparison with other integration methods nor the errors committed by the NN, but we compare the predicted geodesic in the Cartesian plane with the ground truth and we highlight the main differences with the results and the strategies presented in the main text.

In this case we cannot speak about a Hamiltonian Neural Network, since the equations to solve are not the Hamilton equations and the energy conservation cannot be imposed at the level of the loss function. Still, the strategy presented in Section~\ref{sec:HNN} applies to any set of differential equations regardless they are Hamilton or different ones. We will label the results predicted by the Neural Network with \quote{NN}. 

In this case, the NN maps the input time $t$\footnote{Notice that now the equations of motion describe the dynamics in terms of the coordinate time $t$. Nonetheless, the trajectories in the Cartesian plane have to be the same as those that would be obtained by solving the Hamilton equations with respect to the affine parameter $s$. Of course, the distribution of the points along the geodesics is different, but the final result is the same if we take the number of such points much larger than 1 (as we have always done).} to an output layer with two nodes, that are denoted by $O_\rho$ and $O_\phi$. The solution to the system is constructed as

\begin{flalign}
& \widehat{\rho}(t,\textbf{w})=\rho_0 + f(t) O_\rho(t,\textbf{w}),
\\[4pt]
& \widehat{\phi}(t,\textbf{w})=\phi_0 + f(t) O_\phi(t,\textbf{w}).
\end{flalign}

The weights $\textbf{w}$ of the NN are found by minimizing only the loss function $L_\mathrm{dyn}$, since $L_\mathrm{energy}$ in this case is not defined. Therefore, there is no need to tune the parameter $\lambda$. 

We fix the parameters of the geometry to those already used in Section~\ref{results:planar} and we treat the three regimes, over-critical with $b=1.91$, critical with $b=b_c=1.9$ and sub-critical with $b=1.89$. $\rho_0$ and $\phi_0$ remain the same specified in Section~\ref{results:planar} for the three situations. For all the cases we have employed a neural network  consisting in 2 hidden layers with 64 neurons per each. The training is performed over $3\times 10^6$ epochs by varying the learning rate as done in Section~\ref{results:planar}; furthermore the training interval is $\mathcal{T}_\mathrm{train}$ (different for each $b$) and the absence of overfitting is explicitly checked by validating the loss function on $\mathcal{T}_\mathrm{val}$ at the end of the training. 

We move now to to present the results individually for the three cases.

\subsection{\label{subsec:separability_overcritical} Over-critical case}
We set $b=1.91$  and train the neural network over the interval $\mathcal{T}_\mathrm{train}=[0,15.3157]$ divided in $N=800$ points. The loss function is shown in the left-panel of FIG.~\ref{fig:separability_loss}. The reason behind such a particular choice for the upper bound is that the particle motion has an incoming and an outgoing phase. The former is given by considering the negative sign in \eqref{eq:diffeq} and corresponds to a decrease from $\rho_0$ to the minimum value $\rho_+=1.000138$, where $\dd \rho / \dd t \rvert_{\rho=\rho_+}=0$. The latter is given by taking the positive sign and it corresponds to an increase from $\rho_+$ to $\infty$. We obtain the outgoing solution just from the reflection of the incoming one, since the system is symmetric under $\rho\mapsto -\rho$. In order to do so, we need the time $t_+$ such that $\rho(t_+)=\rho_+$ and we obtain the value $t_+=15.3157$ from the $\texttt{RK45}$ method. Notice that for $t>t_+$ the radicand in \eqref{eq:diffeq} becomes negative, we are therefore forced to stop computing the prediction at this time. In the Hamilton case presented in Section~\ref{sec:planar_overcritical} we do not need to choose carefully $\mathcal{T}_\mathrm{train}$ since the information about the incoming and outgoing motion is automatically incorporated into the Hamilton equations.

The predicted $\rho(t)$ and $\phi(t)/2\pi$ are shown in the left panel of FIG.~\ref{fig:separability_variables}, both for the incoming and outgoing phases. The value $t_+$ is highlighted by a vertical black line and we find $\widehat{\rho}(t_+)=1.00020$, which is at $0.006$\% away from $\rho_+$. The behavior in the coordinate time $t$ of the two variables is completely analogous to that with respect to the affine parameter $s$ obtained in FIG:~\ref{fig:Hamilton_overcritical_variables}. 

The geodesic in the Cartesian plane is obtained by using \eqref{eq:cartesian_map} and is compared to the ground truth in the left-panel of FIG.~\ref{fig:separability_cartesian}. The predicted trajectory reproduces perfectly the ground truth.

\subsection{\label{subsec:separability_critical} Critical case}

In order to obtain the critical regime we set $b=b_c=1.9$.  In this case, the coordinate $\rho$ of the particle has to decrease from $\rho_0$ up to reach the critical radius $\rho_c=0.8944$ and then it has to remain on this value. Since there is no growing, we only consider the equation with the negative sign. In this case the radicand in \eqref{eq:diffeq} has only one root, at $\rho_c$, and then it can we written as $(\rho^2-\rho_c^2)^2$. Therefore, it presents no problems and the solution can be computed at arbitrarily large times.   We have trained the neural network in the interval $\mathcal{T}_\mathrm{train}=[0,140]$ and the loss function is shown in the central-panel of FIG.~\ref{fig:separability_loss}. 

The predicted functions $\rho(t)$ and $\phi(t)$ are shown in the central-panel of FIG.~\ref{fig:separability_variables}. The value $\rho_c$ is represented by the black line in the top-panel and, as we can see, $\rho(t)$ remains stable on it. Alike the Hamilton case, the NN confirms to be able to provide stable and accurate predictions for trajectories over long time-scale even without the need to impose the conservation of the energy.

The resulting geodesic in the Cartesian plane is shown in the central-panel of FIG.~\ref{fig:separability_cartesian} and it accurately reproduces the ground truth.

\subsection{\label{subsec:separability_subcritical} Sub-critical case}

\begin{figure*}[h!]
    \centering
    \includegraphics[width=0.32\textwidth]{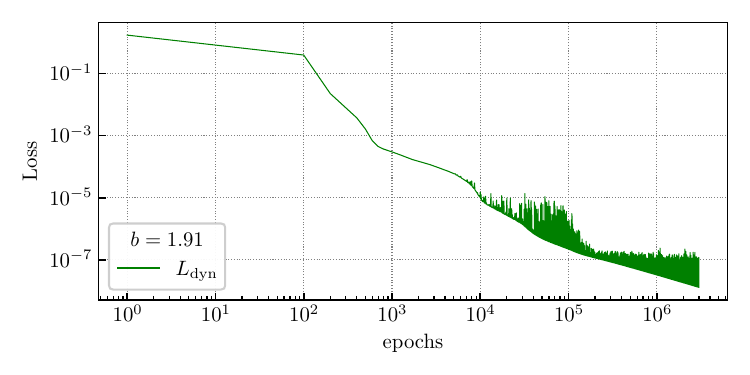}
    \includegraphics[width=0.32\textwidth]{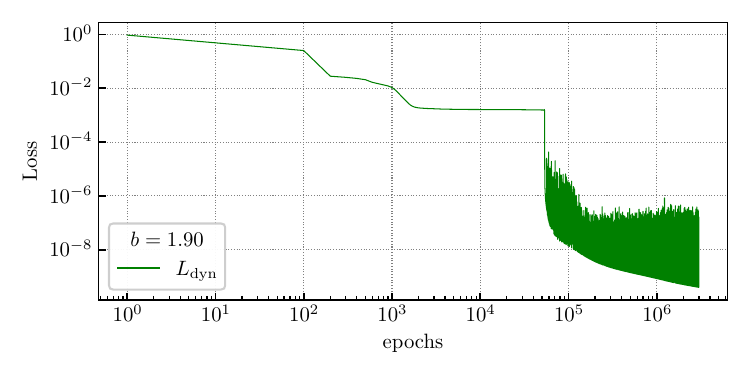}
    \includegraphics[width=0.32\textwidth]{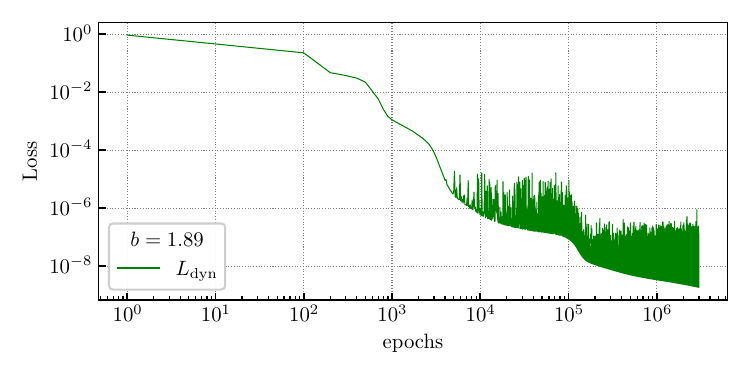}
    \caption{
    \label{fig:separability_loss}
    \footnotesize{
    Dynamical loss functions corresponding to the equations of motion  obtained by exploiting the separability of the geometry. From left to right: over-critical case with $b=1.91$, critical case with $b=b_c=1.9$ and sub-critical case with  $b=1.89$. In all the cases the number of points is $N=800$. 
    }
    }
\end{figure*}

In order to obtain the sub-critical regime we set $b=1.89$ and we train the neural network in the interval $\mathcal{T}_\mathrm{train}=[0,37]$ divided in $N=800$ points.  The loss function is shown in the right-panel of FIG.~\ref{fig:separability_loss}.

In Section \ref{subsec:sub-critical} we pointed out the difficulty in solving the Hamilton equations due to steep decrease of $P_\rho(s)$ approaching the time $\bar s$ and we proposed two different strategies. In this case, there is no dynamical equation for $P_\rho$ thanks to the separability and such difficulty does not arise. In addition, the radicand in \eqref{eq:diffeq} is always positive definite. 

\begin{figure*}[h!]
    \centering
    \includegraphics[width=0.32\textwidth]{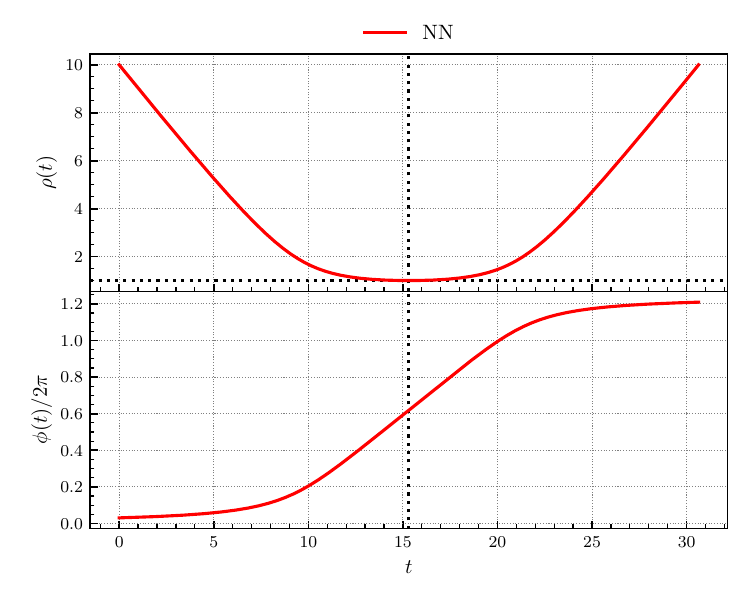}
    \includegraphics[width=0.32\textwidth]{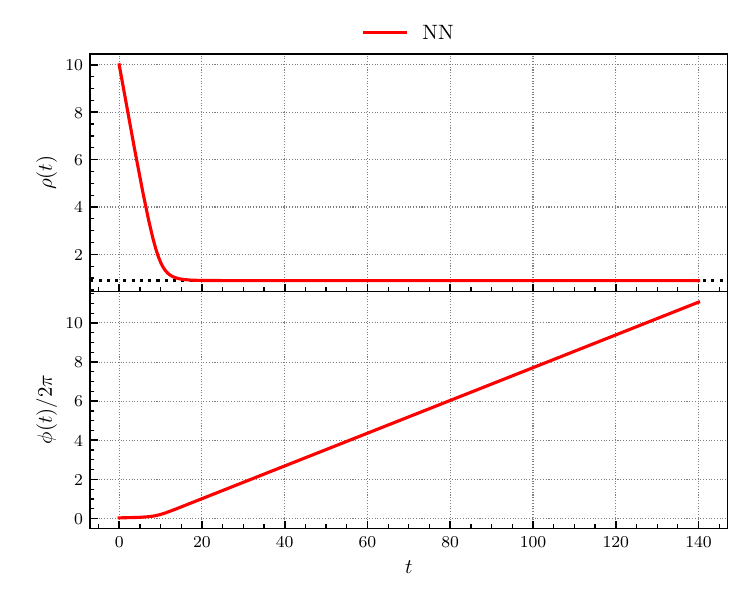}
    \includegraphics[width=0.32\textwidth]{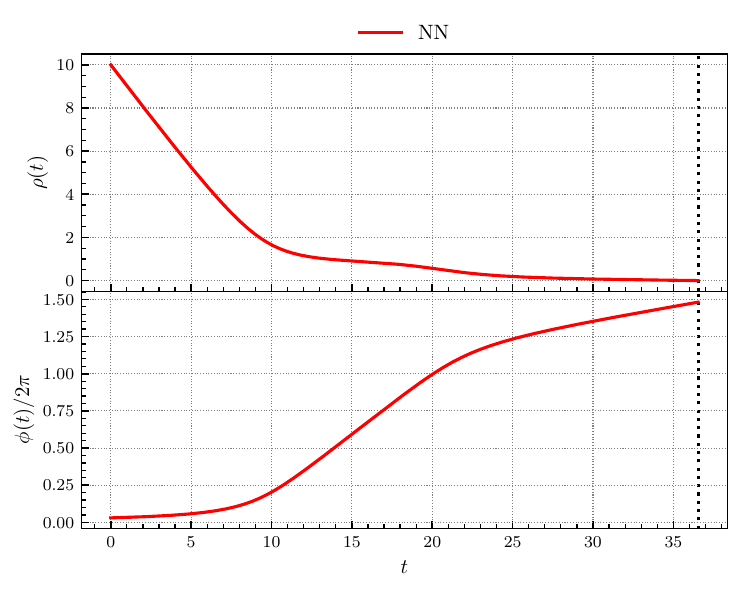}
    \caption{
    \label{fig:separability_variables}
    \footnotesize{
   Predicted $\rho(t)$ (top-panels) and $\phi(t)/2\pi$ (bottom-panels)
 corresponding to the equations of motion obtained by exploiting the separability of the geometry. \emph{Left-panel}: Over-critical case with $b=1.91$. The vertical dashed line corresponds to $t_+=15.3157$. The functions for $t>t_+$ (outgoing motion) are obtained from the reflection of the incoming phase. The horizontal dashed line in the top-panel represents $\rho_+=1.000138$. \emph{Central-panel}: Critical case with $b=b_c=1.9$. The horizontal dashed line in the top-panel represents $\rho_c=0.8944$. \emph{Right-panel}: Sub-critical case with $b=1.89$. The vertical dashed line corresponds to the collision time $\bar t $= 36.5417.
    }
    }
\end{figure*}

The particle trajectory still describes an incoming and outgoing motion and the transition is determined by the minimum value $\rho=0$ that occurs at a given time $\bar t$. We follow the same strategy of Section~\ref{sec:splitting} of determining $\bar t$ by interpolation of $\rho(t)$ with $0$ (recall FIG.~\ref{fig:prediction_interpolation}) and we obtain the outgoing phase by reflection of the incoming one. The functions $\rho(t)$ and $\phi(t)/2\pi$ are shown in the right-panel of FIG.~\ref{fig:separability_variables} and from the aforementioned procedure we find $\bar{t}=36.5417$, which is marked by the vertical black line.

\begin{figure*}[h!]
    \centering
    \includegraphics[width=0.32\textwidth]{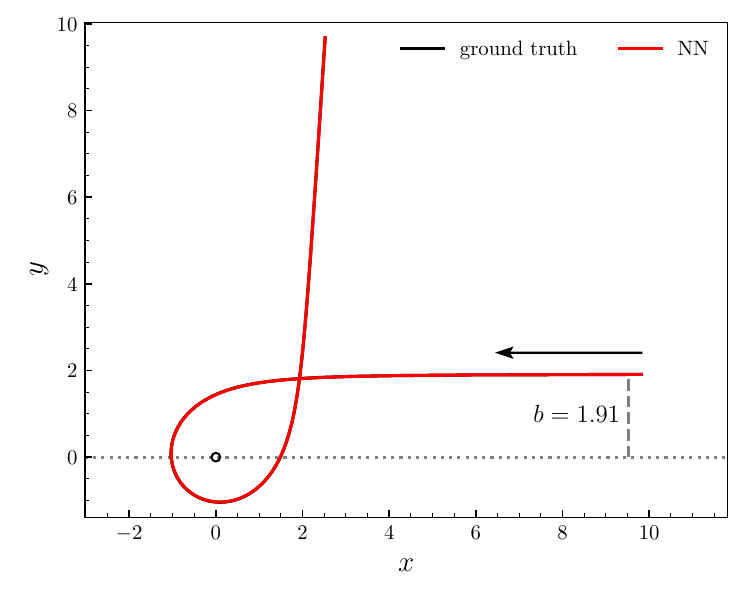}
    \includegraphics[width=0.32\textwidth]{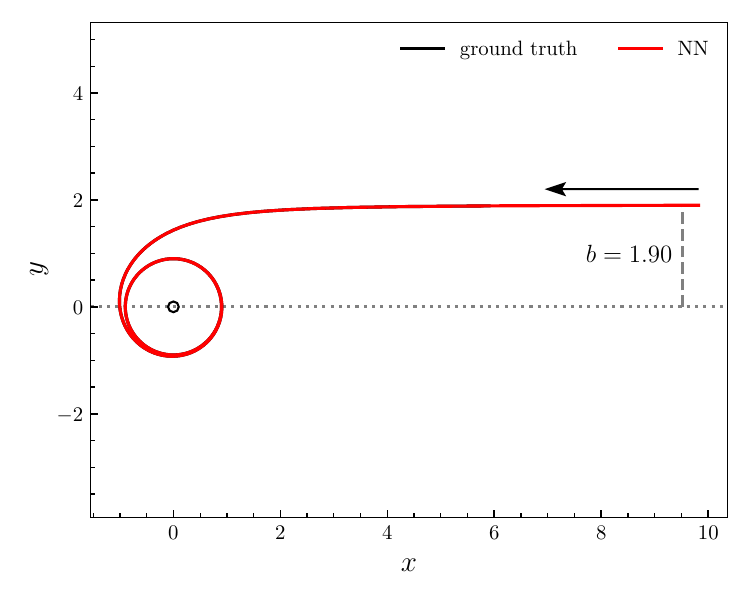}
    \includegraphics[width=0.32\textwidth]{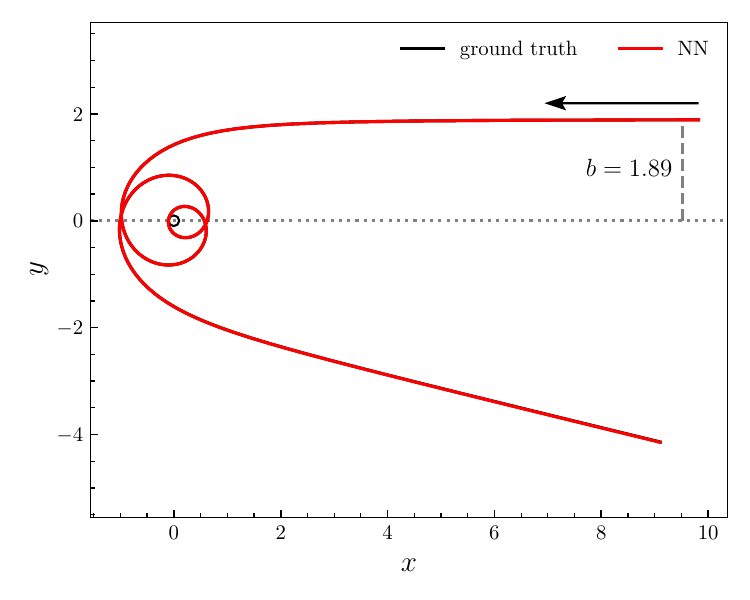}
    \caption{
    \label{fig:separability_cartesian}
    \footnotesize{
    Comparison between the geodesic predicted by the NN and the ground truth in the Cartesian plane. From left to right: over-critical case with $b=1.91$, critical case with $b=b_c=1.9$ and sub-critical case with $b=1.89$. The small black circle has radius $a_f=0.1$ and it represents the fuzzball.
    }
    }
\end{figure*}

Finally, the full geodesic projected onto the Cartesian plane is shown in the right-panel of FIG.~\ref{fig:separability_cartesian} and it is in perfect agreement with the ground truth.


\FloatBarrier

\bibliographystyle{unsrt}
\bibliography{bibHNN}

\end{document}